%% file: main1.tex
\documentclass[12pt,oneside]{book}

\usepackage{setspace}
\usepackage{amsmath}
\usepackage{graphicx}

\usepackage{phdthesis}
\usepackage{multirow}
\usepackage{wrapfig}
\usepackage{ifthen}
\usepackage{subfig}
\usepackage{hyperref}

\begin{document}

\pagenumbering{roman}

\input{frontpage}
\pagebreak

\input{abstract}

\cleardoublepage \pagebreak
\include{chapters/chap1}

\include{chapters/chap2}

\include{chapters/chap3}

\include{chapters/chap4}

\include{chapters/chap5}

\include{chapters/conclusions}

\input{acknowledgement}

\include{Bib}
\end{document}

%% file: frontpage.tex
\thispagestyle{empty} \linespread{1.5}
\begin{center}
\Huge{Phase Transition in a Bond Fluctuating Lattice Polymer}
\end{center}
\mbox{}
\begin{center}
{\textbf{{\large\\ Hima Bindu Kolli\\}}}
{\textbf{{\large and\\}}}
{\textbf{{\large  K.P.N.Murthy\\}}}
\end{center}

\mbox{}

\begin{figure}[h]
\begin{center}

\includegraphics[width=3.5cm]{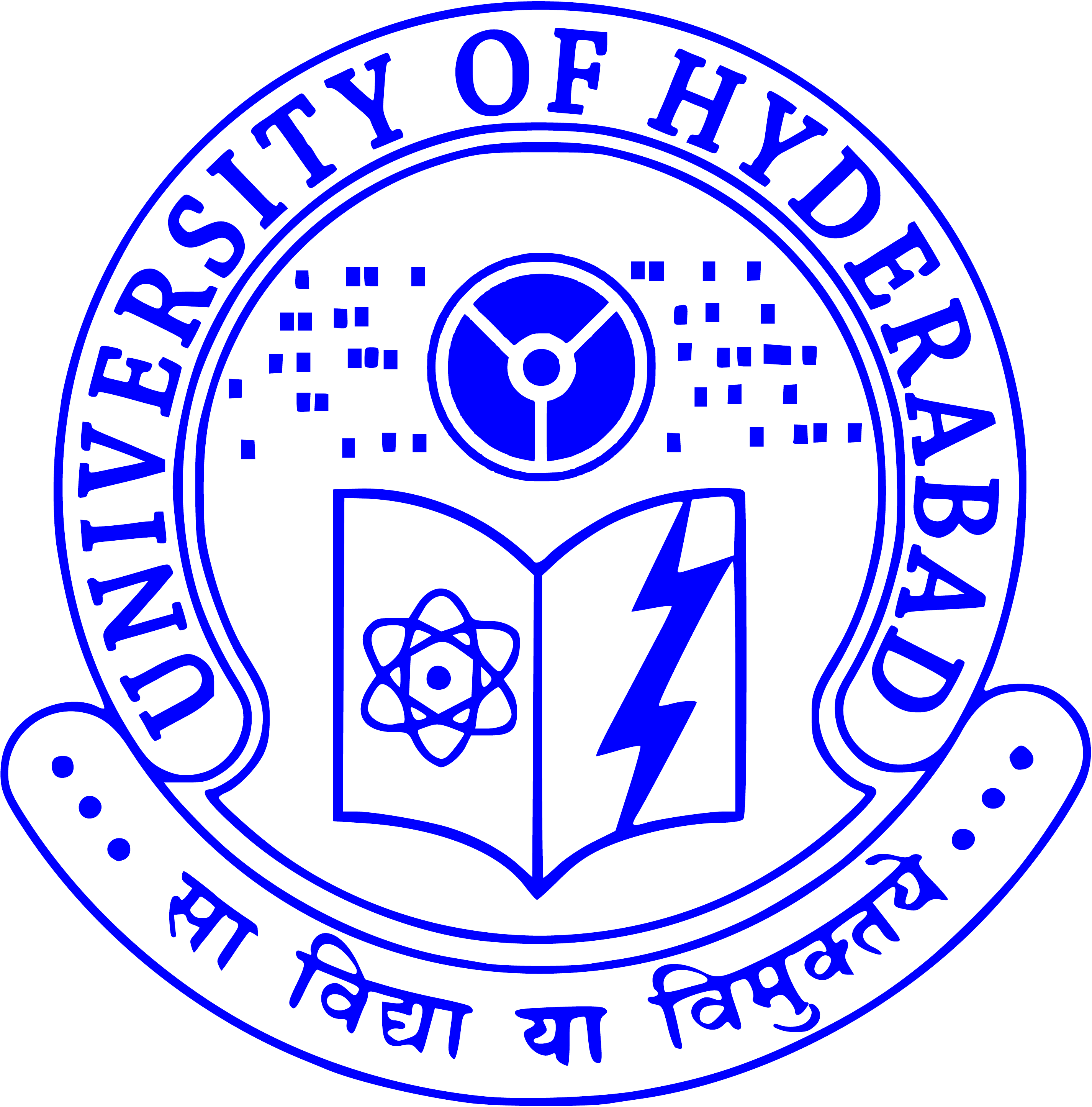}
\end{center}
\end{figure}
\mbox{}
\begin{center}
\large  School of Physics, 
University of Hyderabad\\Hyderabad 500046\\November 2011
\end{center}

%% file: abstract.tex
\begin{center}
\textbf{\large ABSTRACT}
\end{center}
\vskip35pt This report deals with phase transition in Bond
Fluctuation Model (BFM)  of a linear homo  polymer on a two
dimensional square lattice. Each monomer occupies a unit cell of
four lattice sites. The condition that a lattice site can at best be
a part of only one monomer ensures self avoidance and models
excluded volume effect. We have simulated polymers with number of
monomers ranging from $10$ to $50$ employing Boltzmann and
non-Boltzmann Monte Carlo simulation techniques. To detect and
characterize  phase transition we have investigated heat capacity
through energy fluctuations, Landau free energy profiles and
Binder's fourth cumulant. We have investigated (1) free standing
polymer (2) polymer in the presence of of an attracting wall and (3)
polymer confined between two attracting walls.  In general we find
there are two transitions as we cool the system. The first occurs at
relatively higher temperature. The polymer goes from an extended
coil to a collapsed globule conformation. This we call collapse
transition. We find that this transition is first order. The second
occurs at a lower temperature in which the polymer goes from a
collapsed phase to a very compact crystalline phase. This transition
is also discontinuous. We find that in the presence of  wall(s) the
collapse transition occurs at lower temperature compared to a free
standing polymer.

%% file: chapters/chap1.tex
\pagenumbering{arabic}
\chapter{INTRODUCTION}
\section{Polymers}
A polymer is a  long chain of several simple atomic groupings called
{\it monomers}. These monomers are held together by chemical
bonds \cite{P.J.Flory1971}. The process of forming a long chain of
monomers is called {\it polymerization}. The number of monomers in a
polymer is denoted by $N$ and is called the degree of
polymerization. If the available bonds for a monomer are two, then
the process of polymerization leads to a linear polymer. If the
available bonds are more than two, then it leads to a branched and
cross linked polymer. A polymer with same type of monomers is called
a {\it homopolymer} and that with different types of monomers is
called a  {\it heteropolymer}. In this report we shall deal with a
linear homopolymer.
\section{Solvent and Temperature Effect}
 A polymer can exist in several conformations \cite{M.Rubinstein2003}.
  The conformation  has a direct bearing on its physical
properties. The conformational statistics of a polymer chain depends
on the solvent quality. For a good solvent, interactions between
monomers and solvent molecules are more favourable than interaction
between non-bonded monomers. Hence in a good solvent, polymer segments would
prefer to be surrounded by the solvent, leading to a swollen coil
conformation. We call this as extended phase. For a bad solvent, the
interaction between non-bonded monomers are more favourable leading
to a compact globule conformation. We call this as collapsed phase. A good
parameter that quantifies the size of the polymer is the  radius of gyration
\footnote{Radius of gyration is defined as follows. Find out
center of mass of a polymer:$$\overrightarrow{R}=\sum_i
\frac{\overrightarrow{r^i}}{N},$$ where $\overrightarrow{r^i}$ is the
position vector of the $i^{th}$ monomer in the lattice polymer and N
is the number of monomers. Calculate the distance of each monomer
from the center of mass
\begin{eqnarray}
s_i=\sqrt{((\overrightarrow{r^i})_x-(\overrightarrow{R})_x)^2+((\overrightarrow{r^i})_y-(\overrightarrow{R})_y)^2}
\end{eqnarray}
Radius of gyration is given by
\begin{eqnarray}
R_g = \sqrt{\frac{1}{N}\sum_i^N s_i^2}
\end{eqnarray}
} or end-to-end distance.
This quantity will be relatively large for extended conformation and small for collapsed conformations

The quality of solvent is often parameterized by temperature. Low
temperatures correspond to a poor solvent and high temperatures
correspond to a good solvent. As the temperature decreases, a
transition occurs from an extended coil phase to collapsed globule
phase. At very  low temperatures, another transition from globular
phase to crystalline phase takes place if the cooling rate is slow.
Upon fast cooling  glass formation becomes possible at these low
temperatures.

Coarse grained models are often employed in polymer studies. A
coarse grained model, see {\it e.g.} \cite{C.Vanderzande1998,
M.Doi1986, S.M.Bhattacharjee1992}, treats a group of chemical
monomers as a {\it bead} (effective monomer) ignoring the
microscopic degrees of freedom,  which are invariably present; it
retains only the most basic features common to all polymers of the
same chain topology. Such  a model incorporates features such as
chain connectivity, excluded volume effect and monomer-monomer
interactions.

In this work we consider lattice model of  a linear homopolymer. For
an introduction to lattice models of polymers, see {\it e.g}
 \cite{C.Vanderzande1998}. We restrict our attention to a polymer on
a two dimensional square lattice. The trail of a self avoiding
random walk provides a model of a polymer conformation. In this
section, we give a brief description of random walk (RW), self
avoiding walk (SAW) and Interacting self avoiding walk (ISAW).

\section{Random Walk}
A random walk starts from an arbitrary lattice point which is  taken as origin.
It selects one of
the four nearest neighbour sites randomly and with equal
probabilities and steps into it. This process is iterated and we get
a random walk of desired length.
 Thus a
simple random walk generates a chain that can intersect as well as
fold on to itself. Fig. (1.1) shows a random walk of 22 steps.

In a polymer, we have the so-called excluded volume effect. This is
 also known as hard core
repulsion. In a lattice model, the excluded volume effect can be easily
implemented by demanding that a lattice site can at best be occupied by
a single monomer. This leads us to  self avoiding random walk.

\section{Self Avoiding Walk (SAW)}
A self avoiding walk is a random walk that does not visit a lattice
site it has already visited. A self avoiding walk
\cite{N.Madras1993} is an athermal\footnote{Consider an
ensemble of random walks. To each walk we attach an energy as
follows. If the walk intersects itself or if segments of walks
overlap then we say the walk has energy $E=\infty$. The Boltzmann
weight $\exp(-E/k_B T)$ of the walk is zero. If the walk is self
avoiding  then its energy is zero. The corresponding Boltzmann
weight is unity. Thus the Boltzmann weight is either zero when it is
not self avoiding and unity when it is. The Boltzmann weight is not
dependent on temperature. Such an energy is called athermal energy.
In fact, there is no energy scale and the statistical mechanics of
self avoiding walks is completely determined by entropy.} walk.
A self
avoiding walk models a polymer at very high temperatures where
segment-segment interactions are negligible. With lowering of
temperature, the segment - segment interaction comes into play. To
model  segment - segment  interaction we consider Interacting Self
Avoiding Walk (ISAW) \cite{J.Mazur1968}.

\begin{figure}[htp]
\begin{center}
\includegraphics[width=18cm,height=10cm]{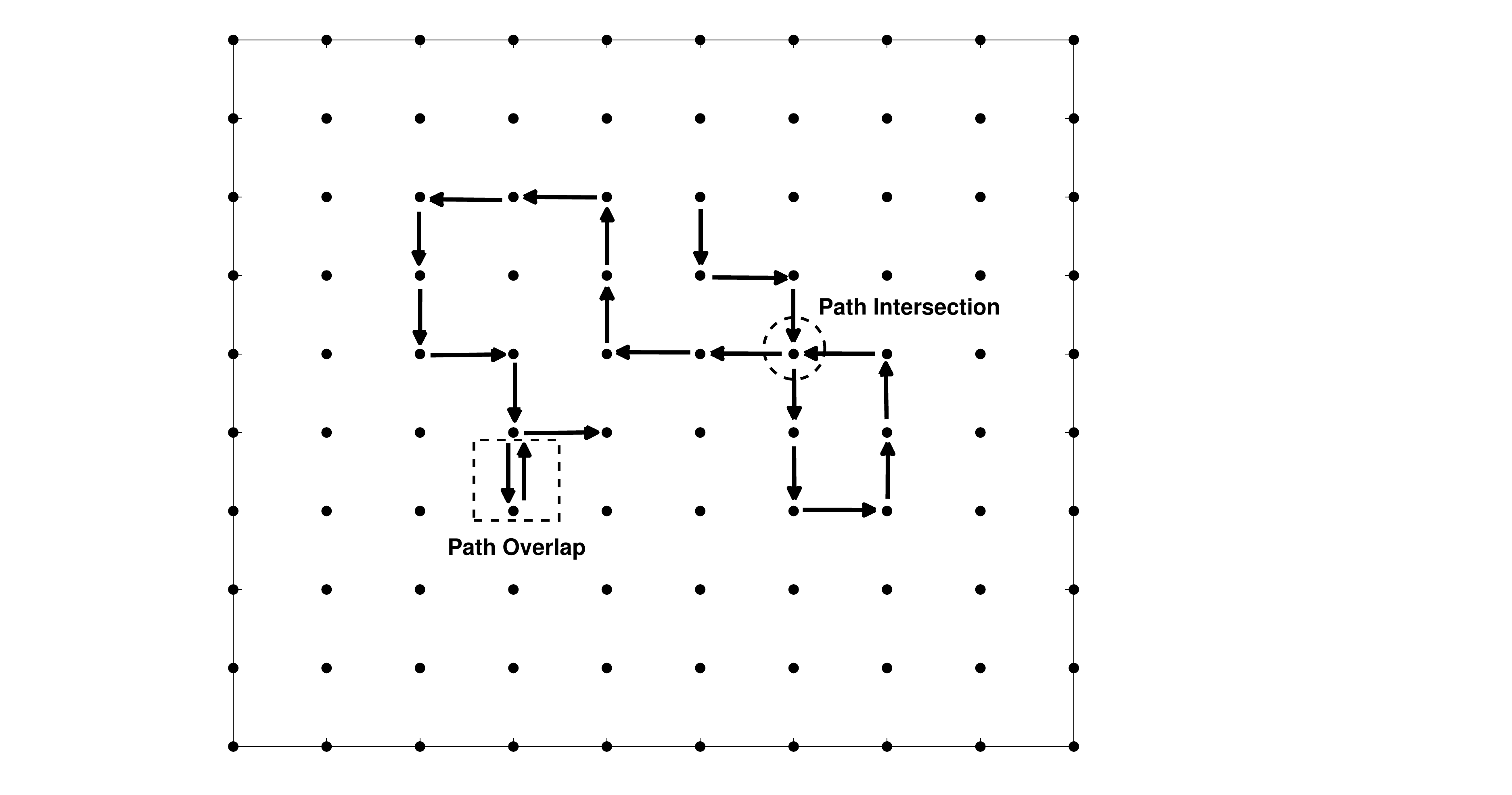}
\caption{{\small A simple random walk of $22$ steps involving intersections
and overlaps of the path.}}
\end{center}
\end{figure}

\section{Interacting Self Avoiding Walk (ISAW)}
We assign an energy $\epsilon$ to every pair of occupied
nearest-neighbour sites which are not adjacent along the walk. Such
a non-bonded nearest- neighbour (nbNN) pair gives rise to an nbNN
contact or simply  a contact. If $ \epsilon\ <\ 0$ the interaction
is attractive. If $\epsilon\ >\ 0$ the interaction is repulsive.
Thus a conformation with $m$ contacts has energy $E=m\epsilon$. A
SAW with energy assigned in this fashion is called an Interacting
Self Avoiding Walk (ISAW). A typical ISAW on a square lattice of
walk length $N=26$ with total number of nbNN contacts $m=6$ is shown
in Fig. (1.2). In this work,  our aim is to study  phase transition
from an extended to a collapsed phase. Hence we take  $\epsilon$ to
be negative. Without loss of generality we set $\epsilon=-1$. In
other words we measure energy in units of $\epsilon$

\begin{figure}[htp]
\begin{center}
\hglue -20mm\includegraphics[width=18cm,height=10cm]{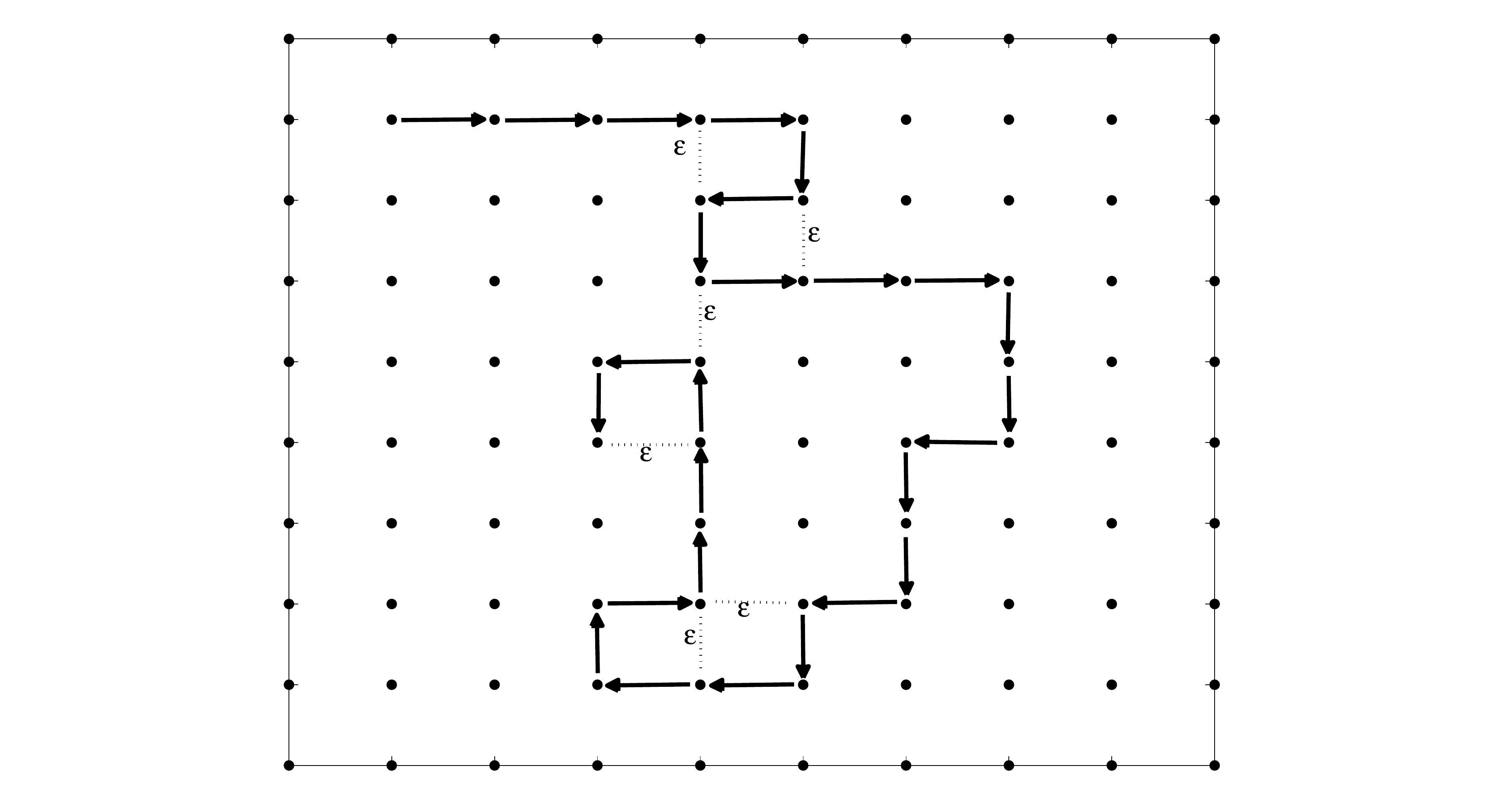}
\caption{{\small A interacting Self Avoiding walk (ISAW) that models a
lattice polymer having $27$ monomers. Two monomers on nearest
neighbour lattice sites but not  connected by a bond constitute a
non bonded Nearest Neighbour (nbNN) contact pair or simply a
contact. The contacts  are indicated  by dotted lines.}  }
\end{center}

\end{figure}
\section{Algorithms to generate ISAW}
In the study of lattice polymers employing self avoiding walks three
approaches have so far been tried out. The first is called
stochastic growth algorithms which include kinetic growth walks. The
second is based on the method proposed by Sokal involving local
changes in a fully grown polymer. The third is called Bond
Fluctuation Model (BFM). We discuss the first two approaches briefly
below. The third model is discussed in detail in the next chapter.
\subsection{Growth algorithm}
An early stochastic growth method on a lattice employed a blind ant
algorithm, where the ant moves blindly to one of the nearest
neighbour sites with equal probability. At any growth step, if the
selected site is already occupied by a monomer then we discard the
whole walk and start all over again. This ensures that all the
random walks are generated with equal probability given by
$(1/4)\times(1/3)^{N-1}$, where $N$ is the number of steps in the
walk. Notice that an $N$-step walk generates a polymer with $N+1$
monomers.
 From the lattice polymers generated by the
blind ant, we can construct micro canonical ensembles by grouping then  in
terms of energies; we can also carry out canonical ensemble averages by
 attaching
a Boltzmann weight to each walk based on its energy. A major problem
with blind ant is sample attrition {\it i.e.}, only a small fraction
of starting chains are finally accepted. Most of the walks
overlap or intersect  before reaching the required chain length; in other words
the walks get terminated early most of the time.
This is called the problem of sample attrition.
Longer the walk, the more is the problem posed by attrition.
Partial remedy is Rosenbluth - Rosenbluth (RR) walk based on
myopic ant \cite{M.N.Rosenbluth1955}. The ant selects one of the
unoccupied nearest neighbour sites randomly and with equal
probability. Sample attrition is considerably reduced, though not
eliminated. Trapping does still occur.

A major problem with a myopic ant is the following.  The walks generated are not
all equally probable. Hence we need Rosenbluth Rosenbluth (RR)
 weights, $W_{RR}$  for calculating
micro canonical ensemble averages. We need RR weights as well as  Boltzmann
weights, see below, for calculating canonical ensemble averages. RR
weight usually fluctuates wildly from one walk to the other.  Besides sample attrition is present,
though less; hence long walks remain difficult to generate.

Kinetic Growth walks (KGW) \cite{I.Majid1984} are RR walks, where,
we ignore RR weights\footnote{This is equivalent to setting RR weight to unity}. The justification is that we are looking at a
polymer that grows faster than it could relax. KGW has been shown to
belong to the same universality class as SAW \cite{KL}. More recently KGW has
been generalized where one selects the available unoccupied sites on
the basis of the number of contacts it would establish
\cite{SLN2001-1,S.L.Narasimhan2001}. For example if the move would increase
the contacts by one then the local Boltzmann weight for that move is
$\exp(-\beta)$. The site for placing the monomer is selected on the
basis of probability constructed from the local Boltzmann weights.
This is called Interacting Growth walk (IGW) and has been shown to
belong to same universality class as ISAW \cite{SLN2001-1}.

Grassberger \cite{P.Grassberger1997} has proposed PERM\footnote{PERM stands for
Pruned, Enriched Rosenbluth Model} algorithms to extract equilibrium
properties from  KGW ensembles. Ponmurugan {\it et al}
\cite{Ponmurugan} have proposed flat-IGW algorithm to calculate
equilibrium averages from IGW ensembles. Thus kinetic and interacting
growth walks
have proved  useful in polymer studies.

\subsection{Sokal algorithm}
The second approach is provided by A D Sokal \cite{A.D.Sokal1995}. We
start with a lattice polymer of a given length. We make local
changes employing moves like pivoting about a chosen monomer,
cranking, rotating about an axis {\it etc.}  in a self avoiding fashion
 to generate a trial configuration. Figs. (1.3, 1.4) depict
 some possible moves. We employ standard
Metropolis algorithm to accept/reject the trial configuration. We
generate a Markov chain of polymer conformations and the asymptotic
part of the Markov chain corresponds to a canonical ensemble.

We calculate the desired properties by averaging over the canonical
ensemble. A problem with Sokal's algorithm is that local changes are
often difficult to make and time consuming especially for long
polymers.

\subsection{Bond Fluctuation Model}
The third approach is the bond fluctuating model \cite{Carmesin1988}.
It combines the advantages of both the growth  and the Sokal
algorithms. In this model we start with a possible polymer
conformation of desired length like in Sokal's algorithm. Hence
there is no problem of sample attrition. A monomer is chosen and
moved to its nearest neighbour site in a self avoiding way. The move
is thus simple and local, like in growth algorithm. In this model,
however,  the bond length fluctuates. Since bond fluctuation model
forms the backbone of this work, it is described separately
and in details in next chapter.

\begin{figure}[hpb]
\begin{center}
\includegraphics[height=5cm]{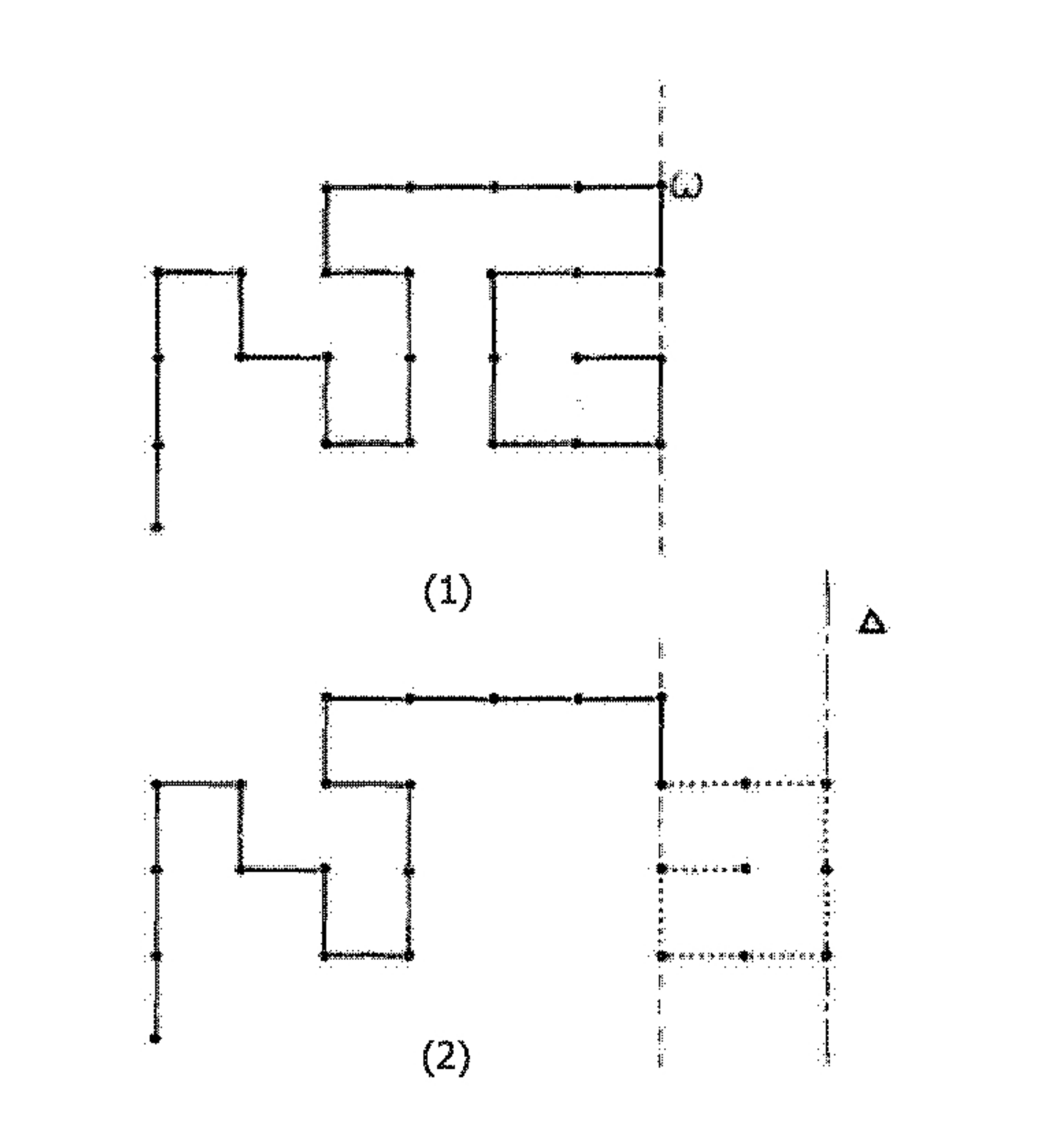}
\caption{{\small A possible move by pivoting, in Sokal's algorithm
(1) Pivot at $\omega$, by reflecting through the dashed line. (2)
The resultant conformation showing the reflected part in dotted
line. (We can now pivot at $\Delta$)}}
\includegraphics[height=5cm]{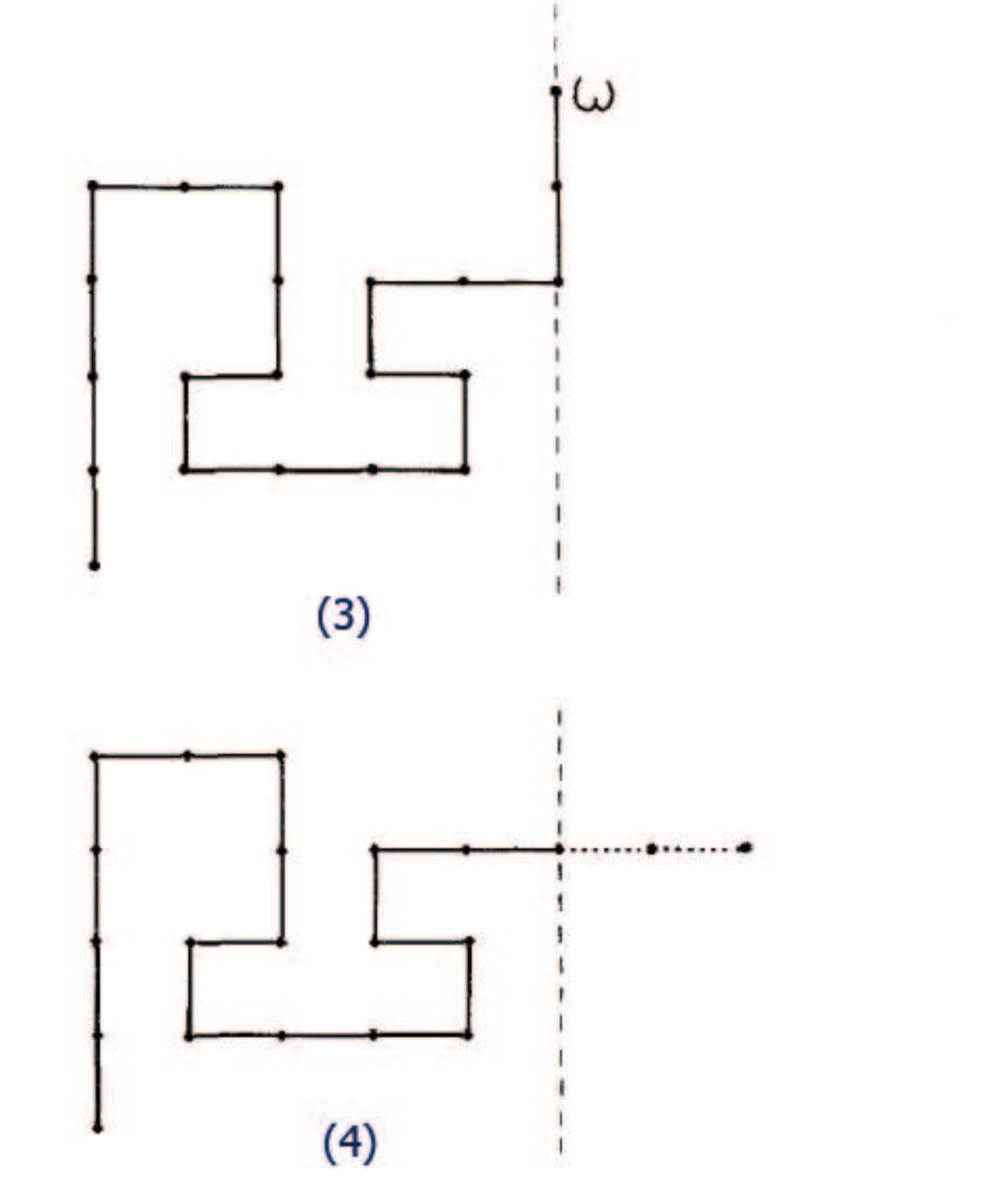}
\caption{{\small A possible move employing rotation, in Sokal's
algorithm (3) Rotate $90^{o}$ at $\omega$. (4) The result.}}
\end{center}
\end{figure}

%% file: chapters/chap2.tex
\chapter{Bond Fluctuation model}
Bond fluctuation model is a lattice model for simulating  polymer
systems. It is useful for obtaining static and dynamic properties of
polymers. According to this model, a trial conformation is generated
by moving a randomly selected monomer from its current location to a
location one lattice spacing away,  along one of the
possible directions chosen randomly and with equal probability. During
the move we must ensure that bond crossing does not occur and self
avoidance condition is satisfied. This model retains the advantages
of both growth algorithms and Sokal's algorithm, see below.
\newline
Like in a  growth algorithm, the move is local {\it i.e.}; a trial
conformation is generated by moving  a randomly
selected monomer.
\newline
Like in Sokal's algorithm, we start with a polymer of required size. A trial conformation,
generated by local rule, is accepted or rejected by Metropolis algorithm. Thus we generate a
Markov chain of lattice polymer conformations; asymptotic part of the chain  constitutes
a canonical ensemble from which the desired macroscopic properties can be calculated.
\section{Basic description of the Bond Fluctuation Model (BFM)}
On  a two dimensional square lattice,  each monomer occupies four
lattice sites of a unit cell \cite{Carmesin1988,K.Binder1995}. Each
lattice site can at best be part of only one monomer by virtue of
self avoidance condition. We implement it on a square lattice with
lattice constant unity. Let $l$ denotes length of a bond. Minimum
bond length is $l=2$. A bond length less than $2$ violates self
avoidance condition. We restrict to bond length less than $4$. This
condition ensures that no bond crossing takes place. The possible
bond lengths\footnote{In three dimensional cubic lattice, each
monomer occupies eight lattice sites of a cube. Bonds may have
lengths ranging between $2$ and $\sqrt{10}$ but bond vectors of the
type $(2,2,0)$ are excluded to avoid bond crossing} less than $4$ are:
$2,\ \sqrt{5},\ \sqrt{8},\ 3,\ \sqrt{10}$ and $\sqrt{13}$.
Figure (2.1) shows a bond fluctuating lattice polymer with all
possible bond lengths. In the present work, we
restrict ourselves to four site model\footnote{In a four site occupancy model,
a monomer occupies four lattice sites.} on a two dimension square lattice.

Instead of four site lattice model one  can  consider one site
model {\it i.e.}, a monomer occupies single lattice site with
possible bond lengths between $1$ and $\sqrt{2}$. But in this model
there exists certain compact conformations that do not evolve; this
renders the model non-ergodic\footnote{For a model to be ergodic we
should be able to reach any conformation from any other conformation
through a series of moves.}.

\newpage
\vglue 30mm
\begin{figure}[htp]
\begin{center}
\includegraphics[height=10cm]{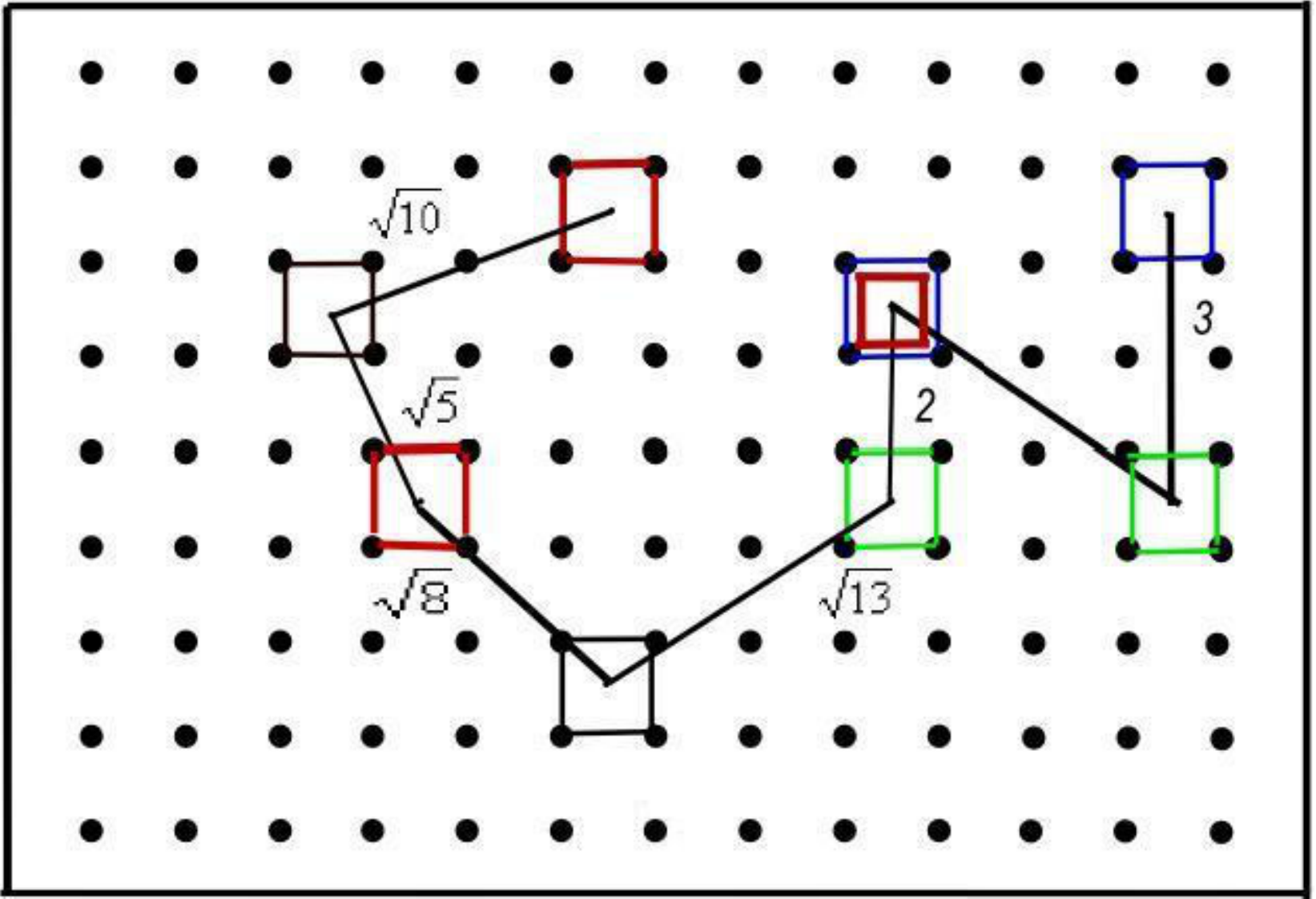}
\caption{{\small A bond fluctuating lattice polymer with all
possible bond lengths less than $4$ lattice units is depicted. Monomers shown in
same color are non bonded nearest neighbours. Two monomers, not connected
by a bond, but separated by a distance $d\ \le\ \sqrt{13}$ constitute a non-bonded
Nearest Neighbour (nbNN) contact. }}
\end{center}
\end{figure}
\vglue 8mm Figure (2.2) shows trapped conformations with monomers
occupying single lattice site. The same conformations with possible
moves in case of four site lattice model are also shown
 in the same figures.
\newpage

\begin{figure}[htp]
\begin{center}
\includegraphics[width=15cm,height=11cm]{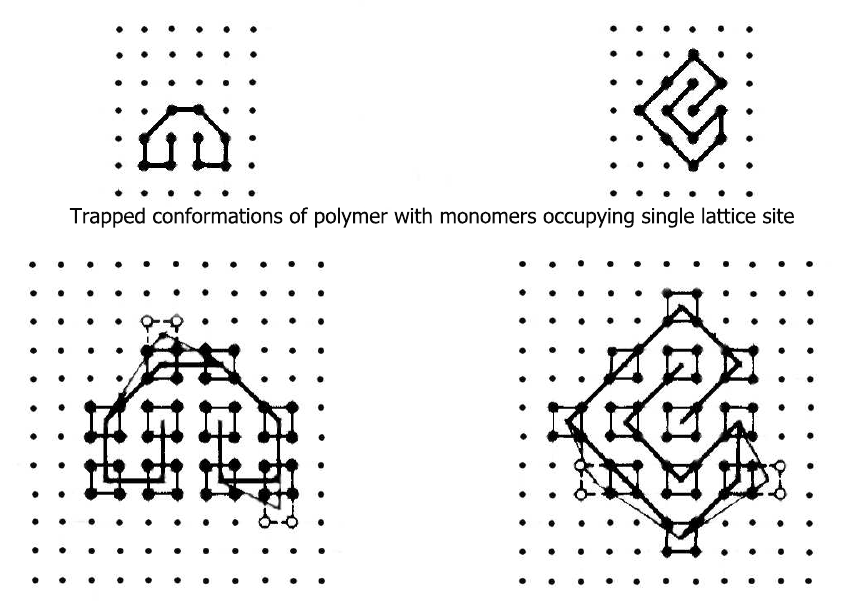}
\caption{{\small Illustration of the
problem with ergodicity in one site model.
The conformations shown do not evolve. However, the same conformations in four site model evolve and hence ergodic.}}
\end{center}
\end{figure}
We are interested in studying low temperature properties of polymer conformations.
Hence we have chosen  four site lattice model, which does not suffer from problems
arising due to of lack of ergodicity.
\newpage
\section{Algorithm} Implementation of Bond Fluctuation model
proceeds as follows.
\begin{itemize}
\item  {\bf Step\ 1:} Start with an initial linear self
avoiding conformation of a lattice polymer consisting of  $N$ monomers.

\item $\bf Step\ 2:$ Select a monomer randomly and select
one of the four lattice directions randomly with equal probability.

\item $\bf Step\ 3:$ Move the selected monomer in the
selected direction by one lattice spacing. Call this a trial move.

\item $\bf Step\ 4:$ Check if the trial move violates
self avoidance condition. If it does, then reject the trial move by
placing the monomer in its earlier lattice position and go to step
2.

\item $\bf Step\ 5:$ Check if trial move increases the
bond length beyond $\sqrt{13}$. If it does, then reject the trial
move by placing the monomer in its earlier lattice position and go
to step 2.

\item $\bf Step\ 6:$ If both requirements self avoidance
and bond length restrictions are met then take the trial
conformation for further processing through Metropolis or entropic
sampling or Wang-Landau algorithms. After processing go to step2.
\end{itemize}

Figure (2.3) depicts moves that are legal and those that are not legal.
~\newpage

\begin{figure}[!h]
\begin{center}
\includegraphics[width=14cm,height=10cm]{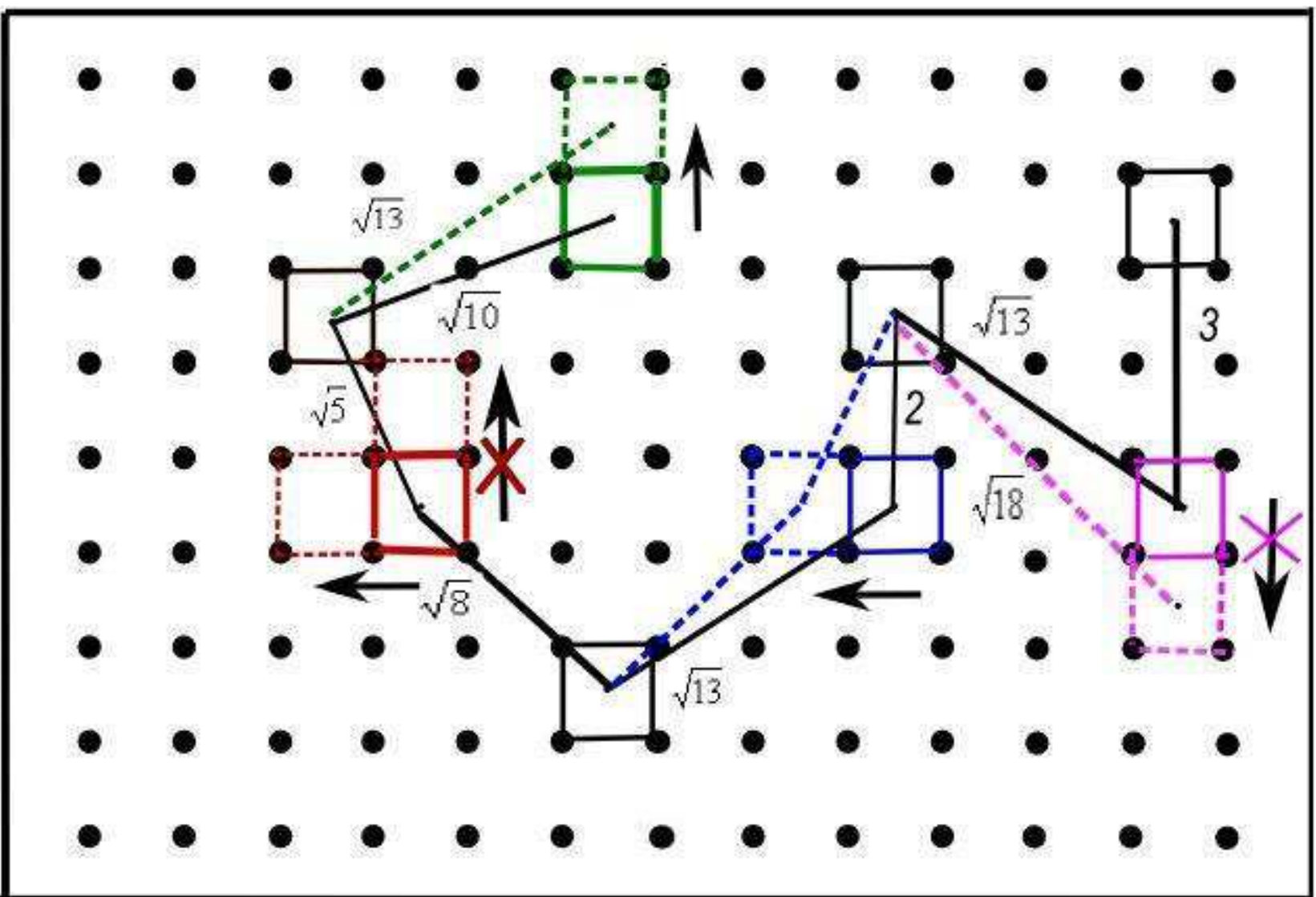}
\caption{{\small Legal moves in Bond Fluctuation Model.
These moves do not cut a bond; obeys self avoidance;
do not stretch the bond length beyond the set limit.
Also shown are moves that are not permitted in the model.} }
\end{center}
\end{figure}

\section{Dynamics of Bond Fluctuation model}
Bond Fluctuation model can be used to find dynamical properties of
polymers since this model has the following characteristics.
\begin{enumerate}
\item The elementary motion is a random local move.
\item  Excluded volume interaction among monomers is modelled through self avoidance
condition.
\item During a move there occurs no bond
intersection.
\item  The algorithm is  ergodic .
\end{enumerate}

We have investigated the dynamics of the entire polymer as follows.
We start with a self avoiding conformation. We select randomly and
with equal probability a monomer and move it by one lattice spacing
in such a way that self avoidance condition is met and no bond
stretches
 beyond the prescribed bond length. We carry out this dynamics for a large number
of Monte Carlo steps. A Monte Carlo step consists of $N$ moves where
$N$ is the number of monomers on the polymer chain. The trace of the
centre of mass is depicted in Fig. (2.4), which resembles that of a
two dimensional Brownian motion.  We have also calculated the  mean
square deviation of the centre mass and found that it diverges
linearly with time. The log-log plot of the mean square deviation is
depicted in  Fig. (2.5).  We get a straight line with slope nearly
unity confirming that the centre of mass executes Brownian motion.

\begin{figure}[!h]
\begin{center}
\includegraphics[width=14.5cm,height=8cm]{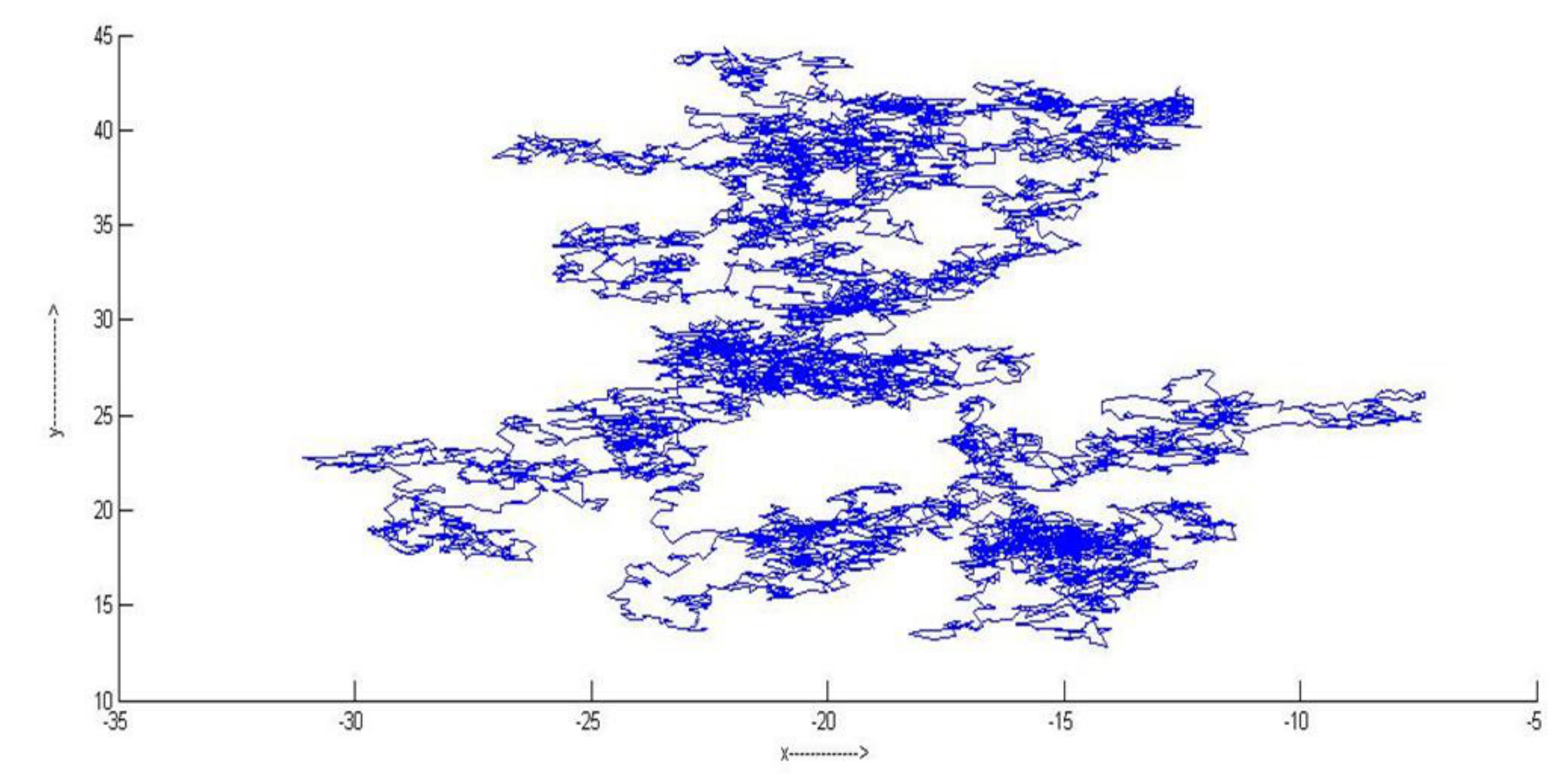}
\caption{{\small Brownian motion of centre of mass of a polymer in
two dimension}}\label{Rouse}
\end{center}
\end{figure}

\begin{figure}[!h]
\begin{center}
\includegraphics[width=15cm,height=10cm]{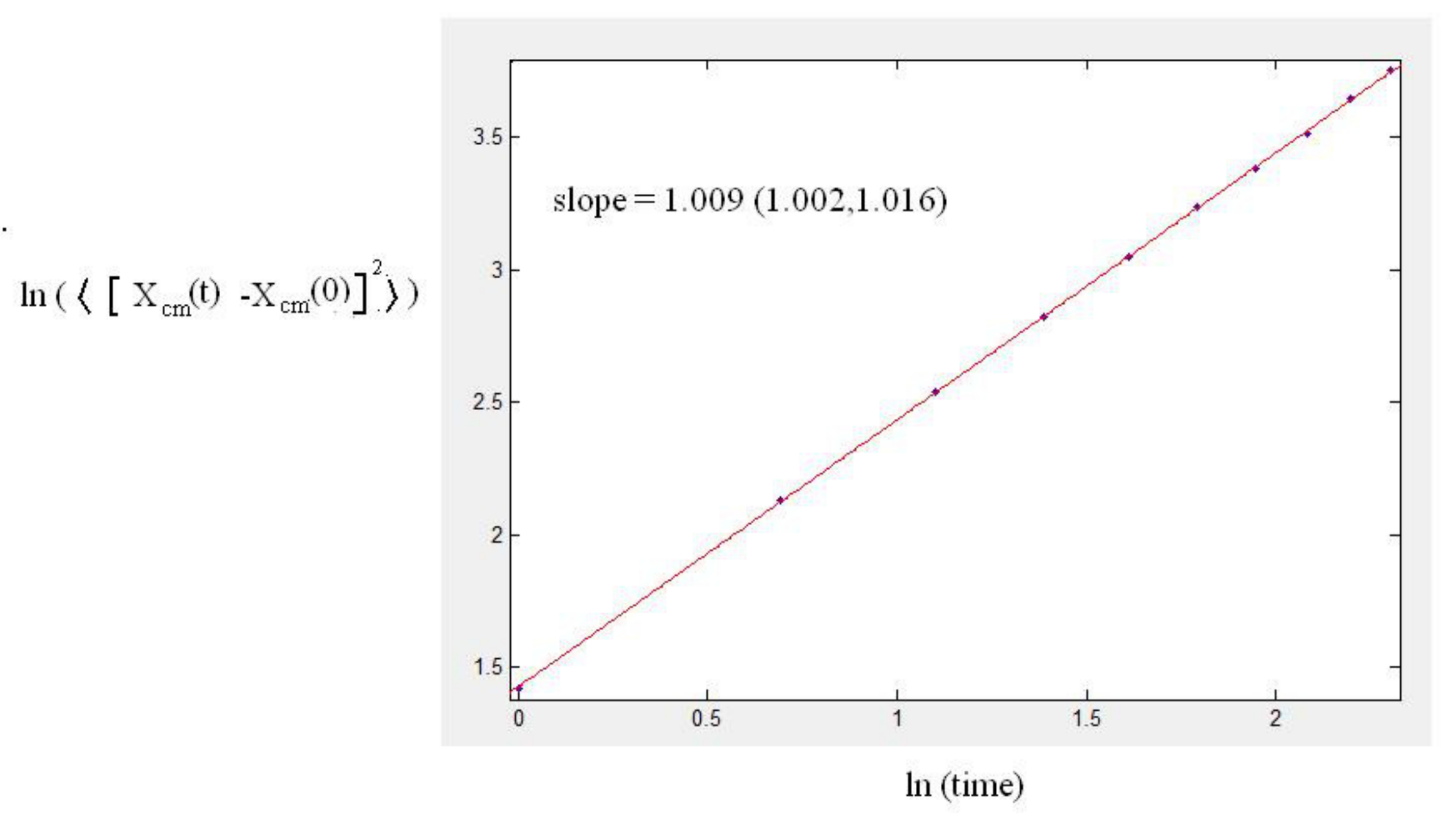}
\caption{{\small Logarithm of mean square deviation of the
$x$-coordinate of centre of mass {\it versus} logarithm of time}}
\end{center}
\end{figure}

%% file: chapters/chap3.tex
\chapter{Monte Carlo Simulation of Bond Fluctuation Model}

Bond Fluctuation model on a two dimensional square lattice discussed
in the previous chapter can be simulated by employing Metropolis
algorithm.  The conformations shown in Figs.~(3.6,~3.7~and~3.9) are
obtained by employing metropolis algorithm on a polymer of length $50$
monomers at low temperature $T=1.2$ and at high temperature $T= 2.2$
respectively. A brief description of the Metropolis algorithm is given
below
\section{Metropolis algorithm}
In order to calculate equilibrium properties like average energy and
average radius of gyration of lattice polymers at a particular
temperature, we generate a canonical ensemble by employing Markov
chain Monte Carlo methods based on Metropolis
algorithm \cite{N.Metropolis1953}. Metropolis algorithm proceeds as
follows.
\begin{itemize}
\item  Let $C_i$ be the current polymer conformation with energy $E_i$
\item Let $C_t$ be the SAW conformation with energy $E_t$ generated
employing bond fluctuating model.
\item If  $E_t$ $\leq$ $E_i$ then accept trial conformation : $C_{i+1}$=$C_t$.
\item If $E_t\ >\ E_i$, calculate the ratio of Boltzmann weights\\
$p = \exp[-\beta(E_t-E_i)]$. Select a random number $\xi$. If
$\xi<p$ accept trial conformation : $C_{i+1} = C_t$;  else reject the
trial conformation and set
$C_{i+1}=C_i$.
\end{itemize}
Iterate the whole procedure and generate a Markov chain. The
asymptotic part of the Markov chain consists of conformations that belong to a
canonical ensemble, at the chosen temperature. The desired
macroscopic properties are calculated by taking arithmetic average
over the  Monte Carlo ensemble.

The conventional Markov chain Monte
Carlo methods based on  Metropolis algorithm are not suitable for
calculating thermal properties like entropy and free
energies\footnote{Also Metropolis Monte Carlo techniques suffer from
critical slowing down. But we shall not be concerned with these
issues in the thesis}. We are interested in studying phase
transition in a polymer. The nature of the phase transitions is
best studied from free energy profiles. To calculate thermal
properties we have to go beyond Metropolis algorithm and resort to
non-Boltzmann Monte Carlo methods.

The first non-Boltzmann Monte
Carlo method called Umbrella Sampling was proposed  by Torrie and
Valleau \cite{G.Torrie1974}.  Several variants of Umbrella sampling
have since been proposed.  They include the multi canonical Monte
Carlo method of Berg and Neuhaus \cite{B.A.Berg1991}, entropic
sampling of Lee \cite{J.Lee1993}, Wang-Landau algorithm
 \cite{F.Wang2001} and several variants of Wang-Landau algorithm
$e.g.$ Frontier sampling \cite{C.Zhou2006}, JSM
technique \cite{D.Jayasri2005} {\it etc.}. For an introduction to
Boltzmann and non-Boltzmann Monte Carlo methods, see $e.g.$
 \cite{K.P.N.Murthy2004}. In the present work, we
 employ Wang Landau algorithm to simulate bond fluctuating lattice
polymer. The algorithm is briefly described below.

\section{Wang-Landau algorithm}
In Wang-Landau algorithm  we bias the Markov chain to move towards
regions of lower entropy. This is carried out as below\\
\textbf{Algorithm}
\begin{itemize}
\item Step1 : Let $g$ denotes the density of states.
In lattice polymer problem energy of the polymer equals negative of the
number of non-bonded
nearest neighbour contacts. The energy is thus an integer.
Hence we calculate
the density of states at discrete energies.
Let $\{ g(E_i)\}$
denote the discrete density of states. Since the density of states
is unknown for the system initially, start with $g(E_k) = 1\ \forall
\ k$ . Take the Wang-Landau factor $f = 2.7183 = e$. We start with an
arbitrary initial conformation\footnote{In all our simulation we
take $C_0\ as\ (0,0),\ (2,0),\ (4,0).....$ The energy of the initial
conformation is zero since there are no non-bonded nearest neighbour
contacts : $E(C_0)=0$.} denoted by $C_0$ and evolve the chain
$C_0\to C_1\to C_2\to \cdots \to C_i\to \cdots$  as per Wang-Landau
dynamics described below.
\item Step 2: Let $C_i$ be the current conformation.  $\{ h(E)\}$
denotes the energy histogram of conformations actually taken  by the lattice polymer
following Wang-Landau dynamics. At the beginning we set $h(E)=0\
\forall\  E$.
\item Step 3:  Employing Bond fluctuating model construct a trial
conformation  $C_t$
\item step 4:   If $g(E(C_t))\le g(E(C_i))$ accept trial conformation.
Update $h(E(C_t) = h(E(C_t) + 1$ and $g(E(C_t))=g(E(C_t))\times f$.
Set $C_{i+1}  = C_t$ and  go to step 3
\item Step 5:  If not, calculate the ratio
$\displaystyle{p=\frac{g(E(C_i))}{g(E(C_t))}}$. Call for a random
number $\xi$. If  $\xi\le p$  then accept trial conformation. Update\\
 $g(E(C_t))=g(E(C_t))\times f\  {\rm and}\  h(E(C_t)) = h(E(C_t)) + 1.$
 Set
$C_{i+1}  = C_t$ and go to step 3
\item Step 6:   If  $ \xi\ >\ p$  reject the trial conformation.  Update
\\ $ g(E(C_{i}))=g(E(C_i)\times f $, $ h(E(C_i))=h(E(C_i))+1$; Set
$C_{i+1}=C_i$ and go to step3
\item Step 7: Sufficiently large number of Monte Carlo sweeps are to
be performed to ensure a flat histogram. Check for the flatness of the
histogram {\it i.e.}, all the
entries in the histogram should be greater than $\delta\% $ of the
average. In all the studies reported in this thesis we have taken
$\delta = 80$.
\item step 8: When the flatness check is passed,
$f$ is modified to $\sqrt{f}$ and then go to step 2. This process is
continued until $f \sim \exp( 10^{-8})$ {\it i.e.,} $f\approx 1$

\end{itemize}

\begin{figure}[!hp]
\begin{center}
\includegraphics[height=20cm]{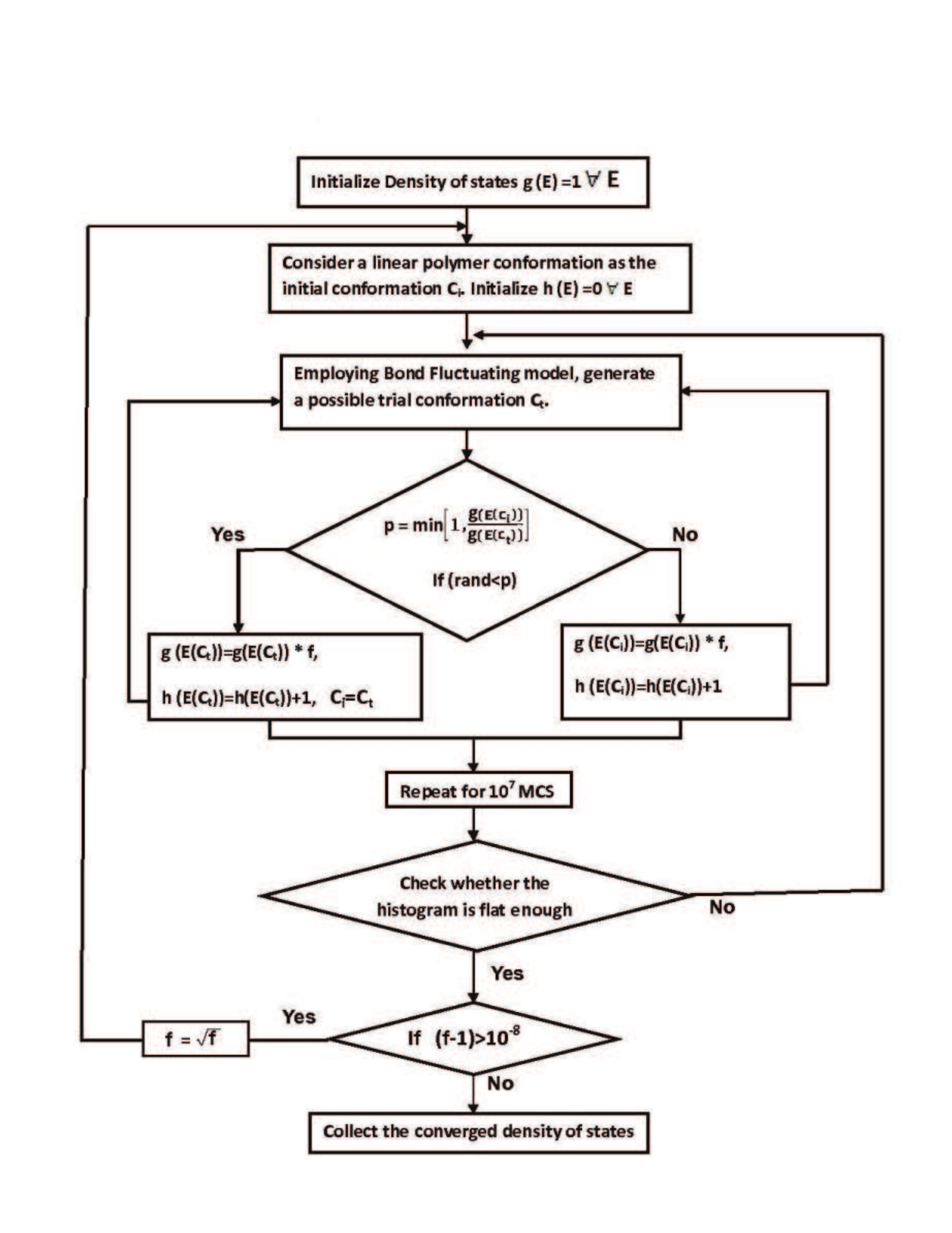}
\caption{{\small Flow chart for the Wang-Landau algorithm}}
\end{center}
\end{figure}

Using  the converged density of states, we can calculate micro canonical
entropy: $S(E)=k_B\ln g(E)$. Employing the standard machinery of
thermodynamics, we can calculate all the desired macroscopic
properties from the micro canonical entropy function. Helmholtz's free
energy can be calculated as
\begin{eqnarray}
F(T)&=& -k_B T \ln(Z)\nonumber \\[8mm]
&=& -k_B T \ln\left(\sum_E g(E)\exp(-\beta E)\right)\\[8mm]
&or& \nonumber \\[8mm]
F(T) &=& E - TS(E)\\[8mm]
T^{-1} &=& \frac{\partial S}{\partial E}\nonumber
\end{eqnarray}

Internal energy can be calculated as
\begin{eqnarray}
U(T)=\langle E \rangle  = \frac{\sum_E\ E\ g(E)\ \exp\left[-\beta
E\right]}{\sum_E\ g(E)\ \exp\left[-\beta E\right]}
\end{eqnarray}
 And the specific heat can be obtained as
\begin{eqnarray}
C_V(T)=\frac{\langle E^2\rangle-\langle E \rangle^2}{k_B T^2}
\end{eqnarray}
Alternately we can generate an entropic ensemble\footnote{An
entropic ensemble is one in which we have equal number of lattice
polymer conformations in equal intervals of energy.} of polymer
conformations $C_i $ employing converged density of states. We can
calculate un-weighting-cum-re-weighting factors\footnote{The
un-weighting and re-weighting factors would be required whenever you
generate one ensemble, say uniform ensemble and would like to
calculate averages over another ensemble, say canonical ensemble.}
for each conformation of the entropic ensemble. It is given by
\begin{eqnarray}
 W(C_i) = g(E(C_i)) \exp [-\beta E(C_i)]
\end{eqnarray}
Using this weight  we can obtain canonical ensemble average of
a macroscopic property say $\eta(C_i)$ at the desired temperature from the entropic ensemble, see below.
\begin{eqnarray}
\langle\eta(\beta)\rangle=\frac{\sum_{C_i}\
\eta(C_i)W(C_i)}{\sum_{C_i} W(C_i)}
\end{eqnarray}
The sum runs over all the conformations of the entropic ensemble.

\section{Results and Discussions}
We have carried out non-Boltzmann Monte Carlo simulation of bond
fluctuation model with $N=10,\ 20,\ 30$ and $50$. Figure (3.2) shows
the energy histogram for a polymer with $50$ monomers. We consider
the energy region : $-130$ to $-20$,  where the histogram is
reasonably flat, consistent with the $80\%$ flatness criterion
described earlier. The micro canonical entropy is given by $S(E)=k_B
\ln g(E)$. Without loss of generality we set the Boltzmann factor
$k_B =1$. In this work, all the calculations were carried out by
taking $\ln(S(E))$ instead of g(E) to avoid overflow problems.
Figure (3.3) shows $\ln(S(E))$ {\it versus} $E$.
\begin{figure}[!hp]
\begin{center}
\includegraphics[height=11cm]{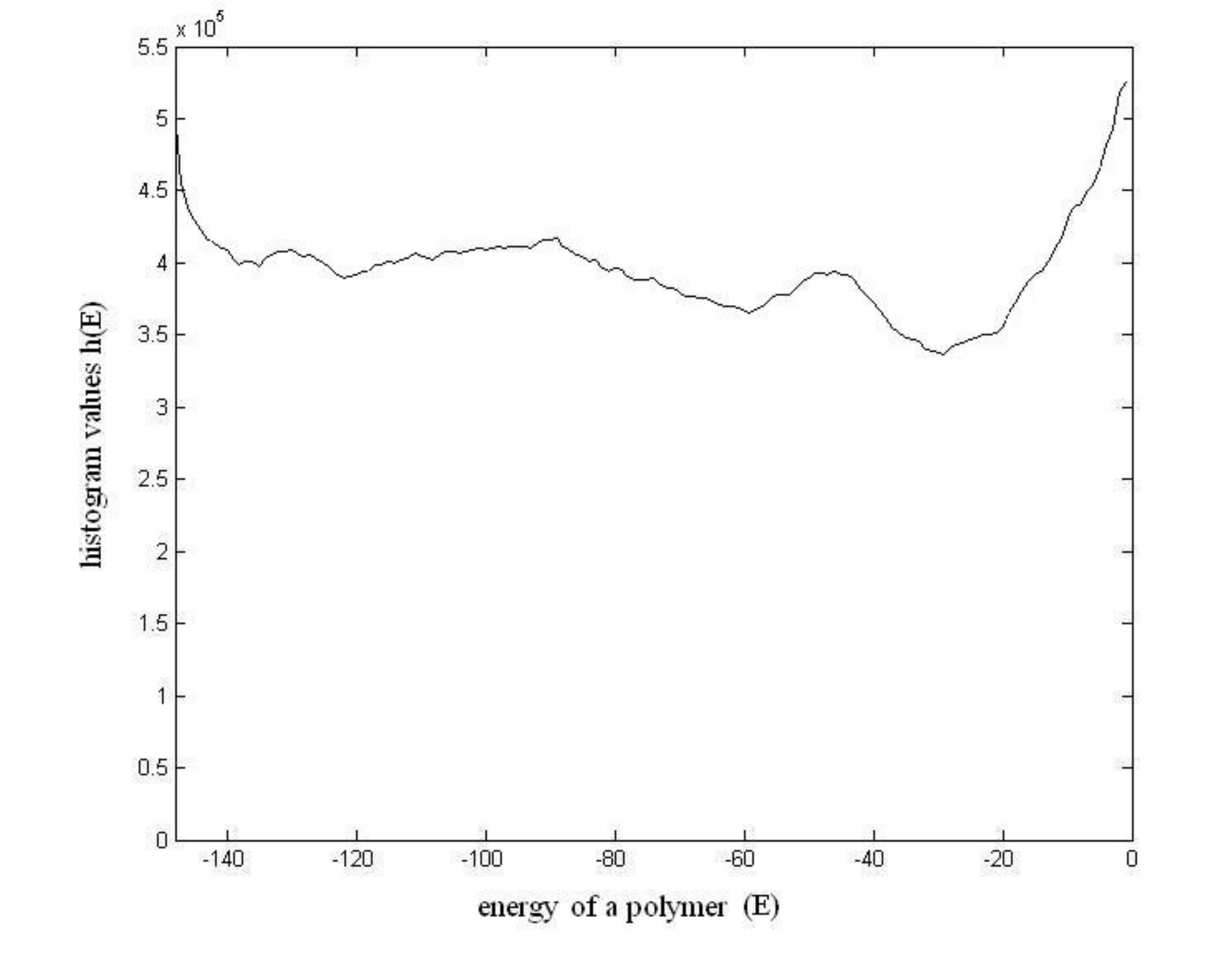}
\caption{{\small Energy histogram, $h(E)$
 of lattice polymer of length $50$
monomers. A range of energy from  $-130$ to  $-20$ is
considered; the  histogram is reasonably flat in this range.} }
\includegraphics[height=7cm]{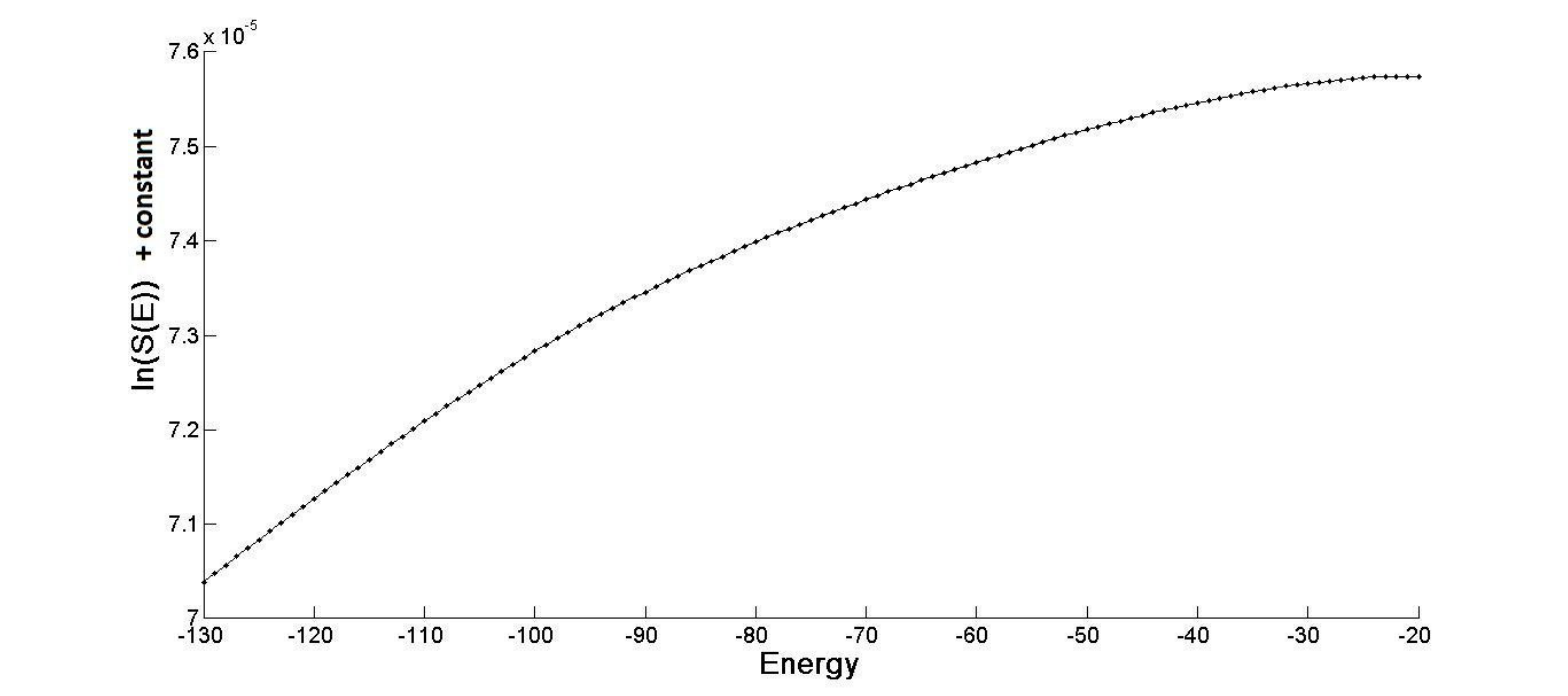}
\caption{{\small Logarithm of entropy as a function of energy in
the  range : $-130$ to $-20$. The value of the constant on Y-axis is -16.4124.}}
\end{center}
\end{figure}

\subsection{Variation of energy with Temperature}

As discussed in the first chapter, a compact globule
structure is an equilibrium structure at low temperature and
an extended coil structure is an equilibrium structure at high
temperature. A compact globule structure contains more nbNN contacts
compared to extended coil structure. Since $\epsilon =-1$ for an
nbNN contact, a compact globule structure has less energy compared
to extended coil structure. So the average energy should be less at
low temperature and high at high temperature. Figure (3.4) shows
average energy calculated from
 both density of states and from the entropic ensemble (generated in the production
 run employing un-weighting and re-weighting factors).

\begin{figure}[!hp]
\begin{center}
\includegraphics[width=18cm,height=14cm]{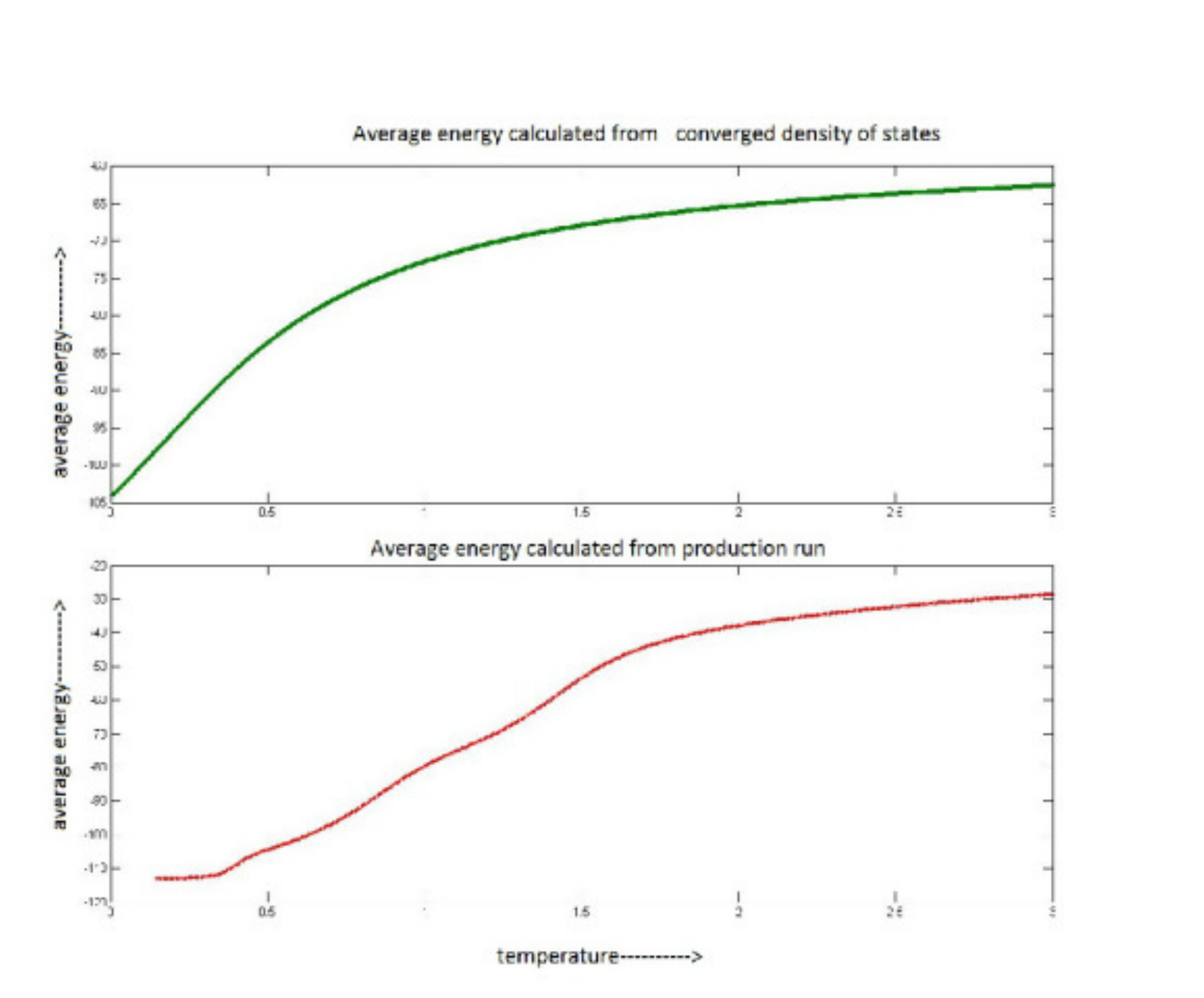}
\caption{{\small {\bf Top}: Average energy as a function of temperature calculated
from the density of states. {\bf Bottom}: Average energy as a function of
temperature calculated over a sample from entropic ensemble. These
calculations are done for a polymer of length $50$ monomers}}
\end{center}
\end{figure}

\subsection{Energy fluctuation}
Energy and energy fluctuations are two important macroscopic
parameters. Energy fluctuation is defined as  $\sigma^2_E = \langle
E^2\rangle-\langle E \rangle^2$, where the angler brackets denote an
average over a canonical ensemble.

Heat capacity is defined as the  thermal energy you need to supply to a
 macroscopic object to raise its temperature by one degree Kelvin.
 Thus $\displaystyle{C_V = \frac{\partial U}{\partial T}}$
where $U$ is the mean internal energy.
  We have
  \begin{eqnarray}
  U\equiv \langle E \rangle = \frac{\sum_r E_r \exp\left[-\beta E_r\right]}{\sum_r \exp\left[-\beta E_r\right]}
  \end{eqnarray}
In the above the sum over $r$ denotes a sum over all the
microstates of the closed system.

If we take the partial derivative of $U$ with respect to $\beta$, we get,
\begin{eqnarray}
  \frac {\partial U}{\partial \beta}  =  -\langle E^2\rangle +
  \langle E \rangle^2
  \end{eqnarray}

  It follows then
\begin{eqnarray}
  \langle(\Delta E)^2\rangle \equiv \langle
  E^2\rangle - \langle E\rangle^2 = -\left(\frac{\partial U}{\partial
  \beta}\right)= k_BT^2\left(\frac{\partial U}{\partial T}\right) = k_BT^2 C_V.
  \end{eqnarray}

\begin{eqnarray}
C_V = \frac{1}{k_B T^2} (\langle E^2\rangle-\langle E \rangle^2)
\end{eqnarray}

At transition temperature specific heat profile has a peak showing the
signature of the phase transition \footnote{Strictly, $C_V$ should diverge at $T\ =\ T_c$. But this would happen only in the thermodynamic limit. Due to finite size, we will get only a peak which will become sharper when the system size is larger.}.

  The results on heat capacity ($C_V$) as a function of temperature
  are depicted in Fig.~(3.5), for $N=10,\ 20,\ 30$ and $50$.
As the system size increases the heat capacity curves peak more and more
  sharply. In the Fig.~(3.5) we  see a sharp peak at $T=T_{C1}=0.83$ and a broad peak
at $T=T_{C2}=1.6\ >\  T_{C1}$. The temperature at which $C_V$ is
maximum
  gives an estimate of the transition temperature $T_C$.

\begin{figure}[!hp]
\begin{center}
\includegraphics[width=14cm,height=10cm]{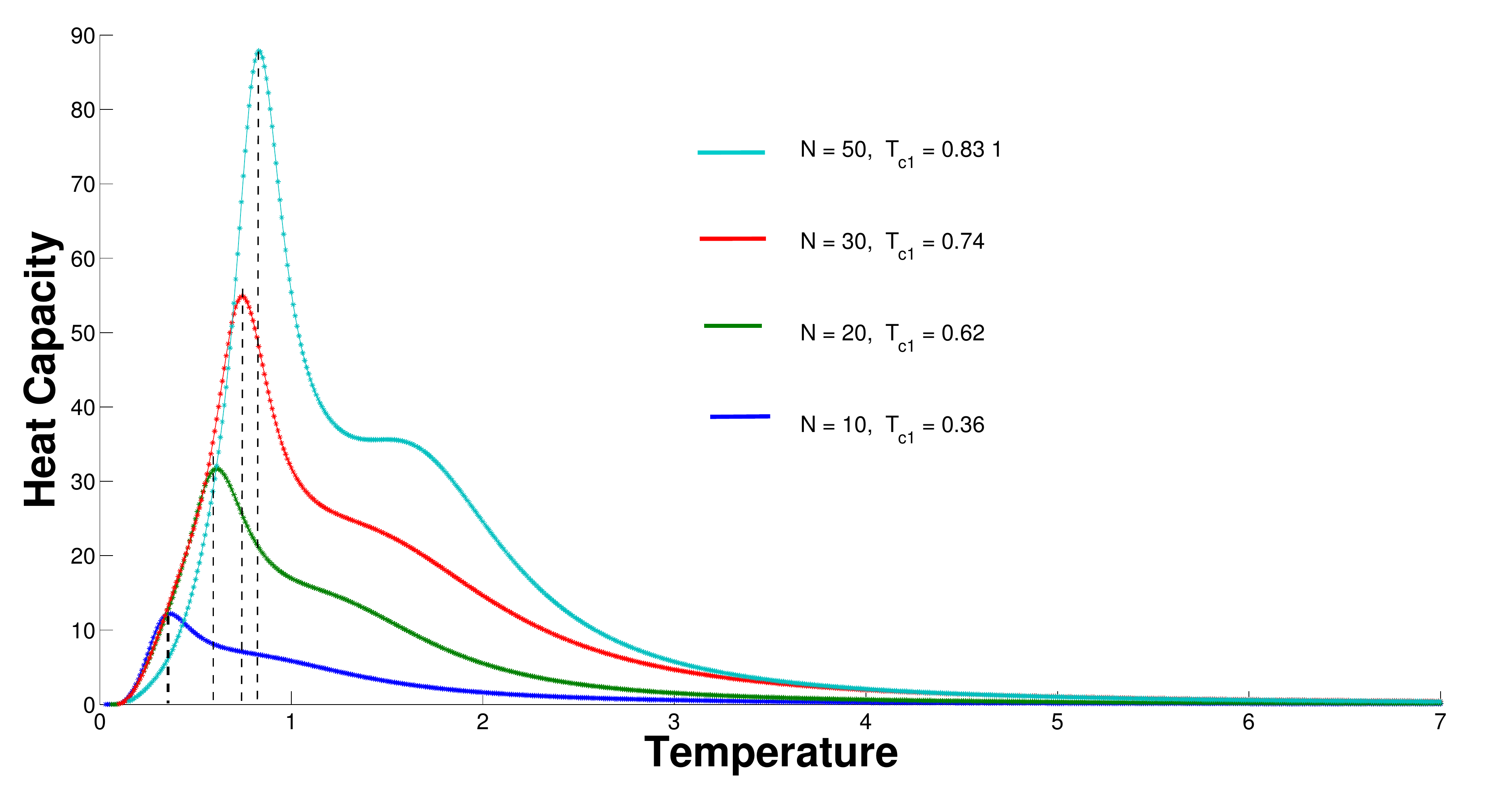}
\caption{{\small Heat capacity as a function of temperature for
polymers of length $N\ =\ 10,\ 20,\ 30$ and $50$ monomers. For
 $N=50$, the sharp peak at $T=T_{C1}=0.83$ corresponds
to crystallization transition and the broad peak at $T_{C2}=1.6$ corresponds to
coil-globule transition.}}
\end{center}
\end{figure}

The broader peak corresponds to a phase transition from extended
coil to compact globule structure. For a self attracting chain, this
transition is caused by the competition between excluded volume
repulsion and attraction due to segment-segment interaction and
configurational entropy. Transition temperature, $T_{C2}$ is found
to be $1.6$. Figures  (3.6 and 3.7) show configurations of a
polymer of length $50$ monomers at
  temperatures  $1.2$, below $T_{C2}$  and  at $2.2$, above $T_{C2}$
respectively. Figure (3.6) shows a compact globule structure of a
polymer of length $50$ monomers whose radius of gyration is $7.6026$ at
at $T=1.2$.  Figure  (3.7) shows an extended coil
structure of a  polymer of length $50$ monomers whose radius of
gyration is $14.3639$ at  $T=2.2$. These
conformations are obtained by simulating bond fluctuation model of
isolated polymer of length $50$ monomers employing Metropolis
algorithm.

\begin{figure}[!hp]
\begin{center}
\includegraphics[width=18cm,height=12cm]{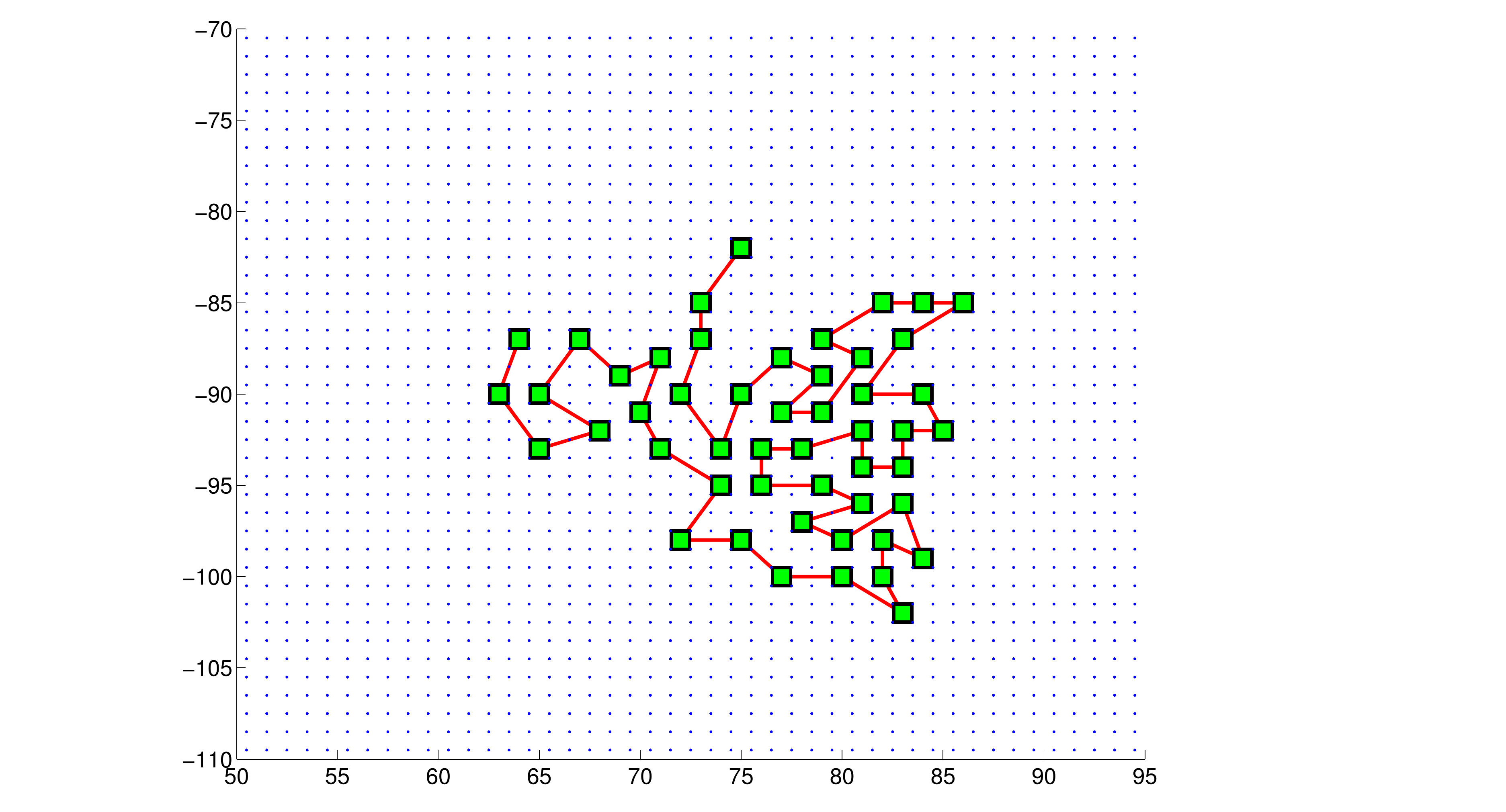}
\caption{{\small Compact globule conformation at $T_{C1}\ < T=1.2\ <\
T_{C2}$. The  radius of gyration for this conformation is $7.6026$.}}
\end{center}
\end{figure}

\begin{figure}[!hp]
\begin{center}
\includegraphics[width=18cm,height=12cm]{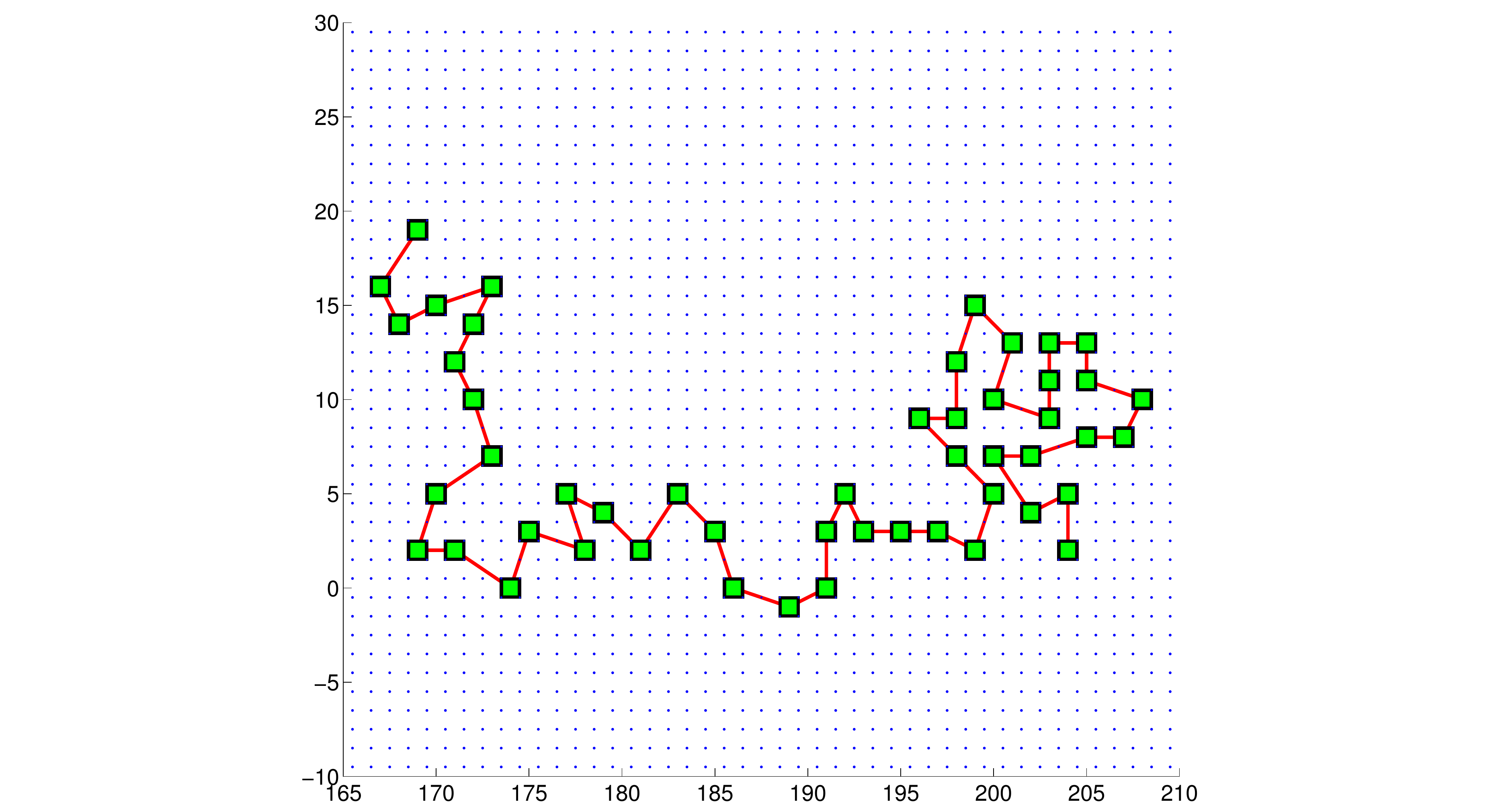}
\caption{{\small Extended coil conformation at $T=2.2\ >\ T_{C2}$.
The radius of gyration for this conformation is $14.3639$.}}
\end{center}
\end{figure}

 Figure (3.8) shows heat
capacity profile of a polymer of lengths $10,\ 30$ and $50$
monomers. The heat capacity profiles have each a peak signalling
coil-globule phase transition. The peaks become sharper as $N$
increases. The calculations were done considering only a narrow
energy region around  the broad peak depicted in Fig. (3.5).
 The transition temperature for a polymer of length $50$ monomers is found to
be $1.96$.

\begin{figure}[!hp]
\begin{center}
\includegraphics[width=14cm,height=12cm]{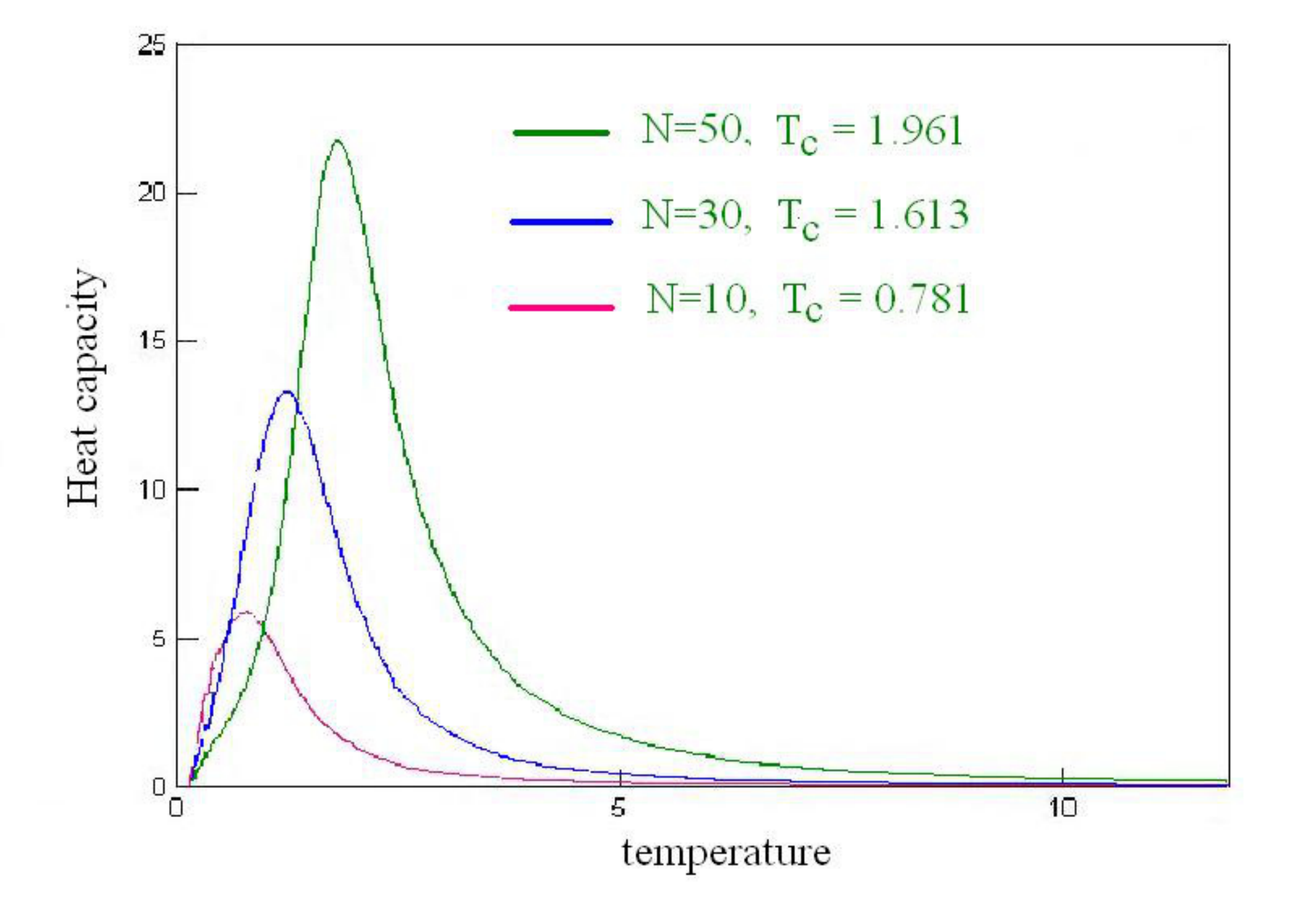}
\caption{{\small Heat capacity curve for polymers of length $10,\
30,\ and\ 50$ monomers when a narrow energy region around the broad
peak in Fig. (3.4) is considered. We see a sharp peak at
$T_{C2}=1.96$ corresponding to coil-globule transition, for
$N=50$.}}
\end{center}
\end{figure}

The sharp peak in the heat capacity profile seen at a lower
temperature in Fig. (3.5), depicts a phase transition which could be
the crystallization as discussed in \cite{F.Rampf2005} or the
solid-liquid transition as described in \cite{D.T.Seaton2009}.

Figure (3.9) shows an extremely compact structure of a polymer of
length $50$ monomers at a  $T=0.4$ whose radius of gyration is
$5.8$. The radius of gyration of this conformation is found to be
less when compared to globular structure at $T=1.2$ whose radius of
gyration is $7.6$.

\begin{figure}[!hp]
\begin{center}
\includegraphics[width=18cm,height=12cm]{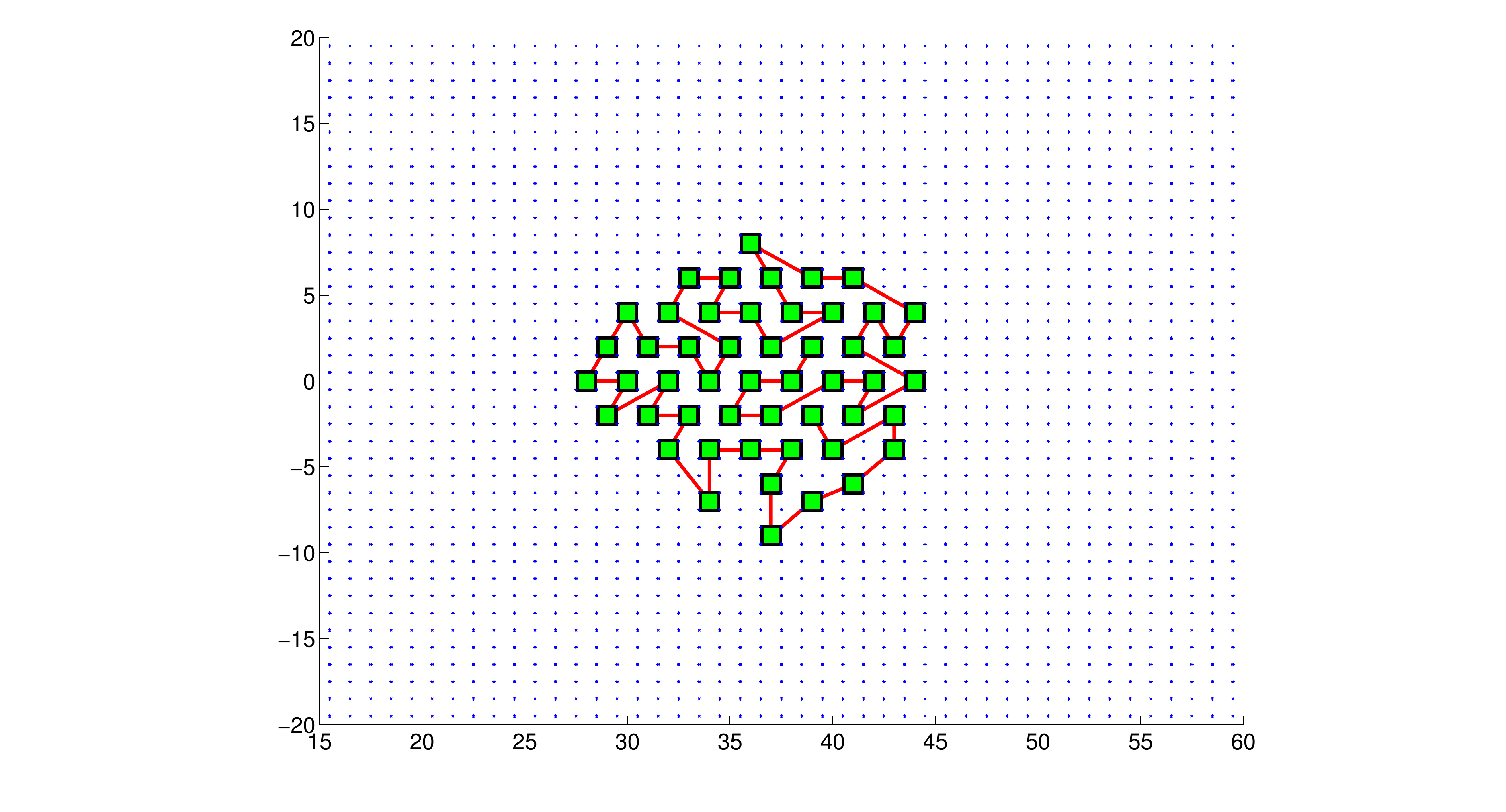}
\caption{{\small Extremely compact conformation at $T=0.4\ <\
T_{C1}$. The radius of gyration for this conformation is $5.7659$.}}
\end{center}
\end{figure}

%% file: chapters/chap4.tex
\chapter{Free Energy and Landau Free Energy}
\section{Introduction}
In this chapter we shall provide a brief introduction to Helmholtz free energy or simply free energy  for an equilibrium system. We shall describe free energy for a closed system (canonical ensemble) and for
an isolated system (microcanonical ensemble). Besides we shall
consider the  phenomenological free energy introduced by Landau for
the bond fluctuating lattice polymer.
\subsection{Free energy for a closed system}
For a closed system, free energy is function of temperature
and other thermodynamic variables
like volume, $V$ and number of molecules, $N$.
It is
written as
\begin{eqnarray}
F(T,V,N)= U(S,V,N)- T S
\end{eqnarray}
In the Right Hand Side (RHS) we eliminate $S$ by expressing it as a function
$T$, $V$ and $N$, see below.
\begin{eqnarray}
T(S,V,N) = \left(\frac{\partial U}{\partial S}\right)_{V,N}.
\end{eqnarray}

In statistical Mechanics, free energy for a closed system is related
to canonical partition function, Q(T,V,N), as shown below.
\begin{eqnarray}
F(T,V,N) &=& -k_B T \ln Q(T,V,N)\\[3mm]
Q(T,V,N) &=&\sum_C \exp\left(-\frac{E(C)}{k_BT}\right),
\end{eqnarray}

Energy of the closed system is given by
\begin{eqnarray}
U(T,V,N) = \langle E\rangle = \frac{1}{Q(T,V,N)}\sum_C E(C)
\exp(-\beta E(C))
\end{eqnarray}
where the sum runs over all microstates of the closed
system.
\subsection{Free energy for an isolated system}

For an isolated system, microcanonical free energy is a function of
energy. We start with $S \equiv S(U,V,N)$. Wang Landau algorithm gives
us an estimate of entropy up to an additive constant. Microcanonical
Free energy is then given by
\begin{eqnarray}
 F(U,V,N) &=& U - T(U,V,N)S(U,V,N)\\[10mm]
\left(\frac{\partial S}{\partial U}\right)_{V,N} &=&  \frac{1}{T(U,V,N)}
\end{eqnarray}

\section{Landau Free energy}
Free energy is either a function of energy (for an isolated
equilibrium system) or a function of Temperature (for a closed
system). For any equilibrium system, isolated or closed, $F$ cannot be
simultaneously a function of both energy and temperature. This is
because,
\begin{itemize}
\item An isolated system with fixed energy has a unique temperature.
\item A closed system at a given temperature has unique energy.
\end{itemize}
Suppose we want to estimate free energy for an energy
different from the equilibrium energy $U$. Let us denote
such a free energy by the symbol $F_L$, see below. Clearly,
\begin{eqnarray}
F_L (E,T,V) \ge F(E=U,T,V),\label{FreeEnergy}
\end{eqnarray}
since equilibrium is characterized by minimum free energy. Notice
that in the above equation we have written $F_L$ as a function of
both $E$ and $T$. This is legitimate since we are inquiring about a
system not in equilibrium. The right hand side of the above equation
is equilibrium free energy:\\ $F_L(E=U,T,V) = F(T,V)$. The
difference
\begin{eqnarray}
\delta F=F_L(E\neq U,T,V)- F(E=U,T,V)
\end{eqnarray}
can be thought of as a penalty we have to incur if we want to keep
the system in a state with energy $E\ne U$. This corresponds to the phenomenological Landau free energy \footnote{originally proposed to describe continuous phase
transition; we also have  Ginzburg-Landau free energy,
proposed in the context
of superconductivity and Landau-de-Gennes
free energy, proposed in
the context of liquid crystals.}, and hence the notation
$F_L$;
see \cite{R.K.Pathria1961} for a
description of Landau/Landau-Ginzburg free energy.

In Eq. (\ref{FreeEnergy}) equality obtains
when $E = \langle E \rangle = U(T)$, and
$\langle E \rangle$ denotes an average  over a canonical
ensemble, at temperature $T$.

Landau free energy can be calculated as follows. In thermodynamics,
start with $S\equiv~S(U)$ and calculate $F_L(U,T) = U - TS(U)$,
assuming $U$ and $T$ to be independent of each other.

In statistical mechanics we define $F(T)$ as
\begin{eqnarray}
F(T) =-k_BT \ln\sum_C \exp[-\beta E(C)],
\end{eqnarray} where the sum
is taken over all microstates of the closed system. However for a
given temperature $T$, the contribution to the partition sum comes
predominantly from those conformations\footnote{This is method of
most probable values often employed in statistical mechanics, see
{\it e.g.} \cite{R.K.Pathria1961}. The partition function can be
written as,
\begin{eqnarray}
Q(T) = \sum_E g(E) \exp(-\beta E)
\end{eqnarray}
At a given temperature the partition sum gets contribution
predominantly from a single value of energy $E = U(T)$. We write,
\begin{eqnarray}
Q(T)&=& g(U)\exp(-\beta U)\\[3mm]
-k_BT \ln Q &=& -T k_B \ln g(U)+U
\end{eqnarray}
We identify $-k_BT\ln Q$ as free energy, F = U-TS.} having energy $E
= \langle E\rangle = U(T)$. Hence we can express free energy as
\begin{eqnarray}
F(T) = -k_B T \ln \sum_C \delta(E(C)- U(T)) \exp[-\beta E(C)]
\end{eqnarray}
In the above if we replace U(T) by E, we get Landau free energy,
\begin{eqnarray}
F_L(E,T) = -k_B T \ln \sum_C \delta(E(C) - E) \exp[-\beta E(C)]
\end{eqnarray}
Note that in the above, the presence of Dirac delta function ensures that
the sum runs only over those microstates for which
$E(C) = E $.

\subsection{Free energy calculations}
Employing Wang Landau algorithm to simulate
bond fluctuation model, we have obtained
converged density of states, $g(E_i)$.  We consider all possible
discrete energies $E_i$
of the lattice polymer.
Microcanonical entropy is given by $S(E_i)=k_B\ln g(E_i)$. Without loss of generality
we take $k_B$=1.
Landau Free energy $F_L(E_i,T)$ can be calculated using any one of the following two equations
\begin{eqnarray}
F_L(E_i,T)&=&E_i-TS(E_i)\\[5mm]
 F_L(E_i,T)&=&-k_BT\ln\left[ g(E_i)\exp(-\beta
E_i)\right]
\end{eqnarray}
We have used both the above expressions and found that they give the same results.
We have calculated Landau free energy for lattice polymers with
$10,\  20,\  30,\  40$ and $50$ monomers. The results  are presented and
discussed below.
\section{Results and Discussions}
From the heat capacity curve for a polymer of length $50$ we found
that there are two transitions. One corresponding to transition from extended coil
phase  to compact globule phase  at a
temperature $T=1.6$. The other corresponds to crystallization transition
at temperature of   $T=0.83$.

We have checked Landau free energy profile, $F_L(T,E)$ {\it versus}
$E$,  for a polymer of length $50$ monomers for different
temperatures. We found that a coil-globule transition occurs at
a temperature of $T_{C2}=1.71$. Figures (4.1, 4.2, and  4.3) correspond
to Landau free energy {\it versus} energy for three values of
temperature, one below, one at,  and one above the transition
temperature of $1.71$.
\newpage

\begin{figure}[!hp]
\begin{center}
\vglue 30mm
\includegraphics[width=15cm]{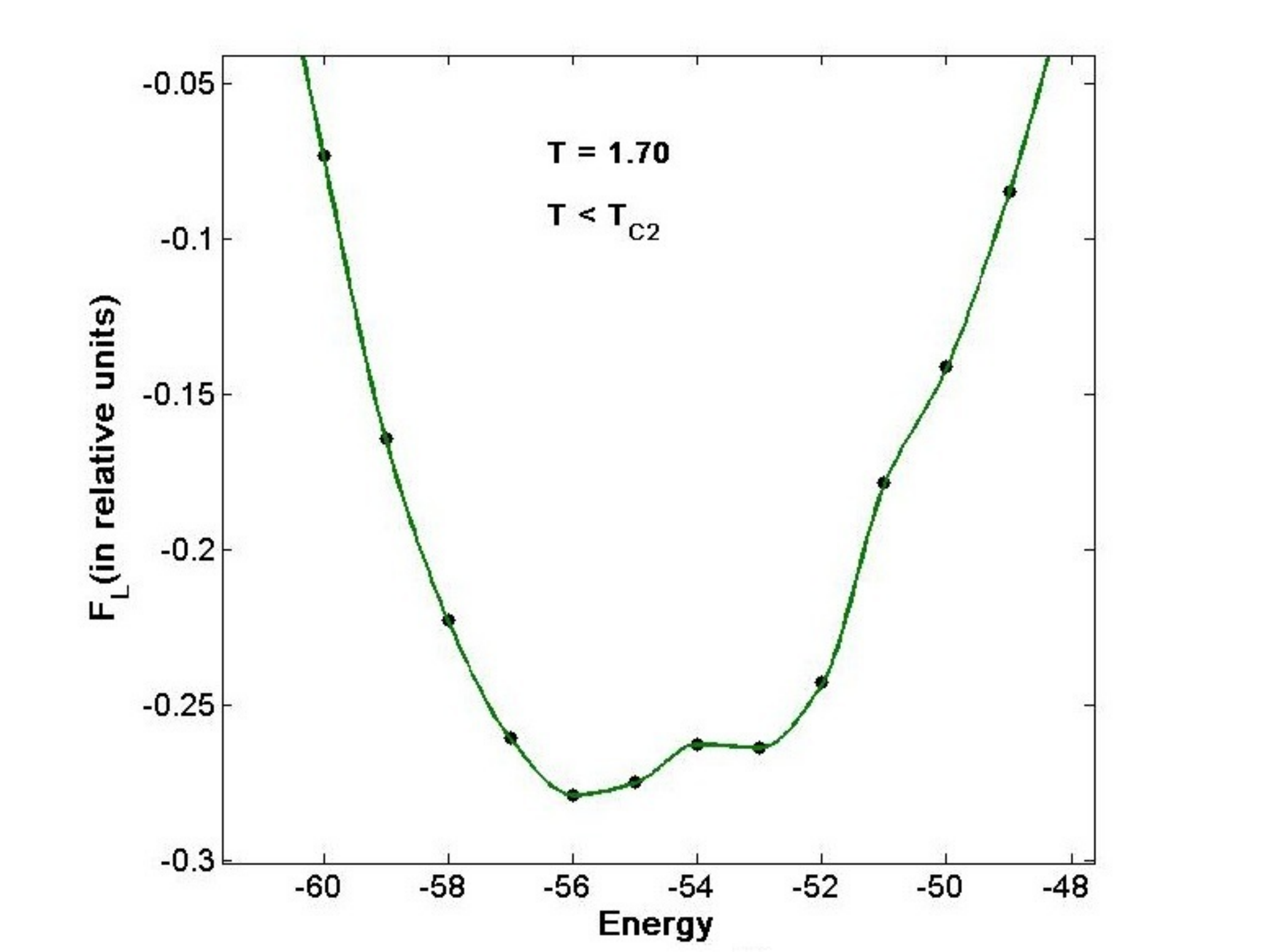}
\caption{{\small Free energy {\it versus} energy for $T=1.70\ <\ T_{C2}$ for an
isolated polymer of length $50$ monomers. A globule structure with low energy
is the
stable state and an extended coil structure with high energy is the
meta stable state. The values shown on the y-axis are $F_L + constant$ (constant=59174). } }
\end{center}
\end{figure}
\newpage
\begin{figure}[!hp]
\begin{center}
\vglue 30mm
\includegraphics[width=15cm]{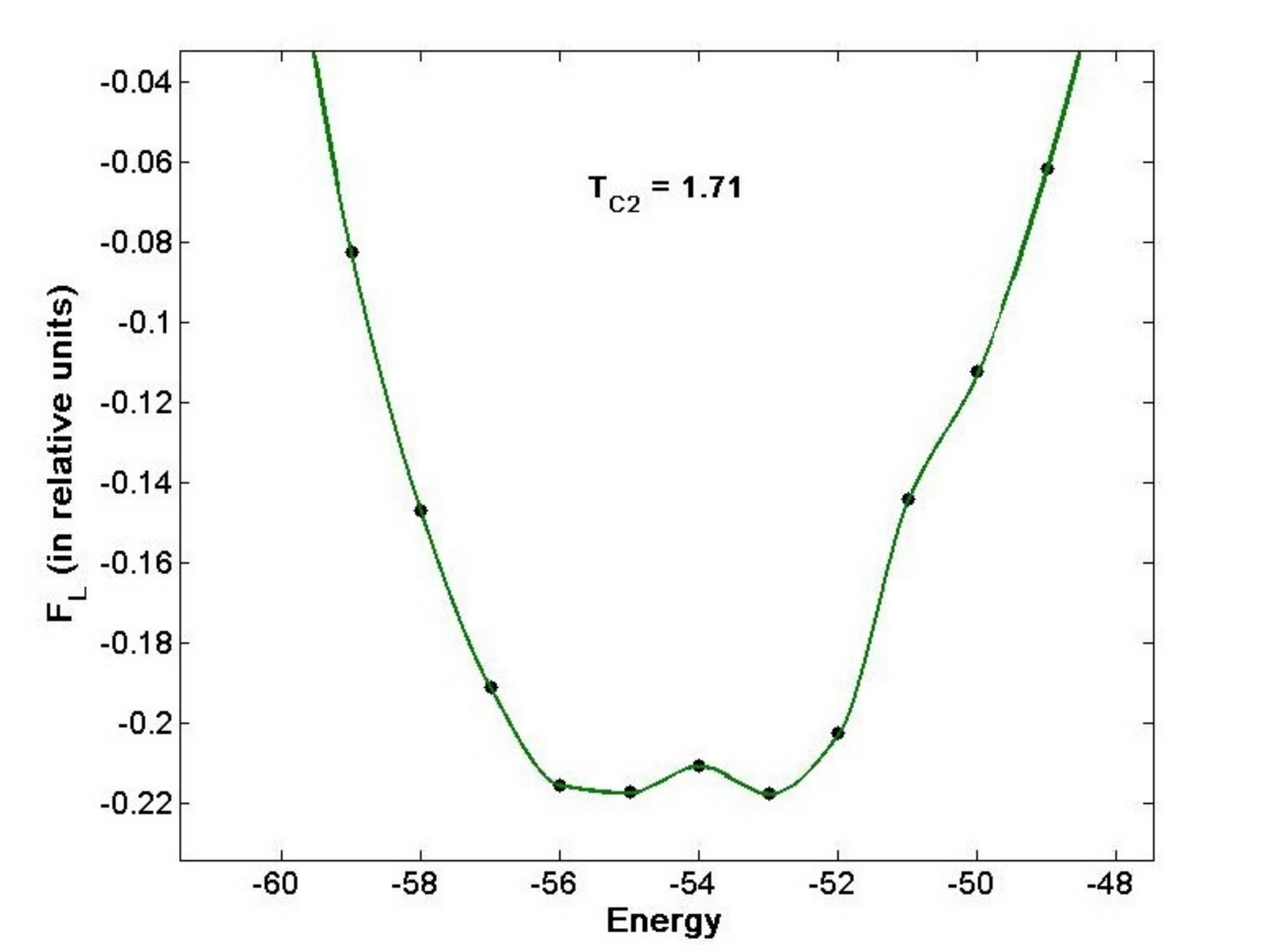}
\caption{{\small Free energy {\it versus} energy for $T = T_{C2}=1.71$  for an
isolated polymer of length $50$ monomers. Both compact-globule and
extended-coil phases are stable. The values shown on the Yaxis are $F_L + constant$ (constant=60066).}}
\end{center}
\end{figure}
\newpage
\begin{figure}[!hp]
\begin{center}
\vglue 30mm
\includegraphics[width=15cm]{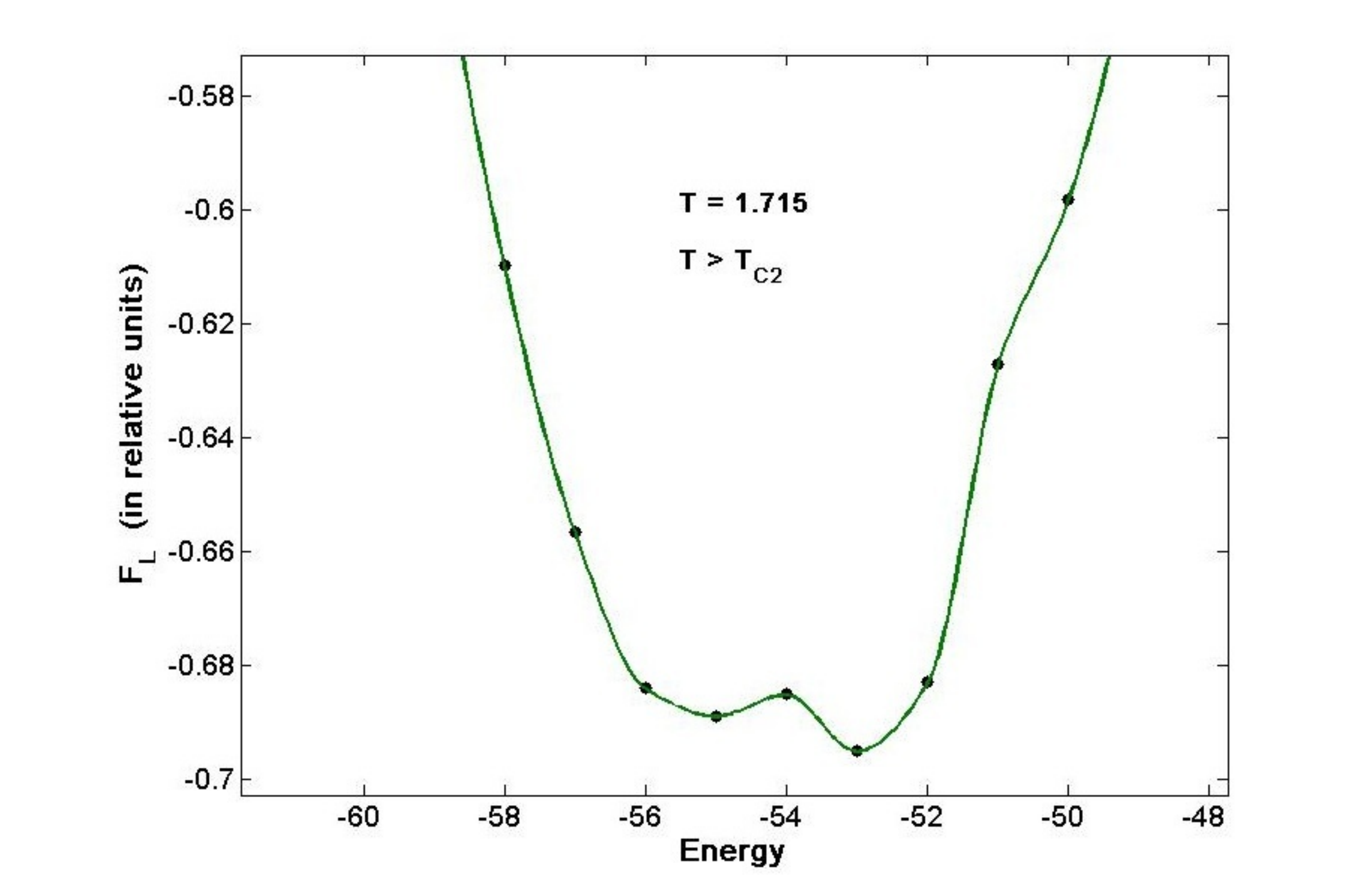}
\caption{{\small Free energy {\it versus} energy for $T=1.715\ >\ T_{C2}$ for an
isolated polymer of length $50$ monomers. An extended-coil phase
with high energy is stable;  a compact-globule phase with low energy
is the meta stable. The values shown on the Yaxis are $F_L + constant$ (constant=60241).}}
\end{center}
\end{figure}
\newpage
From these figures we see that
for $T\ <\ T_{C2}$, there is stable minimum that corresponds to collapsed phase and
 and a meta-stable minimum that corresponds to extended coil phase.
 At $T=T_C$ both these phases co-exist.  For $T\ >\ T_{C2}$
 the extended coil phase is stable and the collapsed globule phase is meta stable.
The phase  transition is first order.
We
find that the free energy barriers are small; this is  due to the fact that
the lattice polymers simulated are not long.

We have calculated  Landau free energy profile for the transition
that corresponds to the peak in the specific heat curve at $T=T_{C1}
< T_{C2}$.  From Fig. (3.4) we find that  $T_{C1} = 0.83$ for a
polymer of size $50$. From Landau free energy curve the transition
temperature is found to be $0.8$. Figures (4.4,\ 4.5,\ 4.6) show
Landau free energy $\it versus$ energy for three values of
temperature, one below, one at and one above the transition
temperature, $T_{C1}=0.8$.  Landau free energy curve exhibits two
minima at all temperatures close to $T_{C1}$. At $T\ <\ T_{C1}$ the
polymer conformations are extremely compact. This phase is usually
referred to as crystalline phase, see \cite{F.Rampf2005}. We find
that this transition is also discontinuous.

\begin{figure}[h]
\begin{center}
\includegraphics[width=15cm]{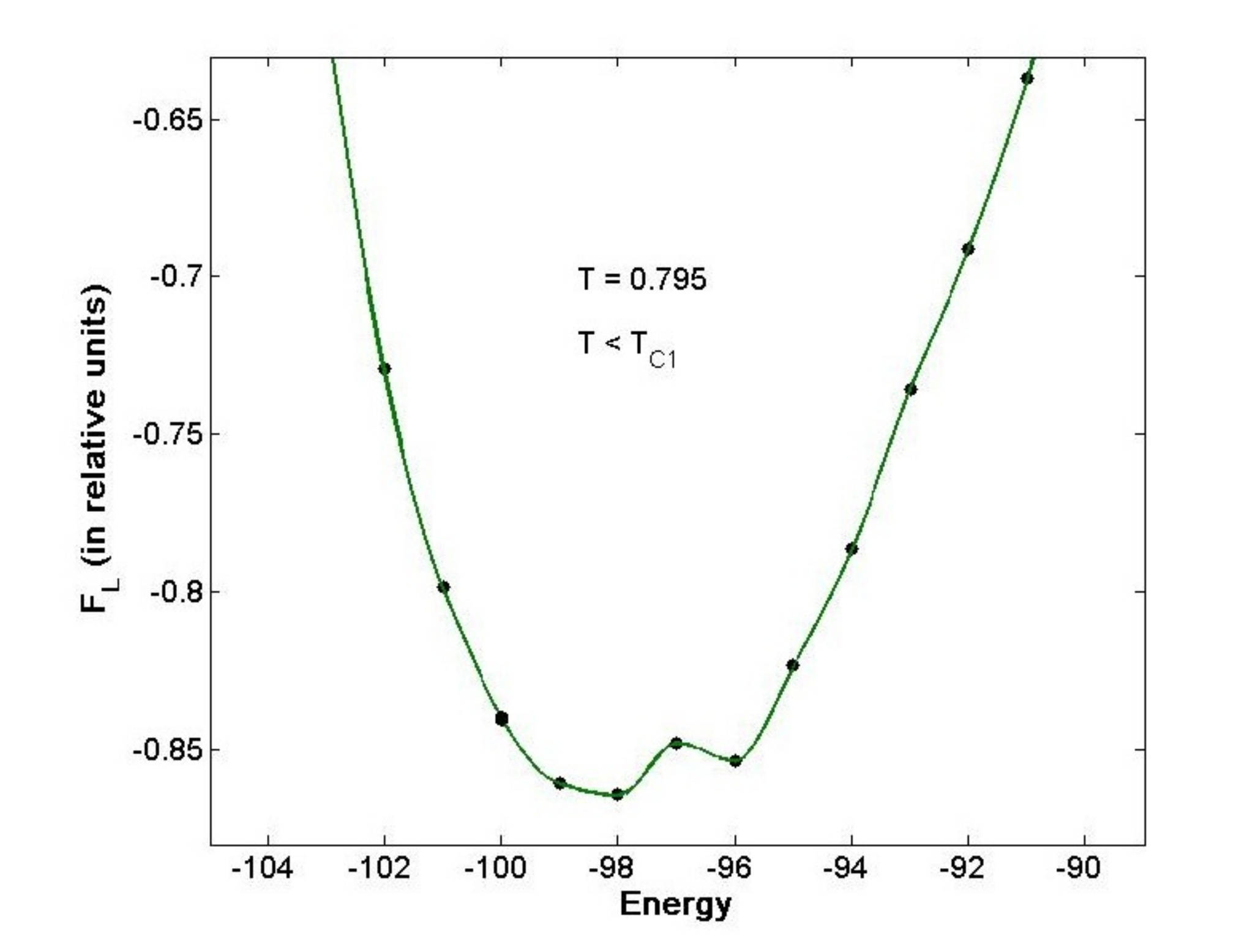}
\caption{{\small Free energy {\it versus} energy for $T =0.795\ <\ T_{C1}$ for an
isolated polymer of length $50$ monomers. Crystalline structure at
low energy is the stable; a globule structure with high energy
is meta stable
state. The values shown on Y-axis are $F_L + constant$ (constant=27965).}}
\end{center}
\end{figure}

\begin{figure}[h]
\begin{center}
\includegraphics[width=15cm]{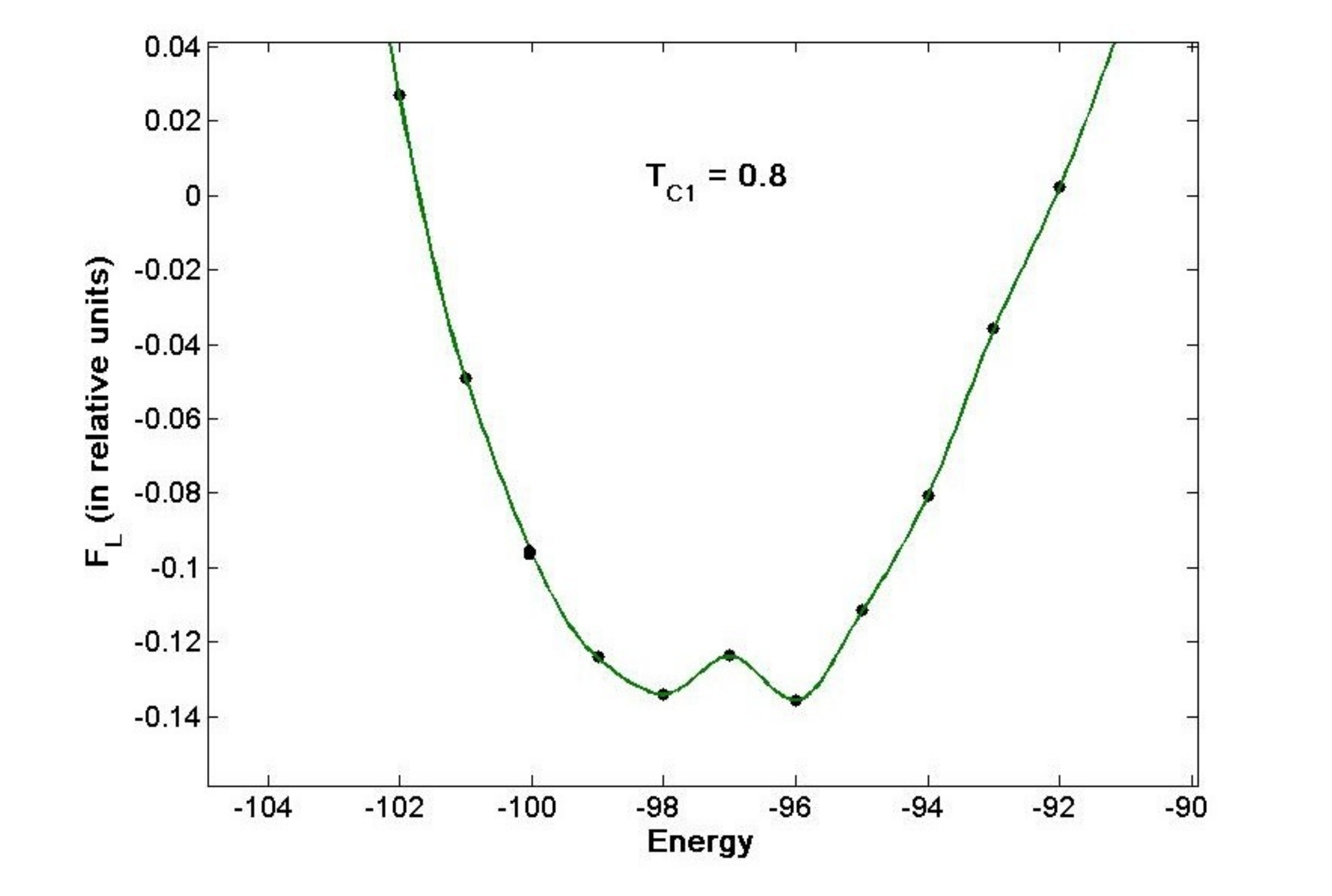}
\caption{{\small Free energy {\it versus} energy for $T = T_{C1}=0.8$ for an
isolated polymer of length $50$ monomers. Both crystalline and
globule phases coexist. The values shown on Y-axis are $F_L + constant$ (constant=28141).}}
\end{center}
\end{figure}
\begin{figure}[h]
\begin{center}
\includegraphics[width=15cm]{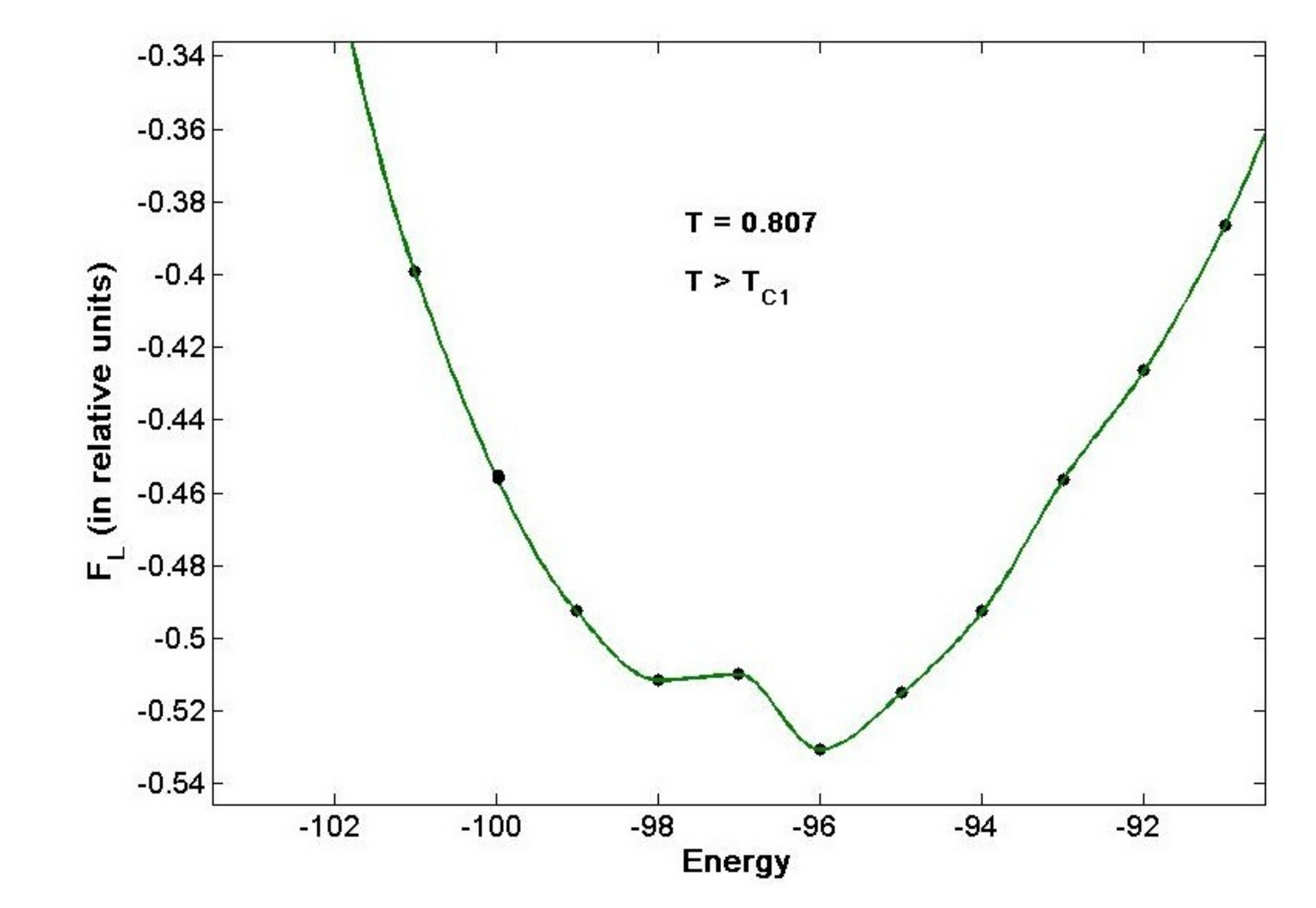}
\caption{{\small Free energy {\it versus} energy for $T=0.807\ >\ T_{C1}$ for an
isolated polymer of length $50$ monomers. Globule phase with high
energy is stable and crystalline phase with low energy
 is meta stable. The values shown on Y-axis are $F_L + constant$ (constant=28386).}}
\end{center}
\end{figure}

The character of the phase transition can also be
analysed using the  fourth order Binder's cumulant, $V_L(T)$ \cite{K.Binder1981},
defined as follows.
\begin{eqnarray}
V_L(T) = 1-\frac{\langle E^4\rangle_L}{3\langle E^2\rangle^2_L}
\end{eqnarray}
where $E$ is the energy.
 Binder's reduced cumulant, $V_L(T)$  are shown in Fig. (4.7) for
 polymers of length $10,\ 20,\ 30$ and $50$. We observe
 that the curves are typical of discontinuous phase transition.

\begin{figure}[!hp]
\begin{center}
\includegraphics[width=14cm,height=12cm]{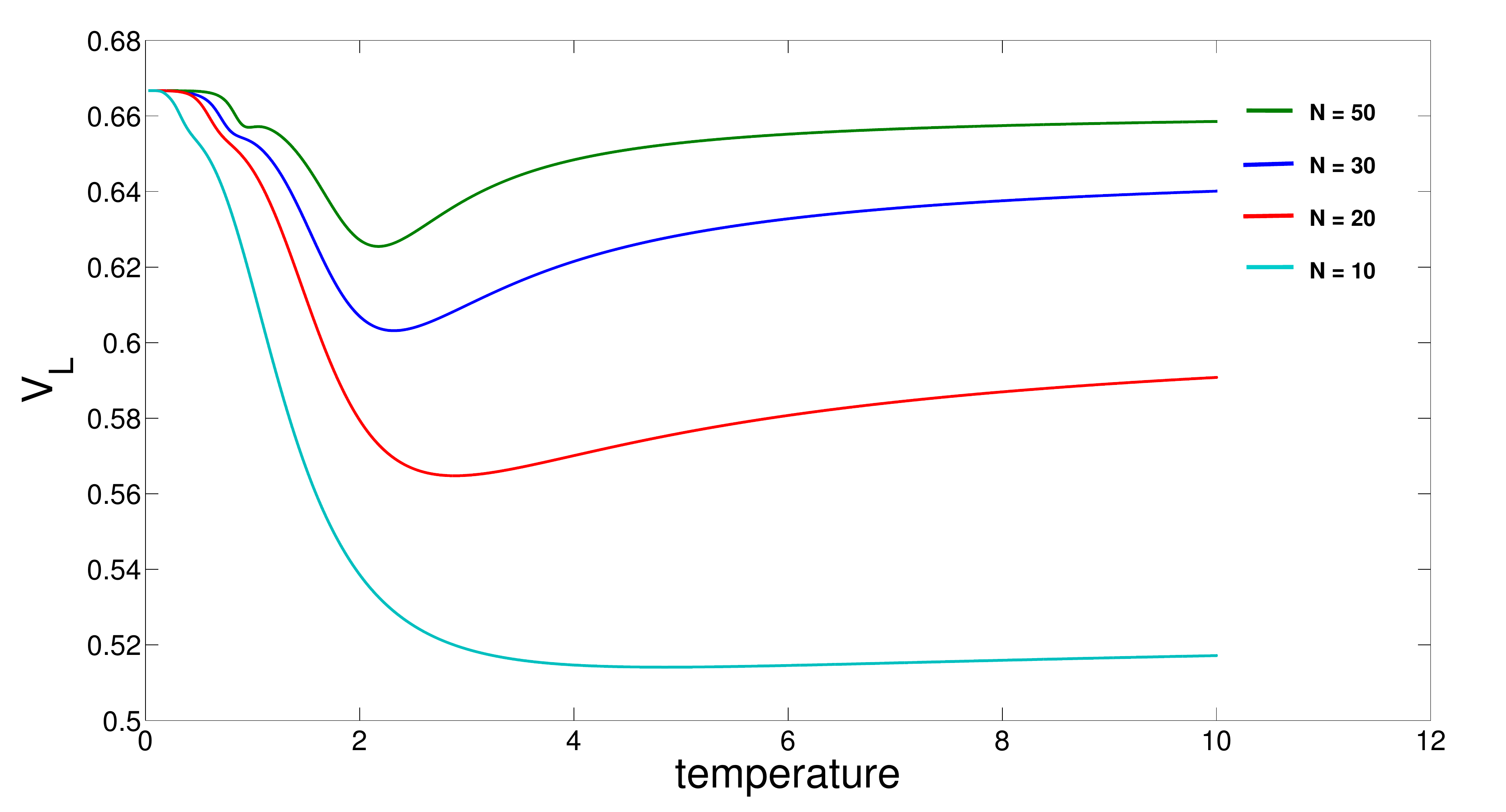}
\caption{{\small Binder's reduced cumulant {\it versus} temperature
for an isolated polymer. The transition is first order.}}
\end{center}
\end{figure}

%% file: chapters/chap5.tex
\chapter{Polymer in the presence of attractive walls}

\section{Polymer near an attractive wall}
 Polymer near an impenetrable attractive wall finds
applications in adhesion, lubrication, colloid stabilization,
chromatography, microelectronic devices, biomedical problems  {\it
etc.}\cite{D.Napper1983}. Since the wall is attractive, the
conformations that make contact with the wall have lower energy as
compared to those away and not making any contact with the wall.
However entropy would be smaller for polymers that make contact with
the wall. These competing processes would result in
adsorption-desorption transition. If, simultaneously, there is also
segment-segment attraction in the polymer, there is a  possibility
of a collapse transition both in the desorbed and in the adsorbed
states \cite{wall2009}. Since the wall is attractive, it contributes
an energy $\epsilon_s = -1$ for each contact the polymer makes with
the wall.

A contact with the wall lowers the energy of the polymer.
However it also leads to decrease in entropy.
Entropy increases rather steeply with increase of energy.
At any given temperature what kind of conformation a polymer takes
is completely determined by the competition between
energy and entropy.
At low temperatures a polymer prefers to get  adsorbed on the surface
since it leads to lower energy. At high temperatures
the polymer would prefer to move away from the wall since that facilitates increase of entropy. Hence desorbed behaviour would prevail at higher temperature.

We model the wall as a long straight polymer
fixed parallel to y-axis. This fixed polymer has fixed bond length of $2$.
Each monomer of the wall occupies four lattice sites see Figs.~(5.2,~\ 5.3).
Let $\epsilon_s$ denote the energy associated with a contact between a
monomer of the polymer and that  of the wall.
Let $N_s$ be the number of contacts between the polymer and the
wall. Let $N_u$ denote the number of nbNN
contacts in the polymer. Total energy of the polymer is thus
given by
 $E = N_s\epsilon_s + N_u\epsilon$.

In our work we have set
$\epsilon_s = \epsilon = -1$. In other words the segment-segment interaction and polymer-wall interaction
are both treated in identical fashion
\footnote{In principle one can take  $\epsilon_s$ different from
$\epsilon$ {\it i.e.} adsorption strength may be less or more compared to self interaction of polymer.}.
We employ Wang-Landau algorithm, as described
chapter (3),  to simulate an entropic ensemble of
polymer conformations. We have also employed Metropolis
algorithm to generate typical equilibrium  conformations
at the desired temperatures.

We have obtained converged density of states for polymer near an
attractive wall. All the standard machineries  we had employed and described
in the third
and fourth chapters are again used here to calculate
 heat capacity,  Landau
free energy and Binder's cumulant.

The heat capacity curves for polymer of length
$N=40$ and $50$ are shown in the Fig. (5.1). The sharp peak corresponds to
transition from adsorbed collapsed state to adsorbed extended state
and the transition temperature is  $T_C=0.88$ . We conjecture
that the broad shoulder corresponds to adsorption-desorption transition.
However we are not able to characterize this transition in
terms of free energy profiles or
Binder's cumulant. May be  we should simulate longer polymers to get
calculate these quantities.

\begin{figure}[!hpb]
\begin{center}
\includegraphics[width=12cm]{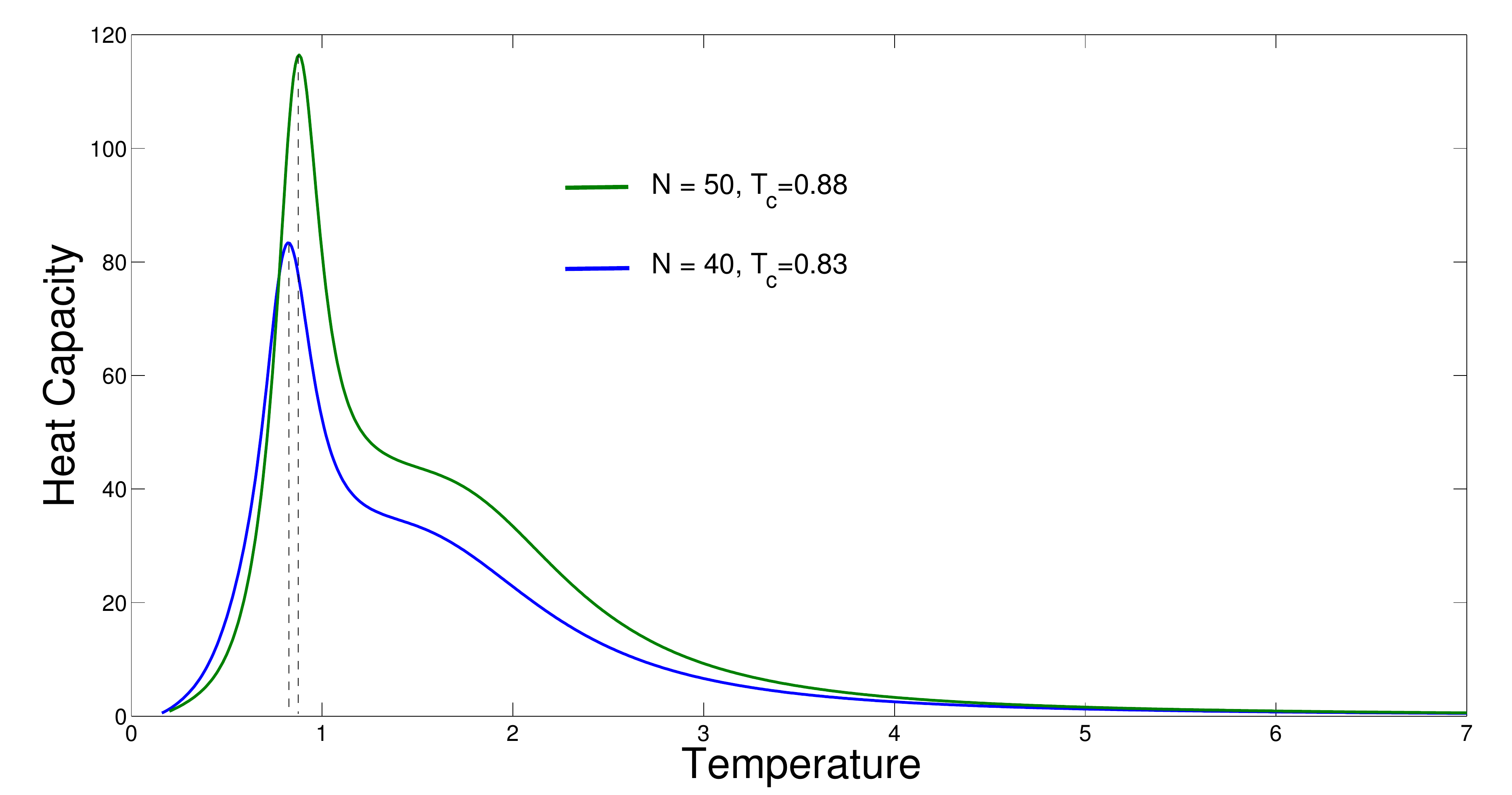}
\caption{{\small Heat capacity curves for  polymers of length $40$ and $50$
monomers in the presence of  an attractive wall. The sharp peak at $T_C$=0.88
corresponds to transition from adsorbed-extended phase to
adsorbed-collapsed phase.
}}
\end{center}
\end{figure}

At low temperature $(T =0.7\ <\ T_C)$, a part of the polymer gets
stuck to the wall and the dangling part forms a layer over it. This
results in a  compact structure adsorbed to the wall with lower
energy. At high temperature \\ $(T=1.3\ >\ T_C)$, the polymer tries
to increase its entropy by extending itself over the wall. At very
high temperatures $(T>2.6)$ the polymer gets detached from the wall
and behaves like an isolated polymer. Figures (5.2 and 5.3) show
conformations of a polymer of length $50$ monomers at a low
temperature $T=0.7$ and at high temperature $T=1.3$ respectively.
Figure (5.4) shows polymer far away from the wall at very high
temperature $T=2.8$.

\begin{figure}[!hp]
\begin{center}
\hglue -10mm\includegraphics[width=18cm,height=11cm]{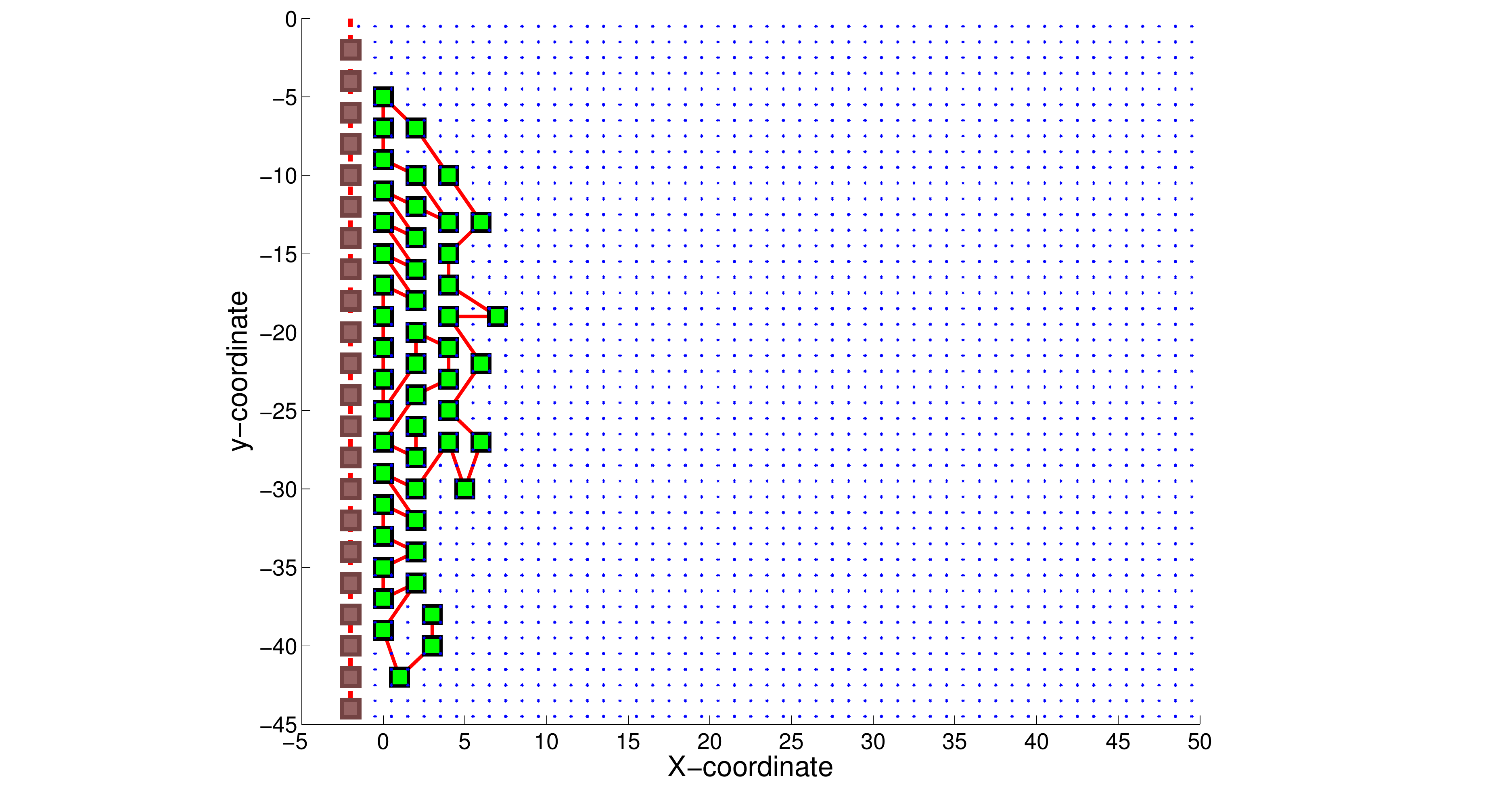}
\caption{{\small Compact conformation of a polymer of length $50$ monomers
in the presence of an attractive wall
at $T=0.7\ <\ T_C$; the radius of gyration
of this conformation is $9.7666$. This figure shows a segment of the polymer adsorbed to
the wall;  the remaining segment forms a layer over the adsorbed
part.}}
\end{center}
\end{figure}

\begin{figure}[!hp]
\begin{center}
\includegraphics[width=14cm,height=15cm]{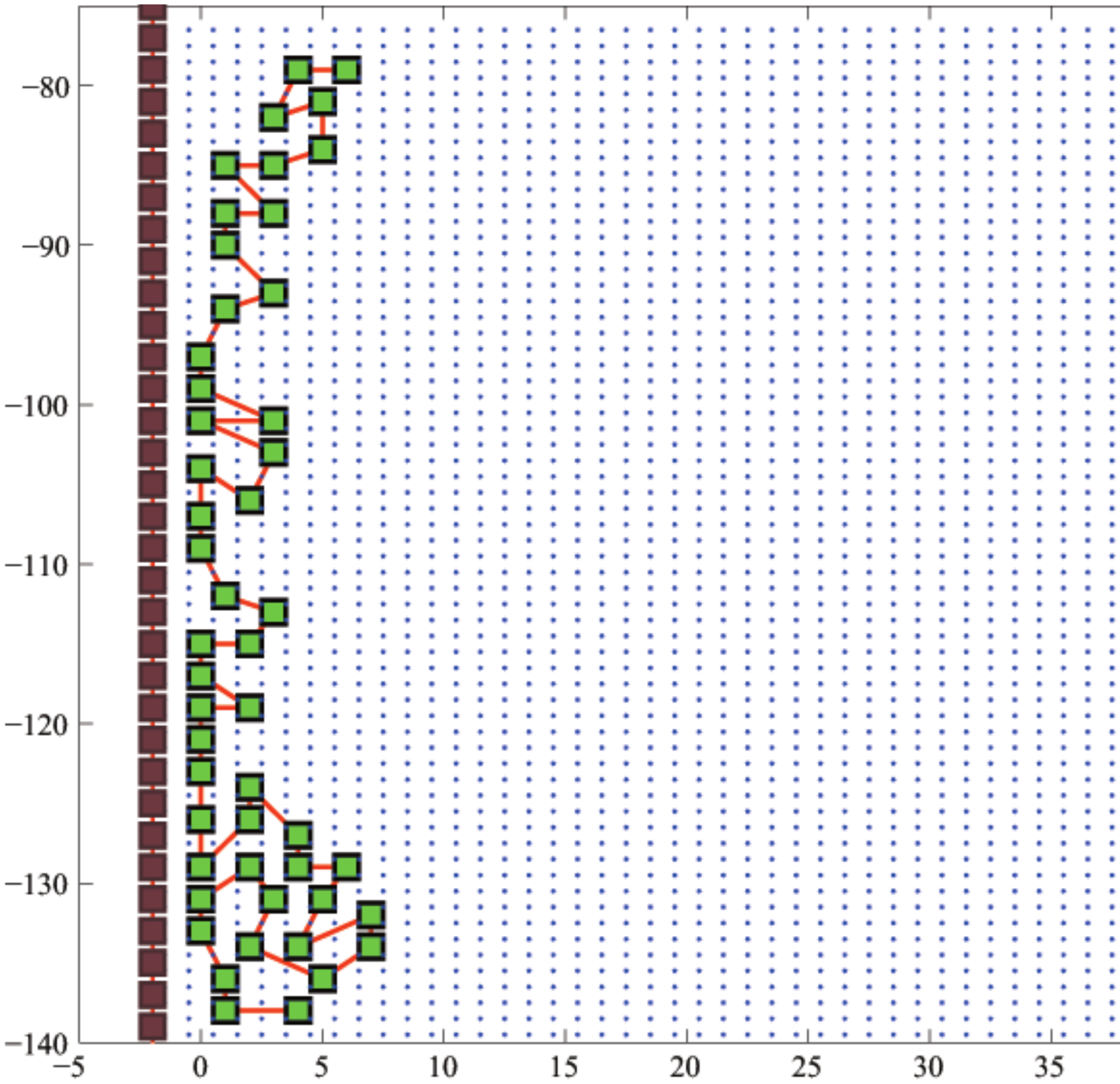}
\caption{{\small Extended conformation of polymer of length $50$ monomers
in the presence of  an attractive wall
at $T=1.3\ >\ T_C$; the  radius of gyration
of this conformation is $18.8566$. This figure shows conformation
extended over the
wall.}}
\end{center}
\end{figure}

\begin{figure}[!hp]
\begin{center}
\includegraphics[width=17cm,height=12cm]{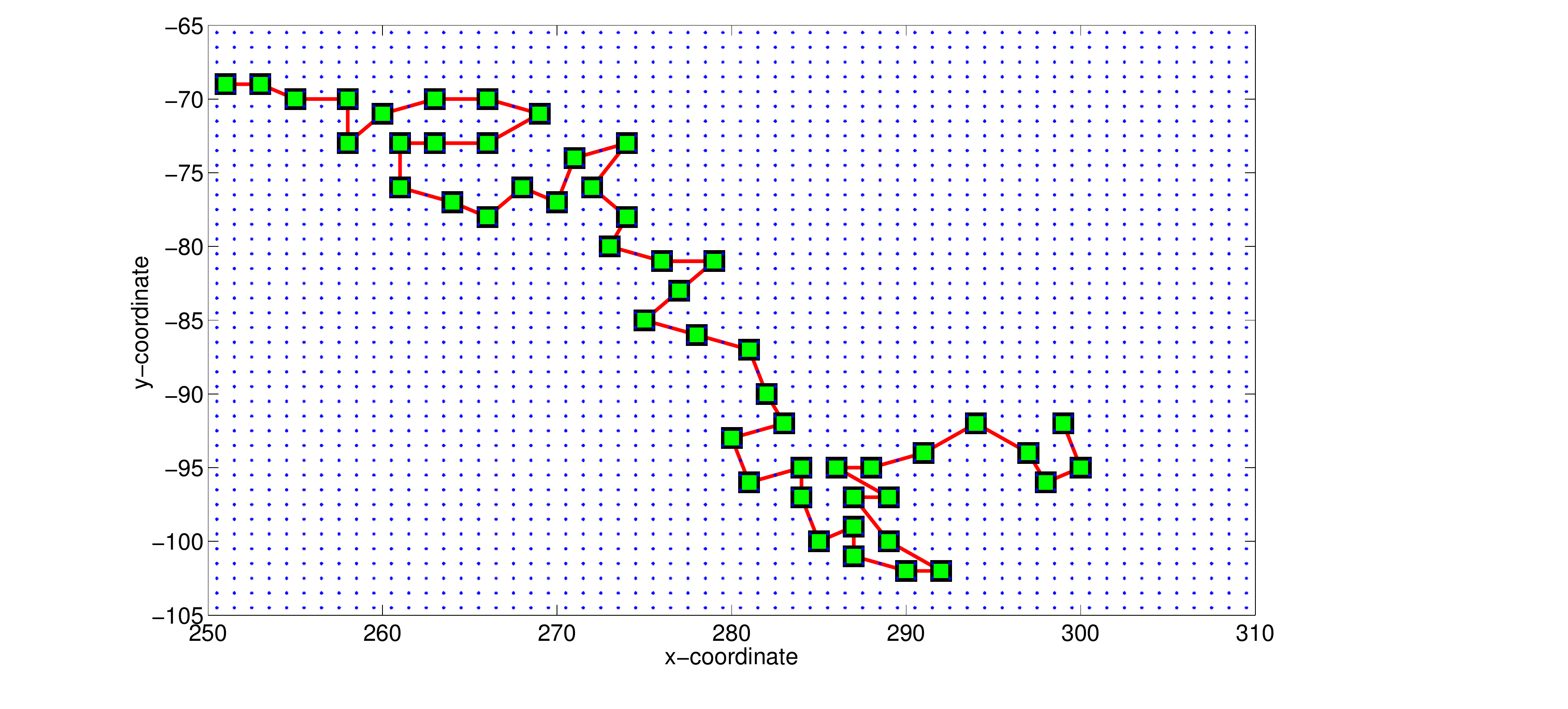}
\caption{{\small Conformation of polymer of length $50$ monomers
detached from an attractive wall at $T=2.8$; the
radius of gyration of this conformation is
$16.9655$ \newline The conformations shown here  are obtained by
simulating bond fluctuation model of a polymer of length
$50$ monomers
near an attractive wall employing Metropolis algorithm.}}
\end{center}
\end{figure}

By analysing a narrow  energy region where the coil
globule transition  occurs, we have obtained specific heat curves
for polymer of length $10,\ 20,\ 30$ and  $50$ monomers in the presence of
the wall. These are shown in Figs. (5.5 - 5.8). For comparison we have reproduced
the specific heat profiles for
a free polymer.
From the figures, $T_C$ for a
polymer near wall is found to be less than $T_C$ for isolated
polymer. This difference is larger for longer polymers.

\begin{figure}[!hp]
\begin{center}
\includegraphics[width=15cm]{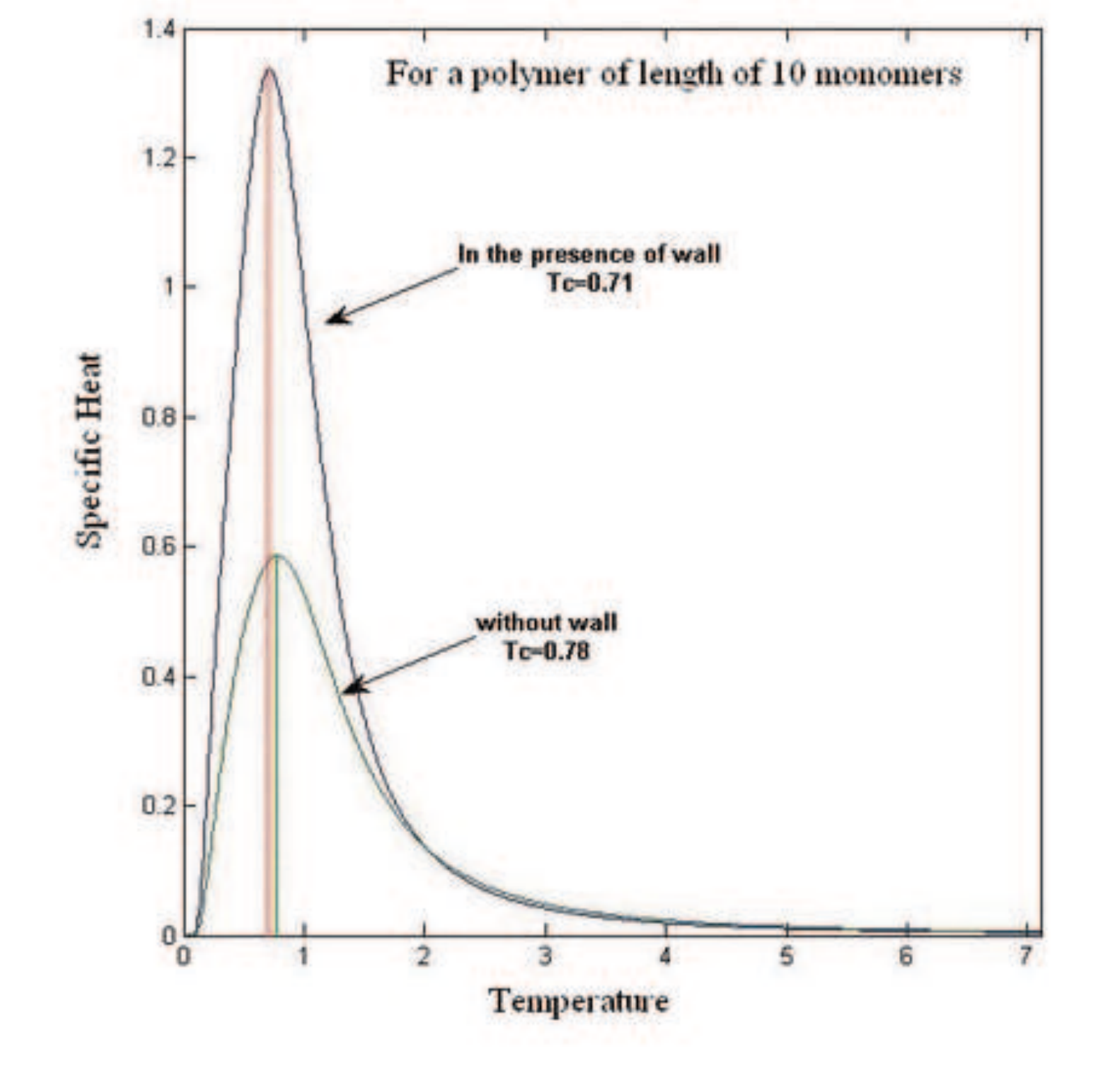}
\caption{{\small Specific heat curves for a polymer of length $10$ monomers
with and without wall.}}
\end{center}
\end{figure}
\begin{figure}[!hp]
\begin{center}
\includegraphics[width=15cm]{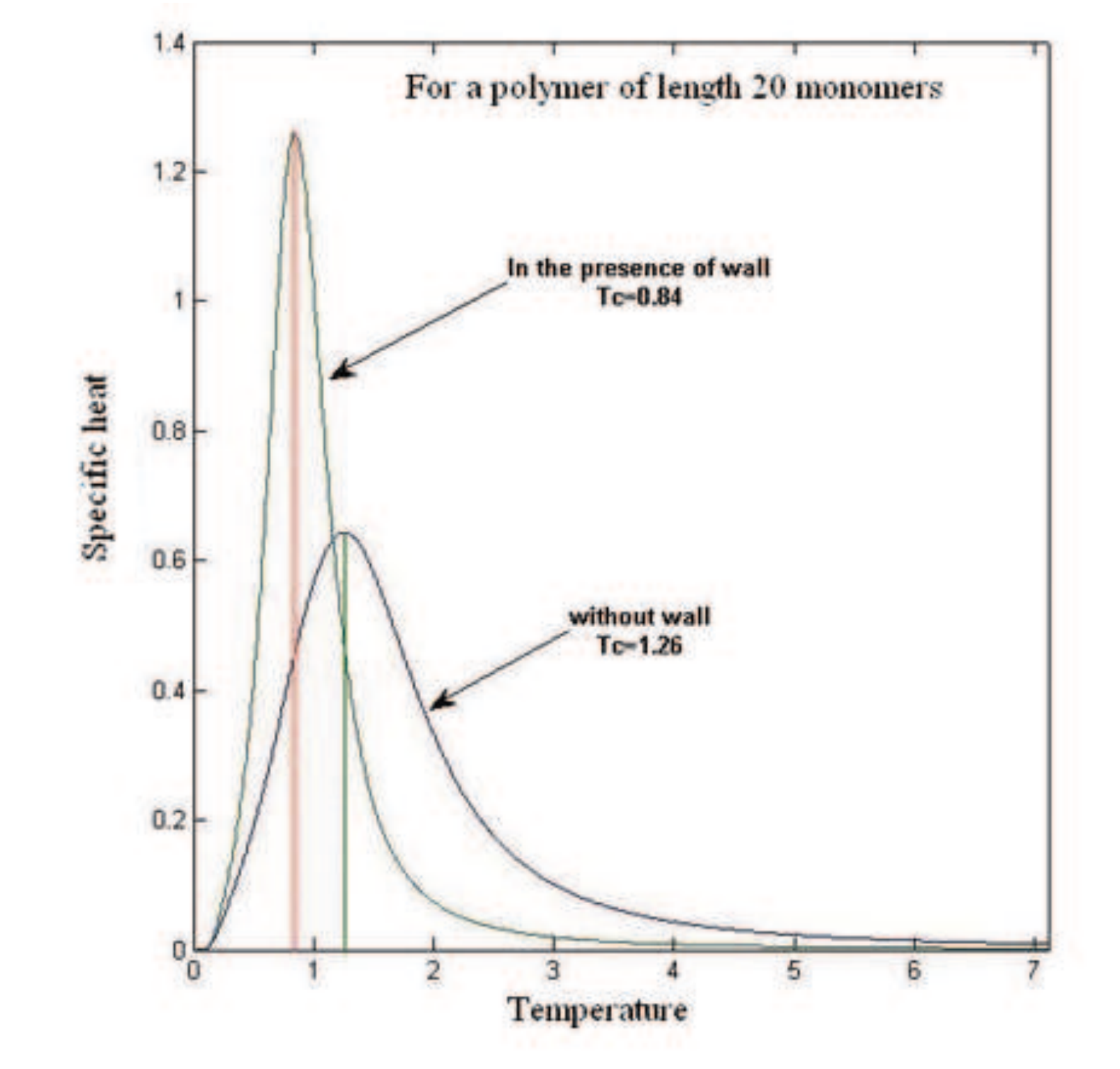}
\caption{{\small Specific heat curves for a polymer of length $20$ monomers
with and without wall.}}
\end{center}
\end{figure}
\begin{figure}[!hp]
\begin{center}
\includegraphics[width=15cm]{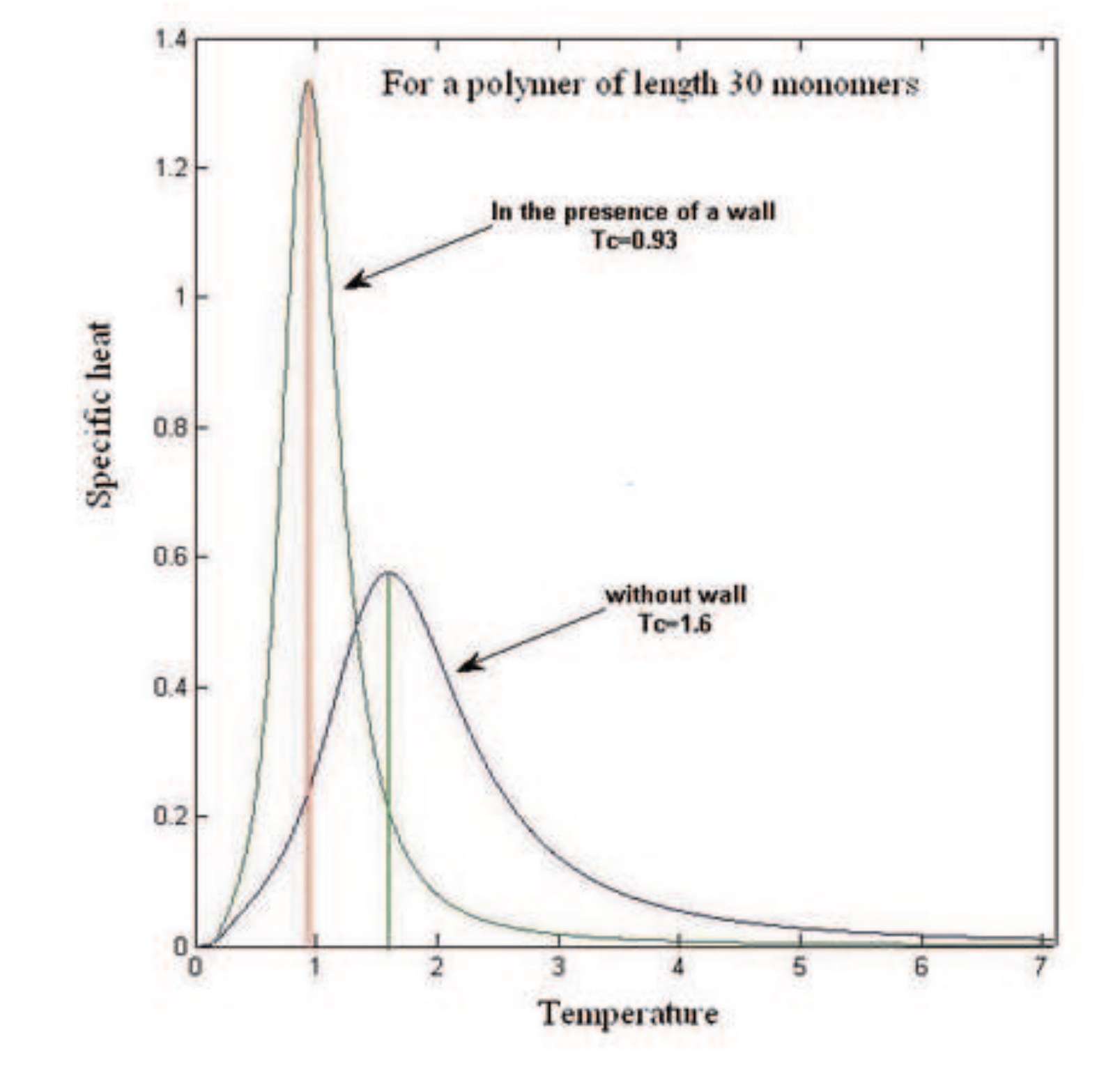}
\caption{{\small Specific heat curves for a polymer of length $30$ monomers
with and without wall.}}
\end{center}
\end{figure}
\begin{figure}[!hp]
\begin{center}
\includegraphics[width=15cm]{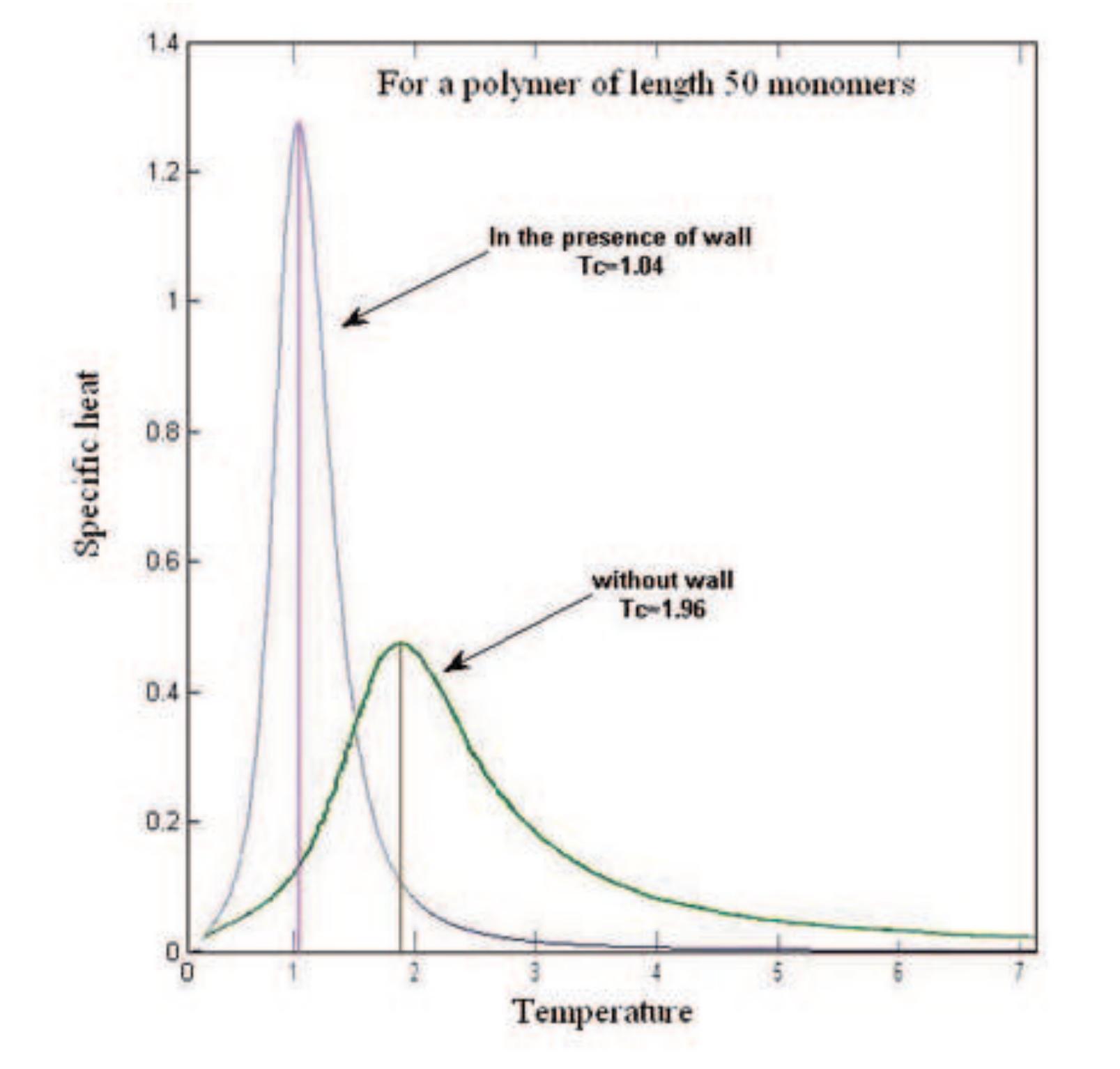}
\caption{{\small Specific heat curves for a polymer of length $50$ monomers
with and without wall.}}
\end{center}
\end{figure}
From specific heat curve of a polymer of length  $50$ in the presence of a wall
the transition temperature is found to be $1.04$. We have obtained
Landau free energy profile for transition from adsorbed-collapsed
state to adsorbed-extended state. These are
shown in Figs. (5.9, 5.10 and  5.11).  From Landau free energy
 profile, critical temperature is found to be $0.91$. The  transition
 from adsorbed-extended to adsorbed-collapsed phase is
 found to be first order. Figure (5.12) shows Binder's fourth
cumulant curves which again confirms that the transition is first order.
\begin{figure}[!hp]
\begin{center}
\hglue -10mm\includegraphics[width=15cm,height=11cm]{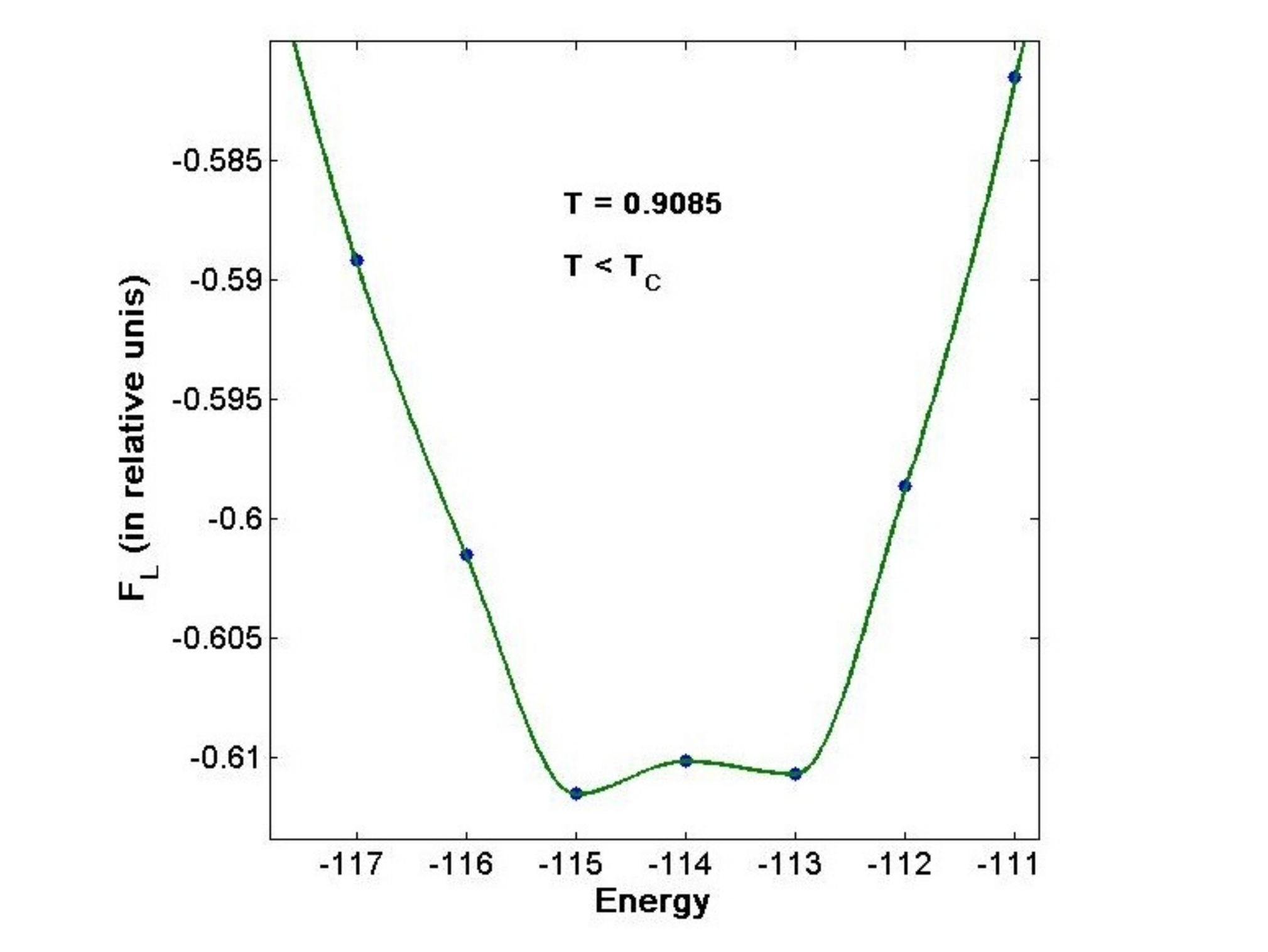}
\caption{{\small Free energy {\it versus} energy at $T=0.9085\ <\ T_C$ for a
polymer of length $50$ monomers near wall. he values shown on Y-axis are $F_L + constant$ (constant=45611864).}}
\end{center}
\end{figure}
\begin{figure}[!hp]
\begin{center}
\includegraphics[width=15cm]{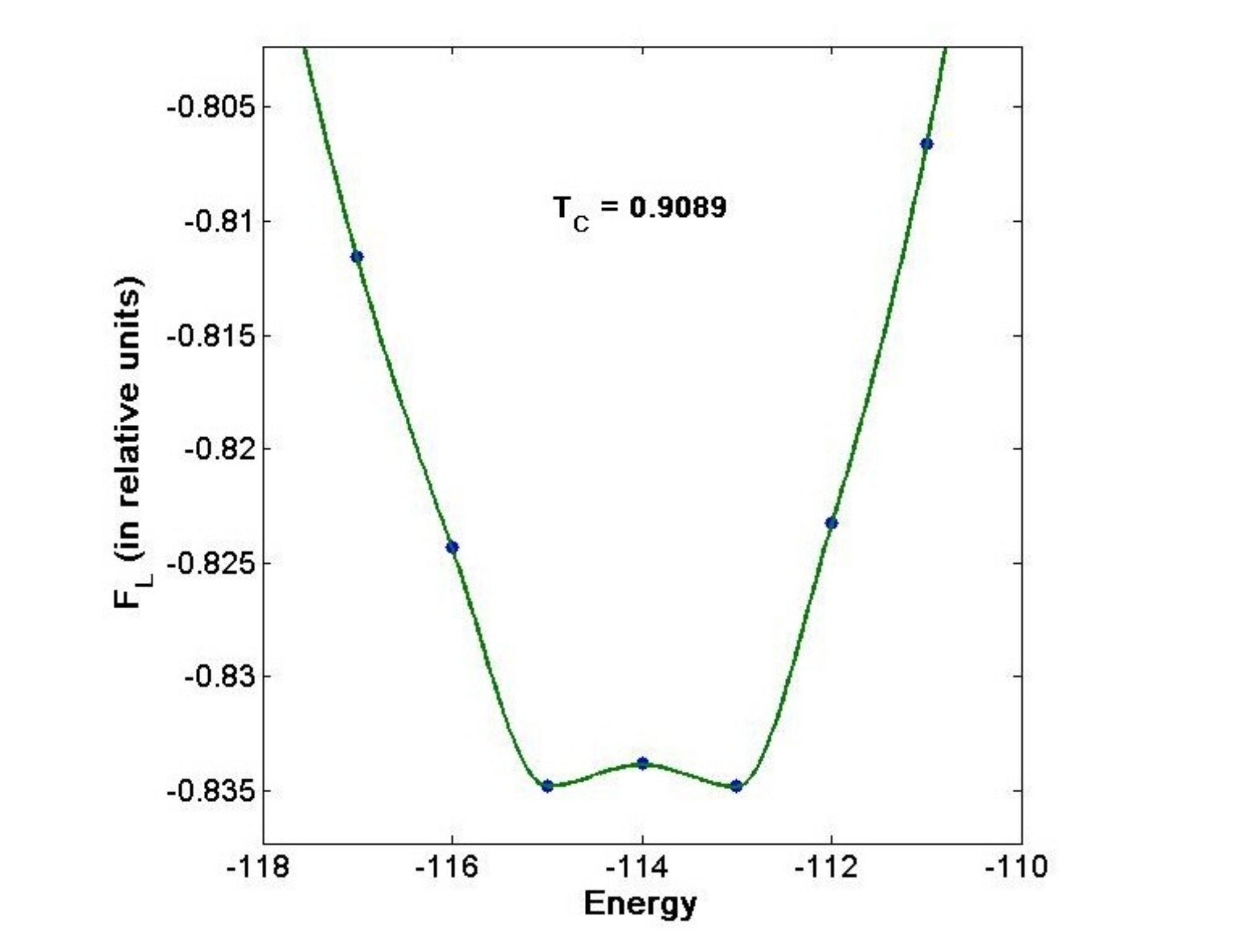}
\caption{{\small Free energy {\it versus} energy at
$T=T_C=0.9089$ for a
polymer of length $50$ monomers near wall. The values shown on Y-axis are $F_L + constant$ (constant=45631946).}}
\end{center}
\end{figure}
\begin{figure}[!hp]
\begin{center}
\includegraphics[width=15cm]{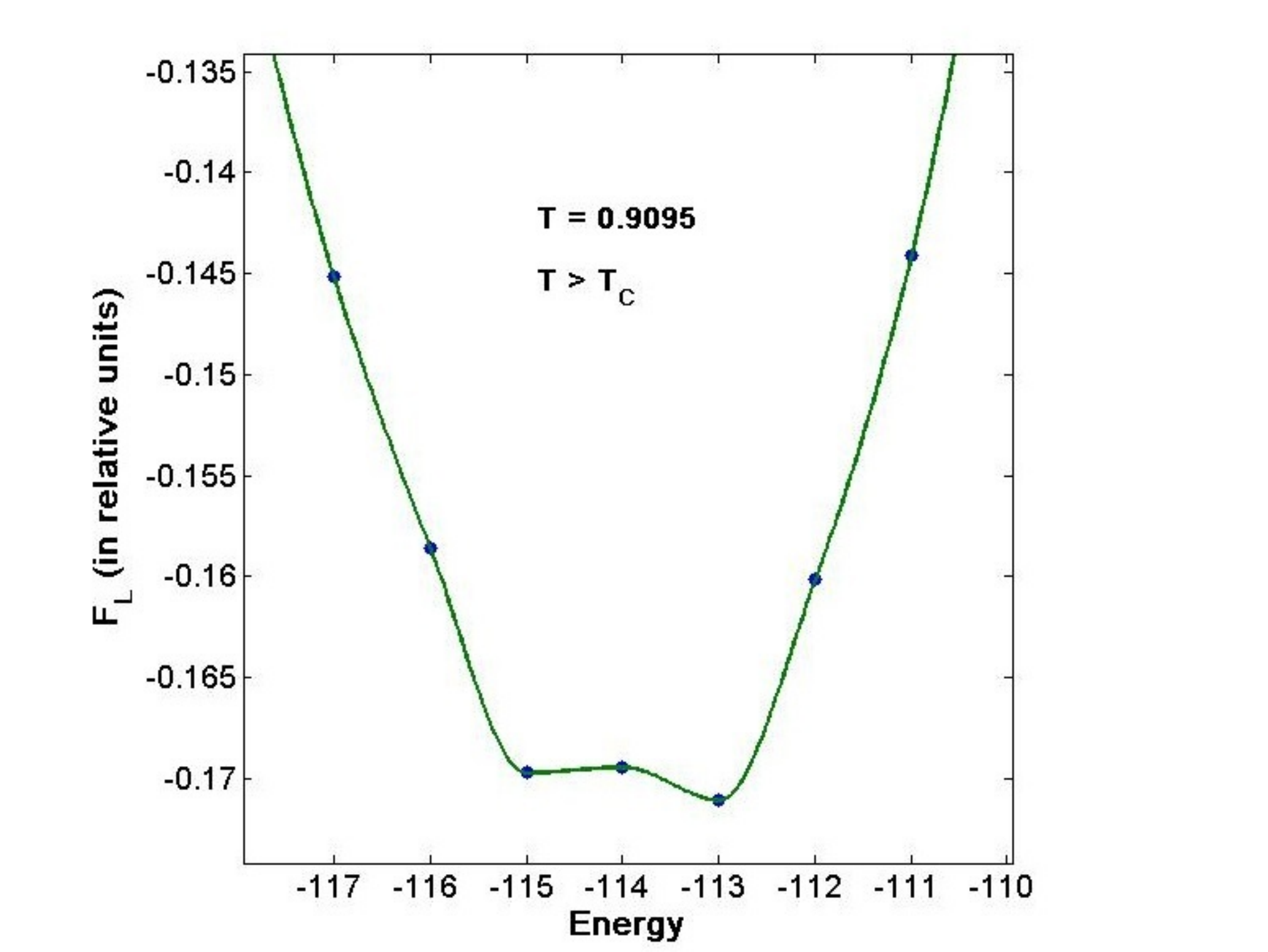}
\caption{{\small Free energy $\it versus$ energy above $T=0.9095\ >\ T_C$ for a polymer of length $50$ monomers near wall. The values shown on Y-axis are $F_L + constant$ (constant=45662070).}}
\end{center}
\end{figure}
\begin{figure}[!hp]
\begin{center}
\includegraphics[width=14cm,height=12cm]{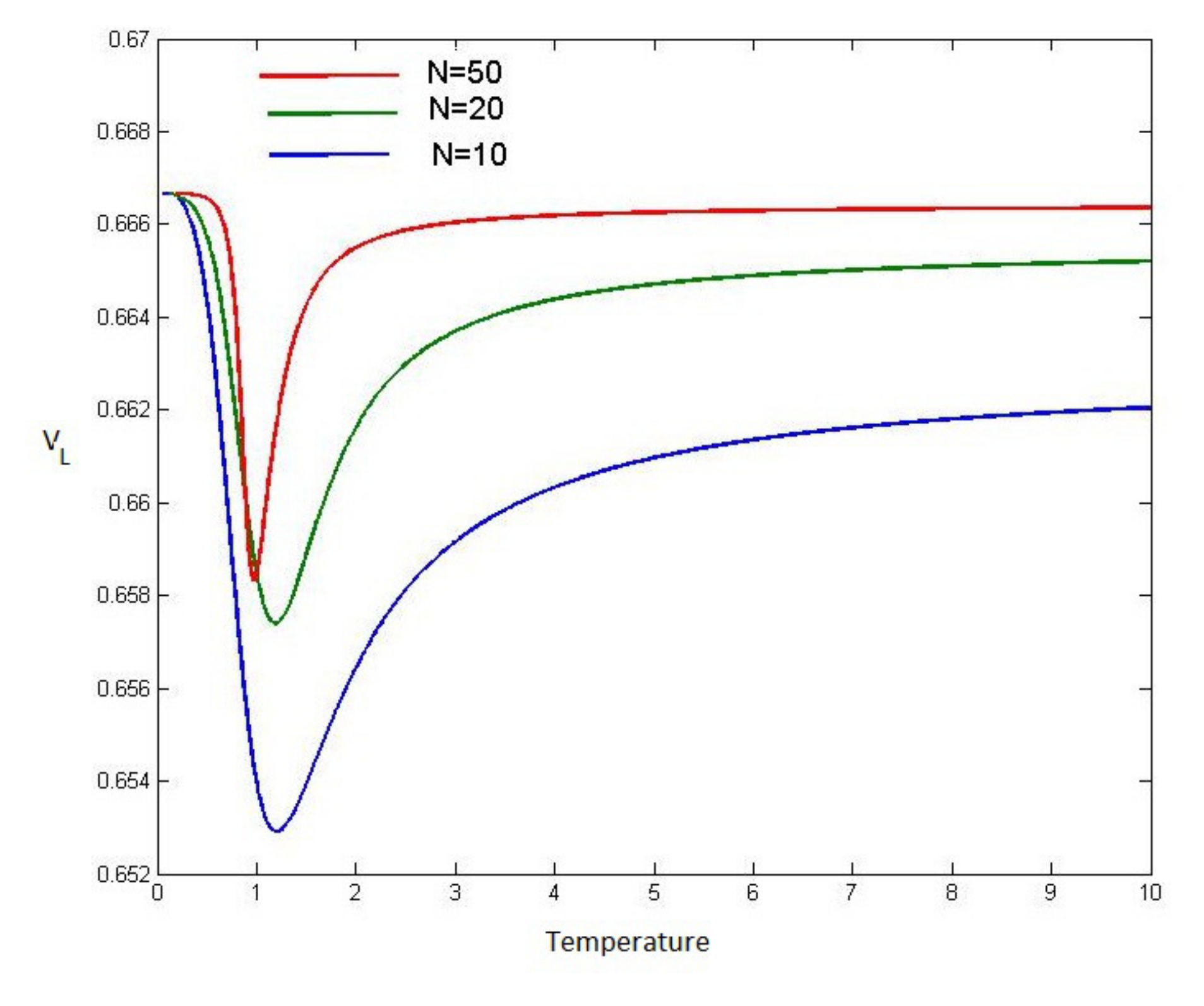}
\caption{{\small Binder's cumulant {\it versus} temperature for a polymer
near an attractive wall. The transition is discontinuous.}}
\end{center}
\end{figure}

\section{Polymer confined between two attractive walls}
In this section we study coil to globule transition of a polymer
confined between two attractive parallel impenetrable walls in two
dimensional lattice \cite{yoshima1999}. We model the walls in
the same way as we did in case of single wall. The interaction with
walls is taken as the same as monomer-monomer interactions. The phase
transition not only depends on the monomer-monomer interaction and
polymer - wall interaction but also on the distance between the
walls. If the walls are far separated they will
 have no effect on the polymer. Hence  in this work
 we have taken the separation distance
 as  $12$ lattice units and simulated polymers of lengths
 $N=10,\ 20,\ 30$ and $40$ monomers.

Employing bond fluctuation model and Wang-Landau algorithm we
get converged density of states of a polymer confined between two
walls.

Heat capacity curve shown in Fig. (5.13) gives the coil globule
transition  temperature as $0.89$ for a polymer of size $40$
monomers. Figure (5.14) shows a compact conformation of the polymer
of length $40$ monomers confined between the walls separated by a
distance of $12$ lattice units at low temperature of $T=0.74\ <\
T_C)$; the radius of gyration of the polymer conformation shown in
the figure is $6.0413$. Figure (5.15) shows
 extended conformation of the same polymer at high temperature of
$T=1.6\ >\ T_C)$; the radius of gyration of this conformation
is  $9.9539$. These
conformations have been  obtained employing Metropolis algorithm.

\begin{figure}[!hp]
\begin{center}
\includegraphics[width=14cm,height=10cm]{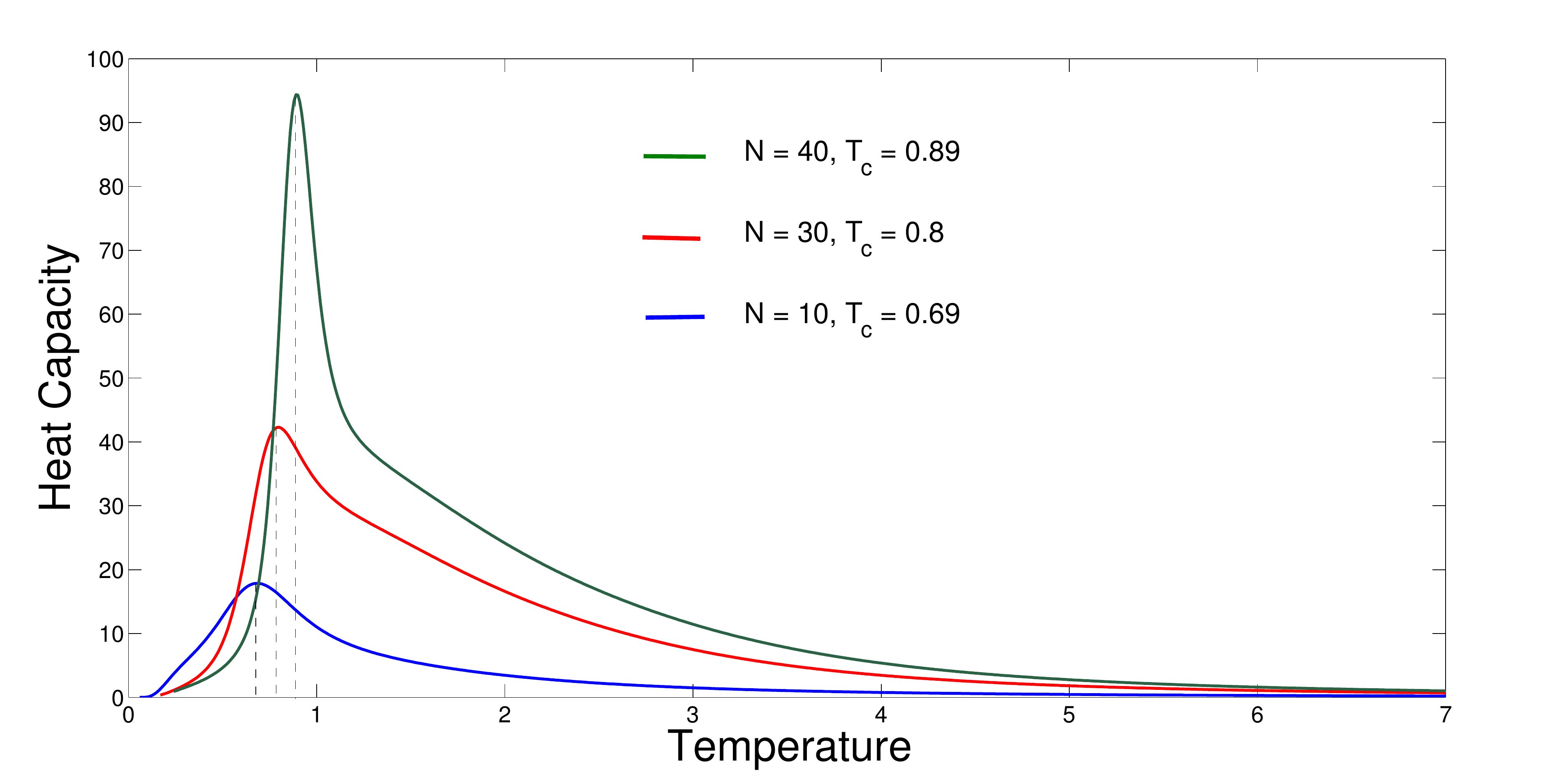}
\caption{{\small Heat capacity profile for polymers of length $10,\
30\ and\ 40$ monomers confined between two walls separated by
distance of $12$ units. Sharp peak at $T_C=0.89$, for a polymer of
length $40$ monomers,  indicates coil-globule transition.}}
\end{center}
\end{figure}

\begin{figure}[!hpt]
\begin{center}
\includegraphics[width=12cm]{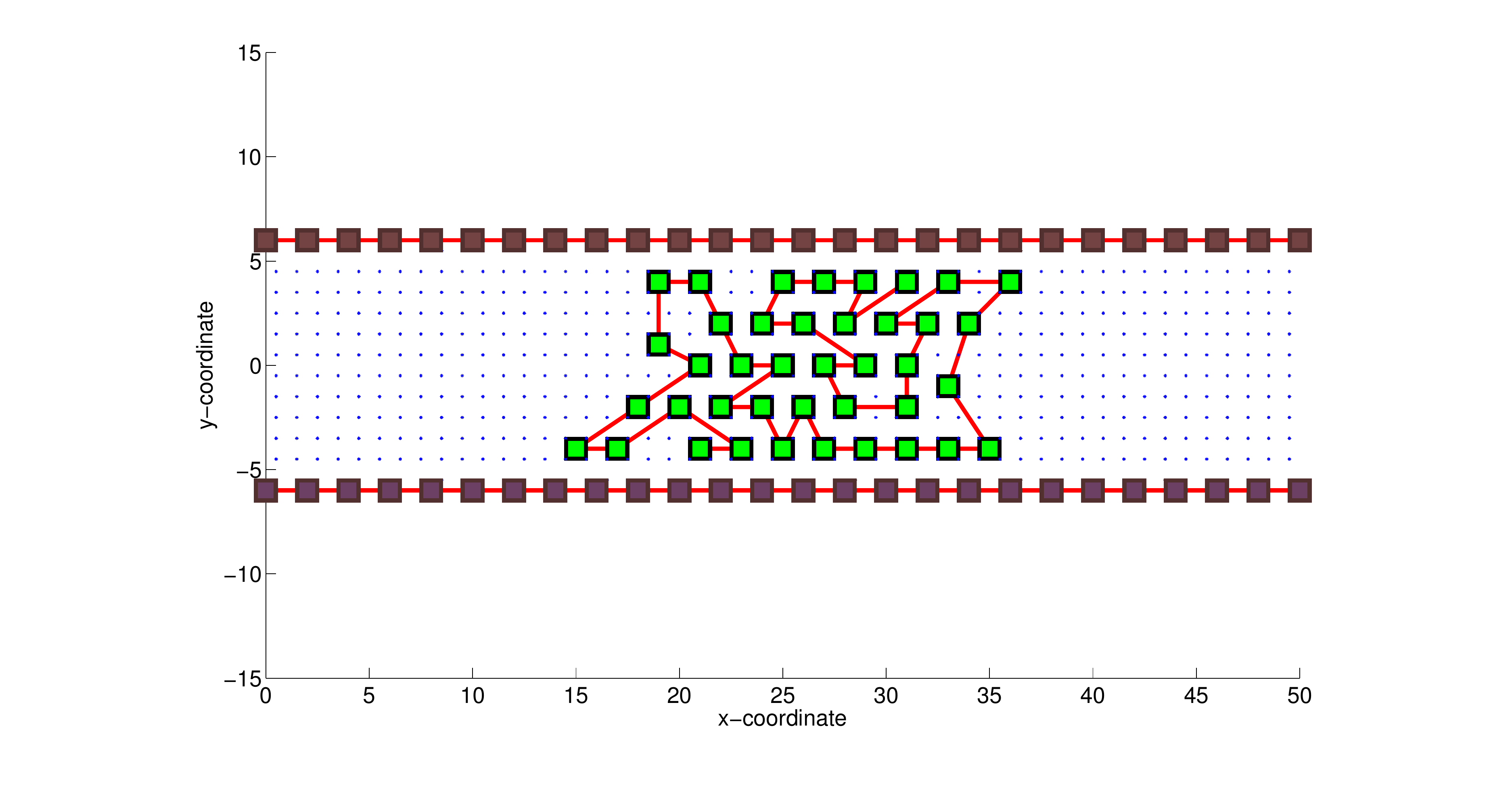}
\caption{{\small Compact conformation of a polymer of length $40$ monomers
confined between two walls separated by distance of $12$ units at
$T=0.74\ <\ T_C$. The  radius of gyration for this conformation is $6.0413$.
This conformation belongs to a canonical ensemble simulated employing
Metropolis algorithm.}}
\end{center}
\end{figure}

\begin{figure}[!hpb]
\begin{center}
\includegraphics[width=12cm]{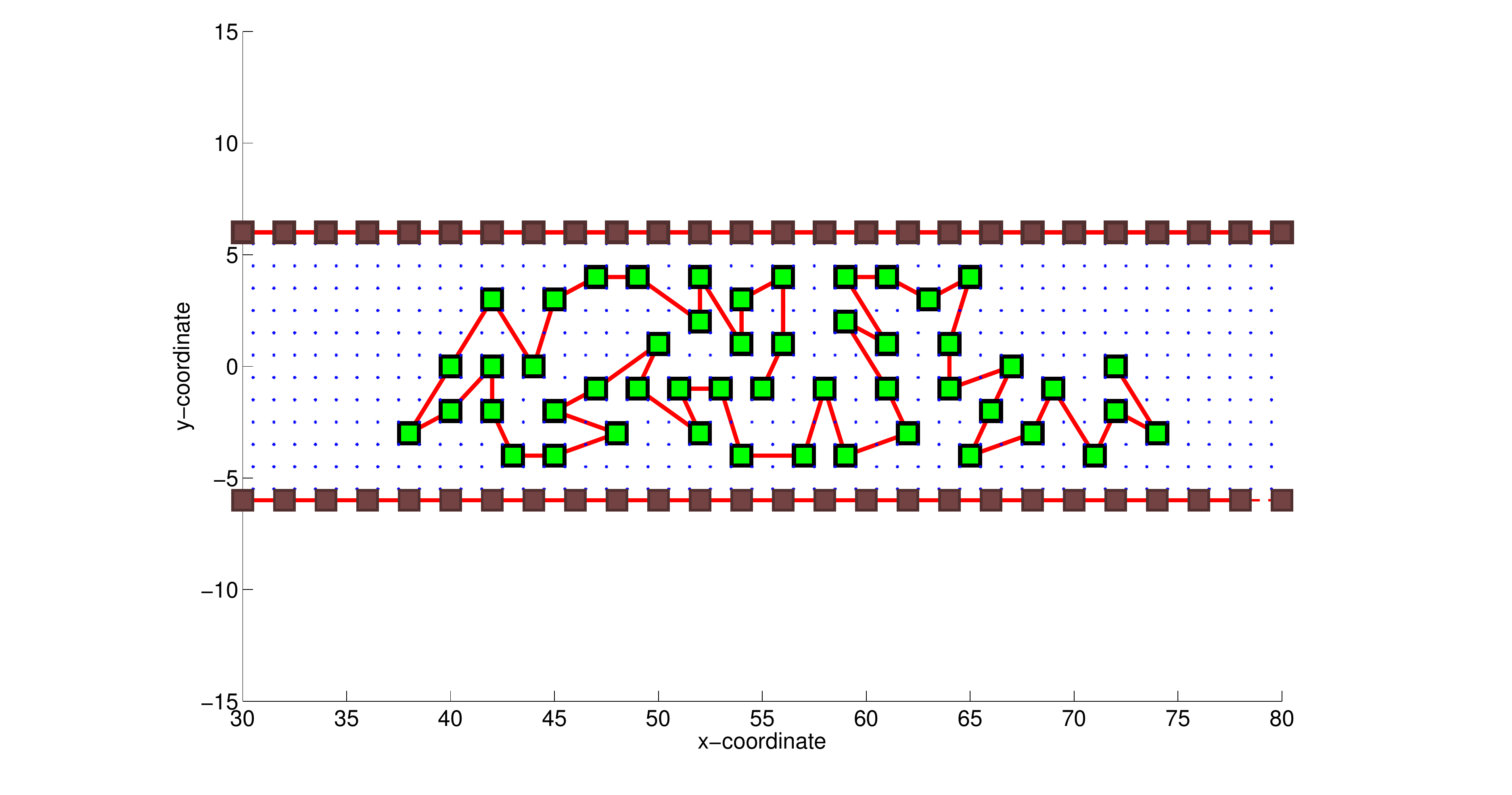}
\caption{{\small Extended conformation of polymer of
length $40$ monomers
confined between two walls separated by distance of $12$ lattice
units at
$T=1.6\ >\ T_C$; the radius of gyration for this conformation is $9.9539$.
This conformation belongs to a canonical ensemble simulated employing
Metropolis algorithm.}}
\end{center}
\end{figure}

We have obtained Landau free energy profile for transition from
extended phase to collapsed phase for a confined polymer of length
$40$. These are shown in Figs. (5.16, 5.17 and 5.18).  From Landau
free energy profile, critical temperature is found to be $0.89$. The
transition is
 first order. Figure (5.19) shows Binder's fourth cumulant
 which confirms that the transition is discontinuous.

\begin{figure}[!hp]
\begin{center}
\includegraphics[width=15cm]{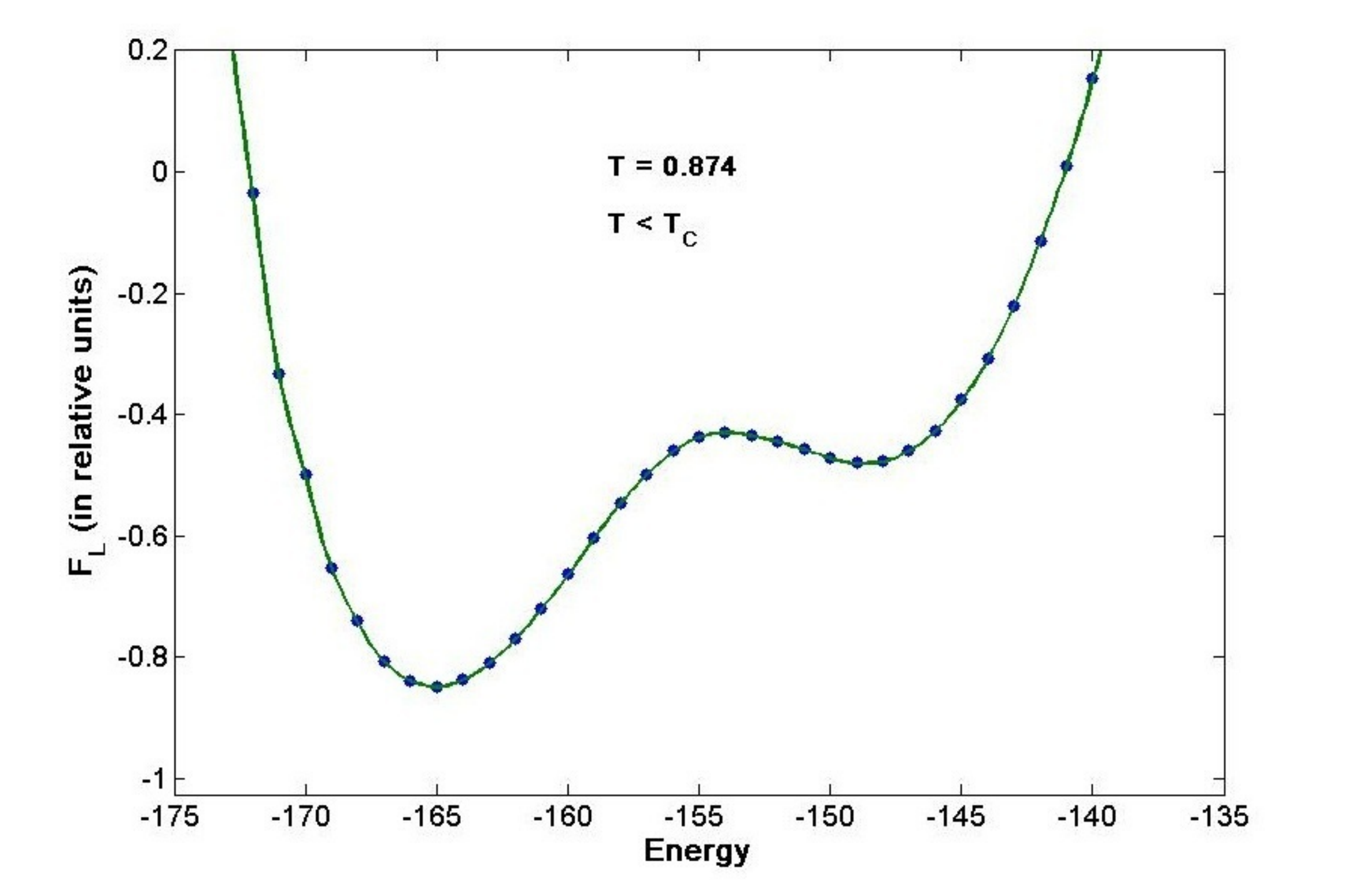}
\caption{{\small Landau free energy {\it versus} energy
for a polymer of length $40$
monomers confined between walls separated by $12$ lattice units; the
temperature is $T=0.874\ <\ T_C)$. The values shown on Y-axis are $F_L + constant$ (constant=49378985). }}
\end{center}
\end{figure}
\begin{figure}[!hp]
\begin{center}
\includegraphics[width=15cm,height=11cm]{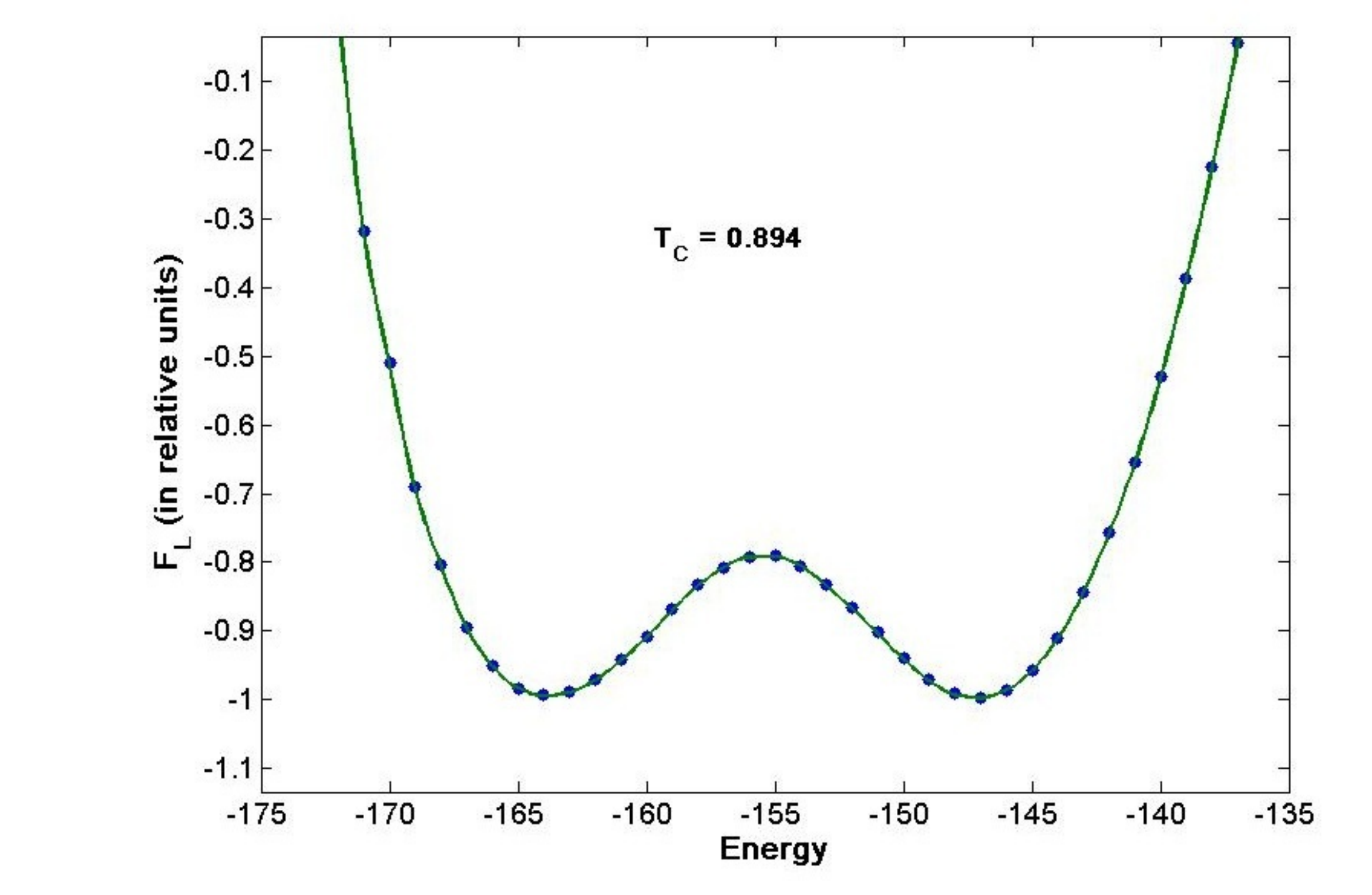}
\caption{{\small Landau free energy {\it versus}
energy for a polymer of length $40$
monomers confined between walls separated by $12$ lattice
units; the
temperature is $T=T_C=0.891$. The values shown on Yaxis are $F_L + constant$ (constant=50508935).}}
\end{center}
\end{figure}
\begin{figure}[!hp]
\begin{center}
\includegraphics[width=15cm,height=11cm]{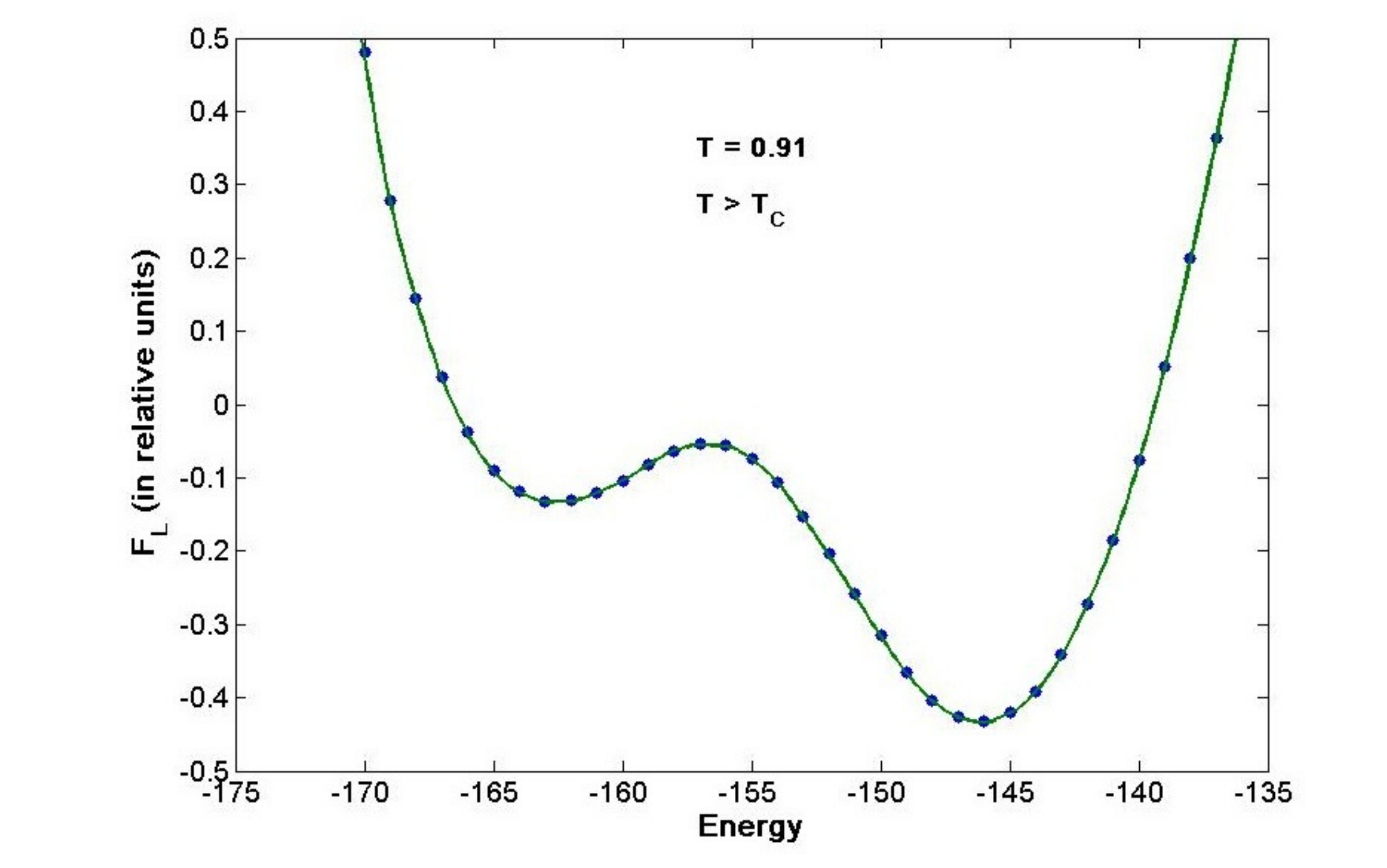}
\caption{{\small Landau free energy {\it versus} energy for a
polymer of length $40$ monomers confined between walls separated by
$12$ lattice units; the temperature is $T=0.91\ >\ T_C$. The values shown on Y-axis are $F_L + constant$ (constant=51412896).}}
\end{center}
\end{figure}

\begin{figure}[!hp]
\begin{center}
\includegraphics[width=15cm]{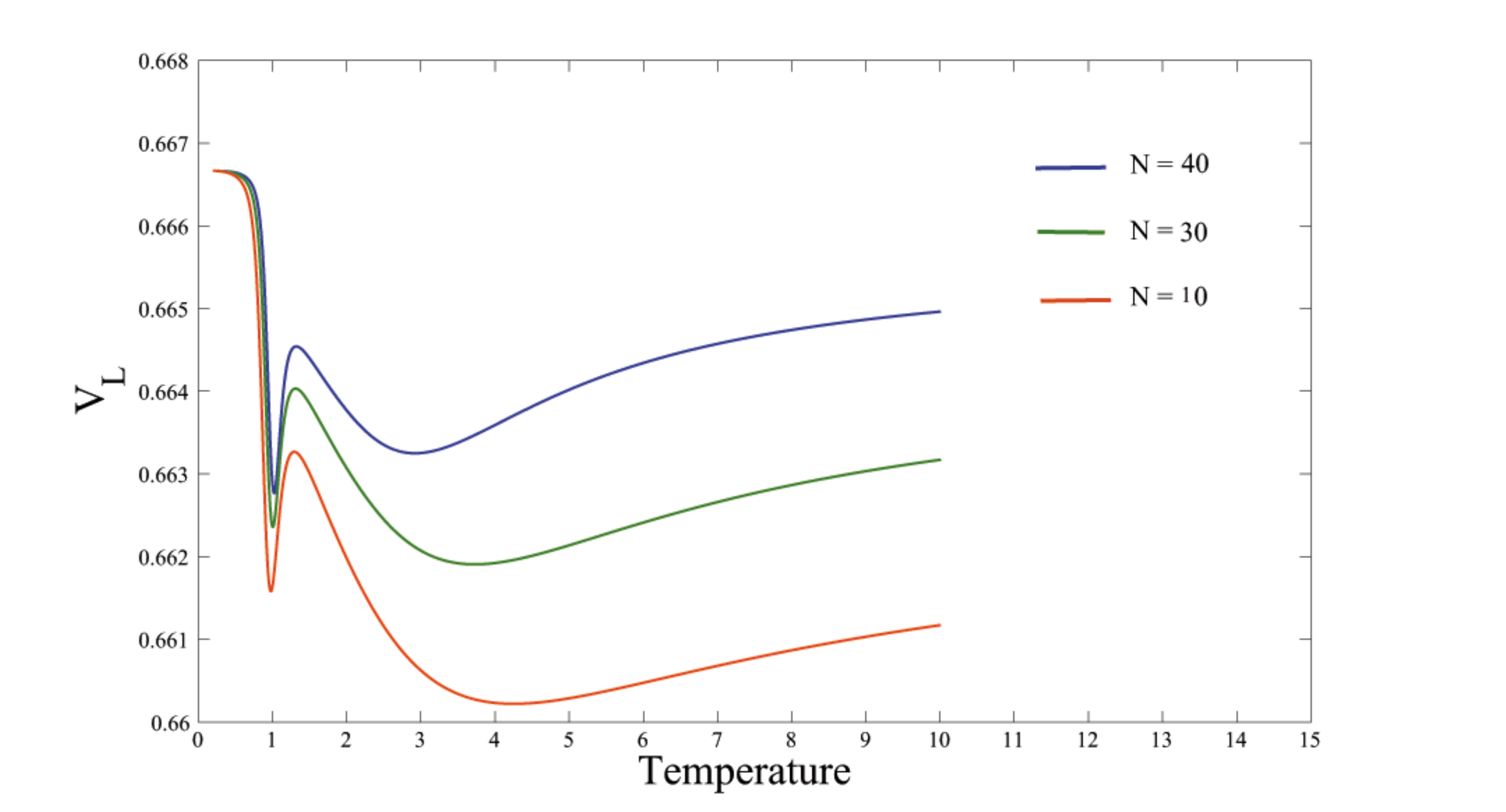}
\caption{{\small Binder's fourth cumulant {\it versus}
temperature for polymer
confined in double wall; the
transition is first order.}}
\end{center}
\end{figure}

%% file: chapters/conclusions.tex
\chapter{Conclusions}

In this report we have investigated the phase behaviour of a linear
homopolymer. We have considered attractive interaction between
segments of the polymer. We have also investigated the polymer in
the presence of a single attracting wall and two walls that confine
the polymer.

We have employed the Bond Fluctuation model on a two dimensional
square lattice. Each monomer occupies four lattice sites. Self
avoidance implies that a lattice site can at best belong to one
monomer. The bond length can fluctuate between $2$ and $4$ lattice
units.

We have employed Wang Landau algorithm for characterizing the phase
transition in the bond fluctuating lattice polymer.
Wang-Landau algorithm simulates an entropic ensemble, which is 
unphysical. Un-weighting and re-weighting of the conformations of 
the entropic ensemble are required for extracting physical canonical
ensemble averages at desired temperatures. The converged density of
states which is responsible for flattening of the energy histogram
can be directly used for estimating microcanonical and canonical
entropies and free energies besides phenomenological Landau free 
energy and Binder's cumulant. We have also employed conventional 
Markov chain Monte Carlo simulation employing Metropolis algorithm
to depict typical conformations belonging to canonical ensemble
at the desired temperatures.  We have simulated 
 polymers with number of monomers ranging from $10$ to $50$.

The principle conclusions of our study are listed below:
\begin{itemize}
\item For a free standing polymer we find there are two transitions.
The one at a higher temperature corresponds to collapse transition.
The one at a lower temperature is crystallization transition. From
Landau free energy profiles and Binder's cumulant we find that both
the transitions are discontinuous.
\item When a wall is present, we find indications of
 two transitions, from results on heat capacity.
However we are unable to characterize the transition at higher
temperature. This, we think, is due to the small size of the 
polymer simulated. The longest polymer we have simulated 
contains $50$ monomers. Our current computational
facilities do not permit study of longer polymers. 

The transition at the
lower temperature is from adsorbed-extended  to
adsorbed-globular phase. This transition is also  first order as indicated by
free energy profile as well as Binder's cumulant.
\item In the presence of two confining walls, 
we are again able to characterize transition at lower temperature only. 
We find that this as collapse
transition and it is discontinuous.
\end{itemize}

There are several issues we have not 
addressed in this report due mainly to want 
 of computational resources. Some of these are listed 
below.
\begin{itemize}
\item There is indeed a need for simulating longer polymers 
 to characterise 
the  phase transitions. 
In fact one should study polymers of 
different lengths, employ finite size scaling and  
obtain equilibrium properties in the thermodynamic limit.

\item There is a need to carry out a more detailed study of 
polymers in the presence of a single  absorbing wall and
a pair of walls that confine the polymer. The relative strengths
of the two interactions - one between two segments of the polymer and the other between the polymer and the wall, should be varied
to investigate their influence on adsorption-desorption and 
coil-globule phase transitions.  

\item Another interesting study that can be taken up is
the transport of a polymer from one side to the other,
 through a hole in a membrane. Bond fluctuation model,
which can also correctly simulate the dynamical
 processes in a polymer  
 is ideally suited for investigating 
this problem.  
\end{itemize}

%% file: acknowledgement.tex
\newpage
\addcontentsline{toc}{section}{\numberline{}ACKNOWLEDGEMENTS}
\thispagestyle{empty}
\begin{center}
\textbf{\large ACKNOWLEDGEMENTS}
\end{center}
\vspace{15mm} 

We are thankful to {\bf Prof. V. S. S. Sastry} for his interest
 in this work and for numerous discussions. We are also thankful to {\bf Ch. Sandhya} for useful discussions.

It is a pleasure to acknowledge that a  good part of the
Monte Carlo simulations reported in this report
has been carried out at the
Centre for Modelling Simulation and Design (CMSD),
University of Hyderabad.